\documentclass{aa}  

\usepackage{graphicx}
\usepackage{txfonts}
\usepackage[]{hyperref}
\begin{document} 

   \title{Kinematic signatures of nuclear discs and bar-driven secular evolution in nearby galaxies of the MUSE TIMER project}
   \titlerunning{Kinematic signatures of nuclear discs and secular evolution}

   \author{Dimitri A. Gadotti\inst{1}\fnmsep\thanks{\email{dgadotti@eso.org}}\and
   Adrian Bittner\inst{1,2}\and
   Jes\'us Falc\'on-Barroso\inst{3,4}\and
   Jairo M\'endez-Abreu\inst{3,4}\and
   Taehyun Kim\inst{5,6}\and
   Francesca Fragkoudi\inst{7}\and
   Adriana de Lorenzo-C\'aceres\inst{3,4}\and
   Ryan Leaman\inst{8}\and
   Justus Neumann\inst{9}\and
   Miguel Querejeta\inst{10}\and
   Patricia S\'anchez-Bl\'azquez\inst{11,12}\and
   Marie Martig\inst{13}\and
   Ignacio Mart\'in-Navarro\inst{3,4}\and
   Isabel P\'erez\inst{14,15}\and
   Marja K. Seidel\inst{16}\and
   Glenn van de Ven\inst{17}}

\institute{European Southern Observatory, Karl-Schwarzschild-Str. 2, D-85748 Garching bei M\"unchen, Germany
\and Ludwig-Maximilians-Universit\"at, Professor-Huber-Platz 2, 80539 M\"unchen, Germany
\and Instituto de Astrof\'isica de Canarias, 38205 La Laguna, Tenerife, Spain
\and Departamento de Astrof\'isica, Universidad de La Laguna, 38206 La Laguna, Tenerife, Spain
\and Department of Astronomy and Atmospheric Sciences, Kyungpook National University, Daegu 702-701, Korea
\and Korea Astronomy and Space Science Institute, Daejeon 305-348, Korea
\and Max-Planck-Institut f\"ur Astrophysik, Karl-Schwarzschild-Str. 1, D-85748 Garching bei M\"unchen, Germany
\and Max-Planck-Institut f\"ur Astronomie, K\"onigstuhl 17, D-69117 Heidelberg, Germany
\and Institute of Cosmology and Gravitation, University of Portsmouth, Burnaby Road, Portsmouth PO1 3FX, UK
\and Observatorio Astron\'omico Nacional, C/Alfonso XII 3, E-28014 Madrid, Spain
\and IPARCOS, Facultad de Ciencias F\'isicas, Universidad Complutense de Madrid, E-28040 Madrid, Spain
\and Departamento de F\'isica de la Tierra y Astrof\'isica, Universidad Complutense de Madrid, E-28040 Madrid, Spain
\and Astrophysics Research Institute, Liverpool John Moores University, 146 Brownlow Hill, L3 5RF Liverpool, UK
\and Departamento de F\'isica Te\'orica y del Cosmos, Universidad de Granada, Facultad de Ciencias, E-18071 Granada, Spain
\and Instituto Universitario Carlos I de F\'isica Te\'orica y Computacional, Universidad de Granada, E-18071 Granada, Spain
\and Caltech-IPAC MC 314-6, 1200 E California Blvd, Pasadena CA 91125, USA
\and Department of Astrophysics, University of Vienna, T\"urkenschanzstr. 17, A-1180 Wien, Austria          
             }

   \authorrunning{D. A. Gadotti \& The TIMER Team}


   \abstract{The central regions of disc galaxies hold clues to the processes that dominate their formation and evolution. To exploit this, the TIMER project has obtained high signal-to-noise and spatial resolution integral-field spectroscopy data of the inner few kpc of 21 nearby massive barred galaxies, which allows studies of the stellar kinematics in their central regions with unprecedented spatial resolution. We confirm theoretical predictions of the effects of bars on stellar kinematics, and identify box/peanuts through kinematic signatures in mildly and moderately inclined galaxies, finding a lower limit to the fraction of massive barred galaxies with box/peanuts at $\sim62\%$. Further, we provide kinematic evidence of the connection between barlenses, box/peanuts and bars. We establish the presence of nuclear discs in 19 galaxies and show that their kinematics are characterised by near-circular orbits with low pressure support, and are fully consistent with the bar-driven secular evolution picture for their formation. In fact, we show that these nuclear discs have, in the region where they dominate, larger rotational support than the underlying main galaxy disc. In addition, we define a kinematic radius for the nuclear discs and show that it relates to bar radius, ellipticity and strength, and bar-to-total ratio. Comparing our results with photometric studies of galaxy bulges, we find that careful, state-of-the-art galaxy image decompositions are generally able to discern nuclear discs from classical bulges, if the images employed have enough physical spatial resolution.
In fact, we show that nuclear discs are typically identified in such image decompositions as photometric bulges with (near-)exponential profiles.
However, we find that the presence of composite bulges (galaxies hosting {\em both} a classical bulge and a nuclear disc) can often be unnoticed in studies based on photometry alone, and suggest a more stringent threshold to the S\'ersic index to identify galaxies with pure classical bulges.
}

   \maketitle

\section{Introduction}

A large number of observational and theoretical studies have been providing mounting evidence of the important physical processes driven by bars in massive disc galaxies. For example, in the area of the disc within the bar radius, stars are continuously trapped by the bar and gas is funnelled to the central region, where new stellar structures such as nuclear discs, inner bars, nuclear rings and nuclear spiral arms are thus built \citep[see][]{SanTub80,SimSuSch80,Kor82,Pre83,LouGer88,ShlFraBeg89,Ath92b,KnaBecHel95,PinStoTeu95,SakOkuIsh99,Gaddos01,KnaPerLai02,SheVogReg05,AllKnaPel06,Woz07,CoeGad11,EllNaiPat2011,deLVazAgu12,deLFalVaz13,AthMacRod13,ColDebErw14,FraAthBos16,KimGadAth16,KruLinBam18,SeoKimKwa19,DonMarJam19}. There is increasing evidence that these processes start to play a major role at redshifts $z\sim1-2$ in the most massive disc galaxies, and at $z\sim0$ these processes take part in the evolution of about 2/3 of disc galaxies \citep[see, e.g.,][]{EskFroPog00,MenSheSch07,SheElmElm08,SheMelElm12,KraBouMar12,SimMelLin14,MelMasLin14,GadSeiSan15,PerMarRui17}.

A major process in the evolution of barred galaxies is the buckling (or bulging) of the inner part of the bar, which grows vertically from the disc plane, creating the so-called box/peanut/X-shaped bulges, such as the one hosted by our own Milky Way \citep[e.g.,][]{ComSan81,deSDos87,Sha87,ComDebFri90,BinOrtSta91,KuiMer95,MerKui99,BurFre99,LueDetPoh00,ChuBur04,BurAth05,BurAroAth06,MarShlHel06,ErwDeb17,FraDiMHay17,FraDiMHay18,KruErwDeb19}. Naturally, these box/peanuts are easier to identify in galaxies close to or at an edge-on projection, but a number of studies have put effort in producing diagnostics able to uncover box/peanuts in inclined galaxies \citep[e.g.,][from photometry]{ErwDeb13}, or even in face-on galaxies \citep[e.g.,][from stellar kinematics]{DebCarMay05,MenCorDeb08,Lok19,FraGraPak19}.

More recently, it has been shown that the inner parts of bars not only expand vertically from the plane of the disc, but also radially, parallel to the disc plane, away from the bar major axis \citep[see][]{LauSalBut05,LauSalBut07,LauSalBut11,AthLauSal15,LauSal16}. When bars are seen {\em face-on}, this additional structure appears as a less eccentric central component, wider and shorter than the main body of the bar \citep[see Fig.\,8 in][]{GonGad16}. This component was named `barlens', but it is important to note that the terms `barlens' and `box/peanut' refer to different projections of the {\em same} stellar structure, namely the inner part of the bar, which expands due to dynamical processes as the bar evolves. It is also very important to highlight that although box/peanuts have a substantial vertical component (and despite their bulge-like morphology), the formation of this bar structure is a process internal to the bar, unlike and unrelated to the violent formation of kinematically hot spheroids: box/peanuts and barlenses are simply the same inner part of the bar seen at different projections.

Photometrically, all these bar-built structures produce an excess of light in the central region of the galaxy on top of the inward extrapolation of the exponential profile of the main disc. They would thus satisfy one of the criteria to identify bulges, even though they are unrelated to the classical picture of a bulge as a merger-built, kinematically hot spheroid with stars in radial orbits. To distinguish bar-built structures from classical bulges the term `{\em pseudo-bulge}' is commonly used \citep[e.g.,][amongst many others]{KorKen04,Gad09b,MenDebCor14,FisDro16,NeuWisCho17}. However, as described above, bar-built structures also come in two different flavours according to their physical nature. Nuclear discs, inner bars, nuclear rings and nuclear spiral arms are all thought to be built mostly from gas brought to the central region where star formation takes place, whereas box/peanuts (and barlenses) are composed by stars that gradually move from one bar orbital family to another. Therefore, using the term `pseudo-bulge' collectively to describe all these structures, these two different flavours of bar-built structures, can be misleading. To minimise confusion, \cite{Ath05b} introduced the term `{\em disc-like bulge}' (which later produced variations such as `discy pseudo-bulge' and `discy bulge') to distinguish nuclear discs and related structures from box/peanuts.

Photometric `disc-like bulges' typically have exponential surface density profiles, and therefore seem to be simply nuclear discs that in contrast to the main galaxy disc are built by bar-driven processes that develop in the main galaxy disc \citep[e.g.,][]{FalPelEms04,FalBacBur06}. These nuclear discs often host inner bars, nuclear spiral arms and nuclear rings. The formation of inner bars and nuclear spiral arms appear to be simply scaled-down versions of the formation of a bar and spiral arms in the main galaxy disc \citep{MendeLGad19,deLSanMen19}, confirming the theoretical work by \citet{Woz15} and \citet{DuSheDeb15}, who also found that nuclear discs and inner bars can be long-lived. Crucially, however, excluding the nuclear disc, the central regions of disc galaxies are dynamically hotter than the regions where the main disc dominates, and thus these processes still need to be better understood.

On the other hand, the formation of nuclear rings is more strongly connected to the properties of the main bar. In fact, a number of studies using hydrodynamical simulations suggest that nuclear rings form close to the Inner Lindblad Resonance (ILR) of the main bar, or in the region where the x$_2$ orbital family of the main bar dominates \citep[e.g.,][and references therein]{KimSeoSto12,LiSheKim15,SorSobFra18}. These theoretical studies show that the radius of the nuclear ring depends on the size and other properties of the main bar. Observationally, \citet{Kna05} and \citet{ComKnaBec10} find indeed corroborating evidence. The interplay and evolutionary connection between nuclear discs and nuclear rings are still unclear, but \citet{ColDebErw14} argue that the nuclear ring is part of the nuclear disc, namely, its outer rim.

It is important to point out that the build-up of nuclear discs in processes unrelated to bars, as via the accretion of external gas onto unbarred galaxies and mergers, has also been explored in numerical simulations \citep[see][]{MayKazEsc08,ChaMayTey13}. This may explain the presence of nuclear discs in unbarred galaxies, but we note that the nuclear discs produced in these unbarred merger simulations are an order of magnitude less extended than those built by bars. In a different study, \citet{EliGonBal11} argued that their idealised, collisionless simulations of minor mergers do create nuclear discs, sometimes without forming a noticeable bar, provided that the satellite galaxy is not too dense and that the primary galaxy has a massive classical bulge before the merger. We will discuss these simulations further below and conclude that they do not reproduce the observed properties of nuclear discs. Another possibility is that nuclear discs would form via gas inflow due to oval distortions in the main disc. This is akin to the bar-driven formation process, only with a weaker non-axisymmetric component. To date, it is still unclear how rare are nuclear discs in unbarred galaxies \citep[but see][]{ComKnaBec10}.

In this paper, we take advantage of the integral-field spectroscopy data from the {\bf T}ime {\bf I}nference with {\bf M}USE in {\bf E}xtragalactic {\bf R}ings (TIMER) project to study the kinematic properties of barred galaxies and nuclear stellar structures with unprecedented spatial resolution. We also take advantage of the vast ancillary data for the TIMER sample to {\bf (i)}, provide evidence that bar-driven processes appear to be the main mechanism responsible for the formation of nuclear discs and related structures; {\bf (ii)}, demonstrate more rigorously the connection between nuclear discs, detected via their kinematic properties, and central exponential components, found via photometric decompositions, and {\bf (iii)}, provide further evidence that nuclear rings are the outer rims of nuclear discs (see also our accompanying paper, Bittner et al. 2020, subm.).

This paper is organised as follows. In the next section we introduce the TIMER project and summarise the main aspects concerning sample selection, observations and data reduction. In Sect.\,\ref{sec:kin}, we discuss the derivation of the parameters characterising the stellar kinematics, and describe the detection of kinematic signatures of nuclear discs, bars and box/peanuts, as well as the observed kinematic properties of galaxies showing barlenses. We connect the kinematic and photometric properties of nuclear discs in Sect.\,\ref{sec:decomps}, and discuss the origin of nuclear discs in Sect.\,\ref{sec:origins}. In Sect.\,\ref{sec:conc} we present a more general discussion and summarise our main conclusions.

\section{The TIMER Project}

The TIMER project is a survey with the Very Large Telescope (VLT) MUSE integral-field spectrograph \citep{BacAccAdj10} of 24 nearby barred galaxies\footnote{Observations are still lacking for three galaxies.} with prominent central structures, such as nuclear rings, nuclear spiral arms, inner bars and nuclear discs \citep[see][hereafter Paper I]{GadSanFal19}. One of the projects' main goals is to study the star formation histories of such structures to infer the cosmic epoch of the formation of the bar and the dynamical settling of the main disc of the host galaxy. The methodology was demonstrated with a pilot study of NGC\,4371 \citep{GadSeiSan15}.

The TIMER sample was drawn from the Spitzer Survey of Stellar Structure in Galaxies (S$^4$G, \citealt{shereghin10}), which includes only galaxies at distances below $40\,\rm{Mpc}$, brighter than 15.5\,B-mag, and larger than $1\arcmin$. The TIMER galaxies are all barred, with stellar masses above $10^{10}\,{\rm M}_\odot$, inclinations below $\approx60^\circ$, and nuclear stellar structures. The presence of the bar and nuclear structures was assessed from the morphological classifications of \citet{ButSheAth15}, who used the S$^4$G images for their work. The TIMER sample is thus biased towards conspicuous bars and nuclear structures, and it is important to keep this in mind when considering the results discussed in this paper.

Most of the observations were performed during ESO Period 97 (April to September 2016) with a typical seeing of $0.8-0.9\arcsec$, mean spectral resolution of $2.65\,\AA$ (FWHM), and spectral coverage from $4750\,\AA$ to $9350\,\AA$. MUSE covers an almost square $1\arcmin\times1\arcmin$ field of view with a contiguous sampling of $0.2\arcsec\times0.2\arcsec$, which corresponds to a massive dataset of about 90\,000 spectra per pointing. The spectral sampling is $1.25\,\AA$ per pixel. The total integration time on source for each galaxy was typically 3\,840\,s.

The MUSE pipeline (version 1.6) was used to reduce the dataset \citep{WeiStrUrr12}, correcting for bias and applying flat-fielding and illumination corrections, as well as wavelength calibration. The exposures were flux-calibrated through the observation of a spectrophotometric standard star, which was also used to remove telluric features. Dedicated empty-sky exposures and a PCA methodology were employed to remove signatures from the sky background. Finally, the exposures were also finely registered astrometrically, so that the point spread function of the combined cube is similar to that in individual exposures. Typically, the averaged signal to noise ratio (S/N) per spectral and spatial pixel at the central spaxels of our fully-reduced data cubes is approximately 100. We refer the reader to Paper I for further details on the sample selection, observations and data reduction.

Within the TIMER collaboration, \citet{MendeLGad19} presented the discovery of the first box/peanut found in an inner bar, and \citet{deLSanMen19} found evidence indicating that inner bars are long-lived. \citet{NeuFraPer20} discovered variations in the stellar population properties across galaxy bars that were predicted in idealised simulations and can be reproduced by state-of-the-art cosmological simulations. Furthermore, \citet{LeaFraQue19}, adding data from ALMA, showed a spectacular example of the effects of bars on the interstellar medium, central star formation and stellar feedback.

\section{Stellar Kinematics}
\label{sec:kin}
\subsection{Methodology}

\begin{table*}[t]
	\centering
	\caption{Systemic velocities and central velocity dispersions. Column (1) gives the galaxy designation and column (2) shows their effective radii $r_{\rm e}$ as derived in \citet{MunSheReg15}. Columns (3) and (4) show, respectively, the tabulated values of systemic radial velocity and the corresponding errors as presented in the Lyon Extragalactic Data Archive (LEDA; \url{http://leda.univ-lyon1.fr/}), whereas our own measurements are presented in columns (5) and (6). In column (7) we show our measurements of the central velocity dispersions as measured within an aperture of $r_{\rm e}/8$, with the corresponding errors shown in column (8). Column (9) shows again our measurements of the central velocity dispersions but now adapted to follow the same aperture corrections as in LEDA. Finally, columns (10) and (11) show the LEDA values of central velocity dispersions and their errors, respectively. See text for further details.}
	\begin{tabular}{lcccccccccc}
		\hline
Galaxy & $r_{\rm e}$ & $v_{\rm LEDA}$ & err($v_{\rm LEDA}$) & $v$ & err($v$) & $\sigma_{r_{\rm e}/8}$ & err($\sigma$) & $\sigma_{\rm corr}$ & $\sigma_{\rm LEDA}$ & err($\sigma_{\rm LEDA}$)\\
\omit & \arcsec & km s$^{-1}$ & km s$^{-1}$ & km s$^{-1}$ & km s$^{-1}$ & km s$^{-1}$ & km s$^{-1}$ & km s$^{-1}$ & km s$^{-1}$ & km s$^{-1}$\\
(1) & (2) & (3) & (4) & (5) & (6) & (7) & (8) & (9) & (10) & (11)\\
		\hline
IC\,1438   &  11.5 & 2616 &  5 & 2618 & 2 & 101 & 2  & 99  & \omit & \omit \\
NGC\,613  &  51.6 & 1484 &  3 & 1506 & 2 & 125 & 3  & 128  & 122   & 18 \\
NGC\,1097  &  58.8 & 1269 &  7 & 1274 & 2 & 196 & 3  & 198  & \omit & \omit \\
NGC\,1291  &  60.7 &  837 &  8 & 858  & 6 & 168 & 7  & 166  & 165   & 10 \\
NGC\,1300  &  71.9 & 1578 &  2 & 1579 & 5 & 100 & 8  & 102  & 218   & 29 \\
NGC\,1365  &  67.2 & 1638 &  4 & 1633 & 7 & 157 & 10  & 160  & 141   & 19 \\   
NGC\,1433  &  67.5 & 1076 &  2 & 1086 & 4 &  95 & 13  &  94  & \omit & \omit \\  
NGC\,3351  &  64.5 &  778 &  1 & 791  & 4 & 98 & 8  & 98  & 116   & 10 \\
NGC\,4303  &  56.0 & 1567 &  2 & 1577 & 3 &  79 & 8  &  79  &  95   &  8 \\ 
NGC\,4371  &  33.5 &  913 &  4 & 972  & 8 & 132 & 12 & 131  & 129   &  2 \\
NGC\,4643  &  24.2 & 1328 &  2 & 1341 & 2 & 133 & 3  & 131  & 147   &  3 \\
NGC\,4981  &  29.9 & 1678 &  2 & 1688 & 2 &  95 & 3  & 95  & \omit & \omit \\
NGC\,4984  &  18.1 & 1215 & 10 & 1271 & 2 & 113 & 3  & 109  & \omit & \omit \\
NGC\,5236  & 145.9 &  508 &  2 & 527  & 2 &  75 & 6  &  75  & \omit & \omit \\   
NGC\,5248  &  45.7 & 1152 &  2 & 1168 & 4 &  91 & 7  &  90  &  99   &  9 \\
NGC\,5728  &  28.8 & 2788 &  4 & 2773 & 5 & 160 & 7  & 161  & 197   & 14 \\
NGC\,5850  &  49.4 & 2546 &  3 & 2558 & 2 & 123 & 3  & 128  & 140   &  4 \\
NGC\,6902  &  24.0 & 2793 &  4 & 2799 & 2 & 119 & 2  & 120  & \omit & \omit \\
NGC\,7140  &  38.1 & 2978 &  4 & 2982 & 2 &  98 & 3  &  100  & \omit & \omit \\
NGC\,7552  &  12.2 & 1609 &  5 & 1612 & 3 &  84 & 6  &  81  &  98   & 19 \\
NGC\,7755  &  32.2 & 2960 &  3 & 2952 & 2 & 114 & 3  & 116  & \omit & \omit \\
		\hline
	\end{tabular}
\label{tab:velos}
\end{table*}

The technical details behind our derivation of the stellar kinematics in TIMER were extensively presented in Paper I and \citet{GadSeiSan15}. Here we simply summarise the essential aspects. The stellar line of sight velocity distributions (LOSVDs) were parameterised as Gauss-Hermite functions with four parameters (following \citealt{vanFra93}): velocity ($v$), velocity dispersion ($\sigma$), and the h$_3$ and h$_4$ higher-order moments. The h$_3$ and h$_4$ parameters can be used to examine in further detail the orbital structure of the stellar systems in question. For instance, near-circular orbits, with a distribution of $v/\sigma$, produce h$_3$ values that are anti-correlated with $v$, whereas elongated orbits result in a {\em correlation} between $v$ and h$_3$. High values of h$_4$ suggest the superposition of structures with different LOSVDs \citep[see, e.g.,][and references therein]{BenSagGer94}. For the derivation of 2D maps of $v$, $\sigma$, h$_3$ and h$_4$, the spectra from the data cube of each galaxy were spatially binned to ensure a minimum S/N of approximately 40 per spectral pixel. This was done using the Voronoi binning technique as described in \citet{CapCop03}. To derive systemic velocities and central velocity dispersions, we combined all spectra within a circular aperture. The spectra from each Voronoi bin or aperture were fitted with the {\tt pPXF} code of \citet[][see also \citealt{CapEms04} and \citealt{CapEmsKra11}]{Cap17}. The entire ensemble of procedures was performed using the GIST\footnote{Available at \url{ascl.net/1907.025}.} pipeline \citep[which employs {\tt pPXF};][]{BitFalNed19}.

The only fundamental difference between the kinematic maps presented in Paper I (where we presented the maps corresponding to NGC\,1097 and NGC\,4643 only) and the ones derived here is the rest-frame wavelength range employed. In Paper I we opted to use the range between 4750 and 5500\,\AA\ after checking that similar results are obtained when the full MUSE wavelength range is employed. That approach has the advantage of necessitating less computation time. In addition, it avoids complications arising from bright emission lines, particularly if they are not masked, but modelled. Some emission lines are nevertheless present in the restricted wavelength range, and those were masked before the fitting procedure. However, in this paper, we used  an extended wavelength range (4800 to 8950\,\AA) that is close to the full wavelength range provided by MUSE.
A comparison between the maps derived employing these two different approaches shows that the results are qualitatively identical and quantitatively very similar. However, the maps derived using the extended wavelength range reveal some results in a more enhanced fashion. For example, regions with elevated absolute values of h$_3$ and h$_4$ appear more conspicuously and sharply defined. This seems to be related to the fact that the instrumental spectral resolution increases with wavelength, and thus lower values of velocity dispersion, as well as deviations from a pure Gaussian function for the LOSVDs (i.e., values of h$_3$ and h$_4$ different from zero), can be more robustly measured. Further, the signal is boosted by the addition of several absorption features, including the calcium triplet. When using the extended wavelength range, for two galaxies (NGC\,7140 and NGC\,7755) small differences in the continuum shape between the galaxy spectra and the input library of synthetic models required more than a multiplicative low-order Legendre polynomial included in the fit (as done in Paper I). We thus included in the fit an 8th-order multiplicative polynomial plus a 4th-order additive Legendre polynomial to account for such small differences. In addition, in this paper, we also employed a non-constant line spread function (LSF) to account for the wavelength dependence of the instrumental spectral resolution. We adopted the LSF derived in \citet[][their equation 8]{BacConMar17}.

\begin{figure*}[t]
\begin{center}
	\includegraphics[width=1.7\columnwidth]{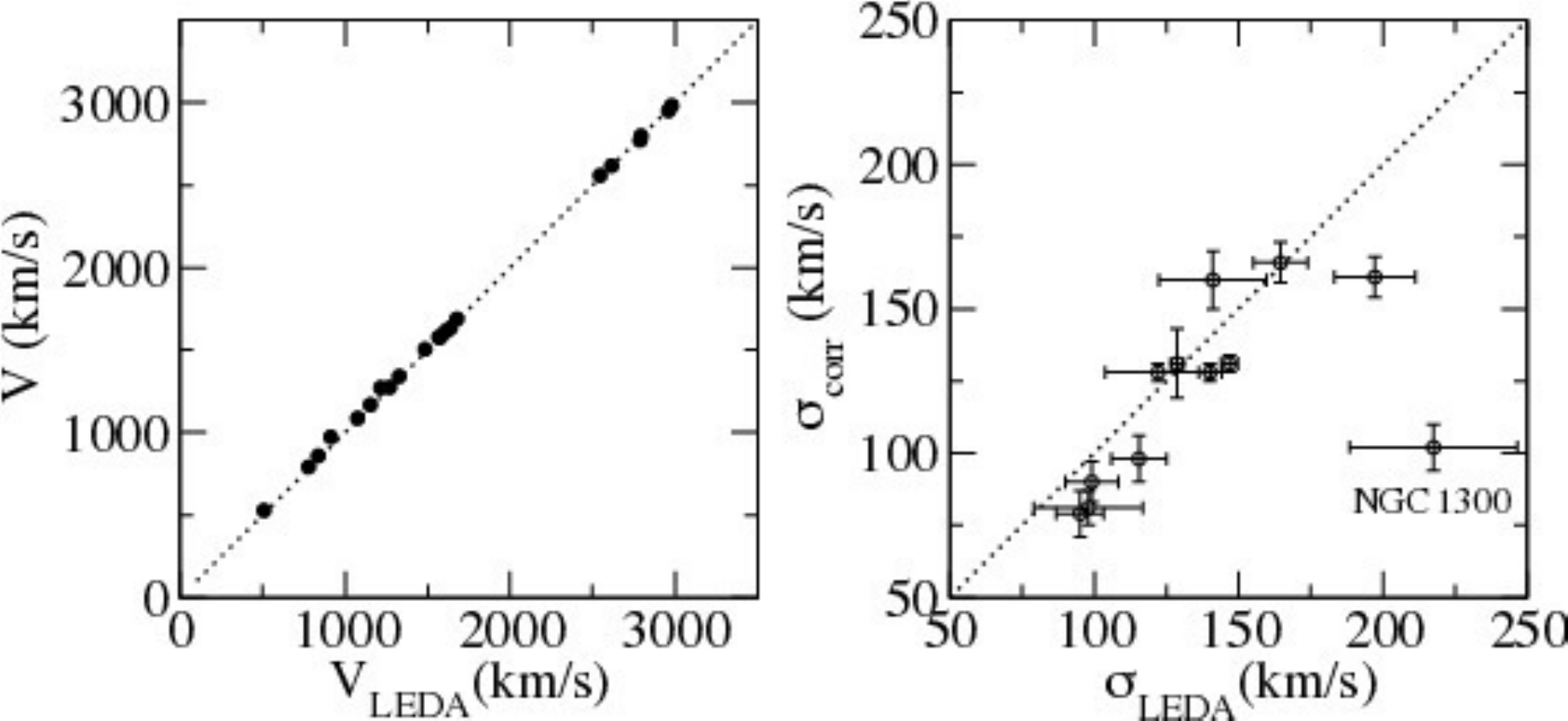}
\end{center}
    \caption{{\em Left}: TIMER measurements of systemic velocity against the values available at LEDA. {\em Right}: TIMER measurements of central velocity dispersion against the LEDA values. The TIMER velocity dispersions were adapted to follow the same aperture corrections of LEDA, and the error bars in both panels correspond to $1\sigma$ errors. NGC 1300 is the only significant outlier but this is due to a mistake in LEDA (see text for further details; the LEDA team has been informed).}
    \label{fig:velos}
\end{figure*}

\subsection{Systemic velocities and central velocity dispersions}

For each galaxy, the heliocentric systemic radial velocity and central velocity dispersion were measured by combining all central spectra within a circular aperture corresponding to one eighth of the galaxy effective radius, $r_{\rm e}/8$. For $r_{\rm e}$ we use the measurements derived by \citet{MunSheReg15} using S$^4$G $3.6\mu$m growth curves. These measurements are presented in Table \ref{tab:velos}, along with error estimates and the corresponding values available at the Lyon Extragalactic Data Archive (LEDA). Our error estimates were derived from 30 Monte Carlo realisations. Note that to properly compare the measurements of central velocity dispersion we adapted our measurements to follow the same aperture corrections of LEDA, which correspond to an aperture of 0.595 h$^{-1}$\,kpc.

In Fig.\,\ref{fig:velos} we compare our measurements with the LEDA values. The agreement in the measurements of systemic velocity is excellent. Concerning the central velocity dispersion, although there are no available measurements in LEDA for nine galaxies, the agreement is very good, with only one significant outlier, NGC\,1300. The LEDA values for the central velocity dispersion in NGC\,1300 are a factor two larger than ours. Curiously, most of the LEDA measurements are quoted as from \citet{DavBurDre87} but an inspection of the article shows that NGC\,1300 is not part of the sample of galaxies in that study. In fact, the measurements by Davies et al. correspond to NGC\,2300. Another measurement quoted in LEDA is from an unpublished study that reports $145\pm22$\,km\,s$^{-1}$, a value closer to our measurement ($102\pm8$\,km\,s$^{-1}$). Part of the differences between our measurements and the values quoted in LEDA can be due to differences in seeing and pixel size amongst the different observations.

\subsection{Kinematic maps}

\begin{figure*}
\begin{center}
	\includegraphics[clip=true, trim=20 20 20 20, width=0.85\columnwidth]{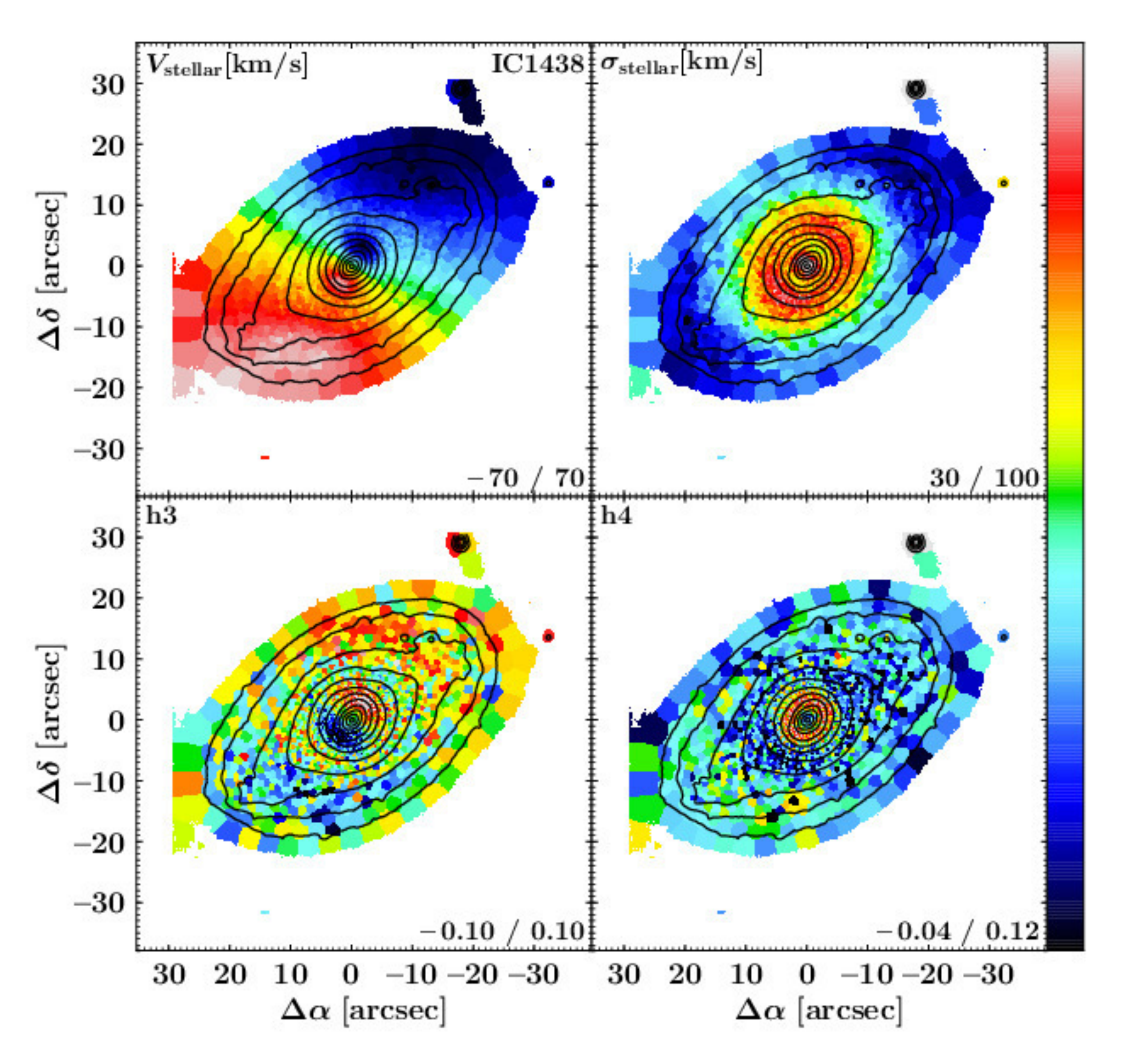}\hskip0.5cm
	\includegraphics[clip=true, trim=20 20 20 20, width=0.85\columnwidth]{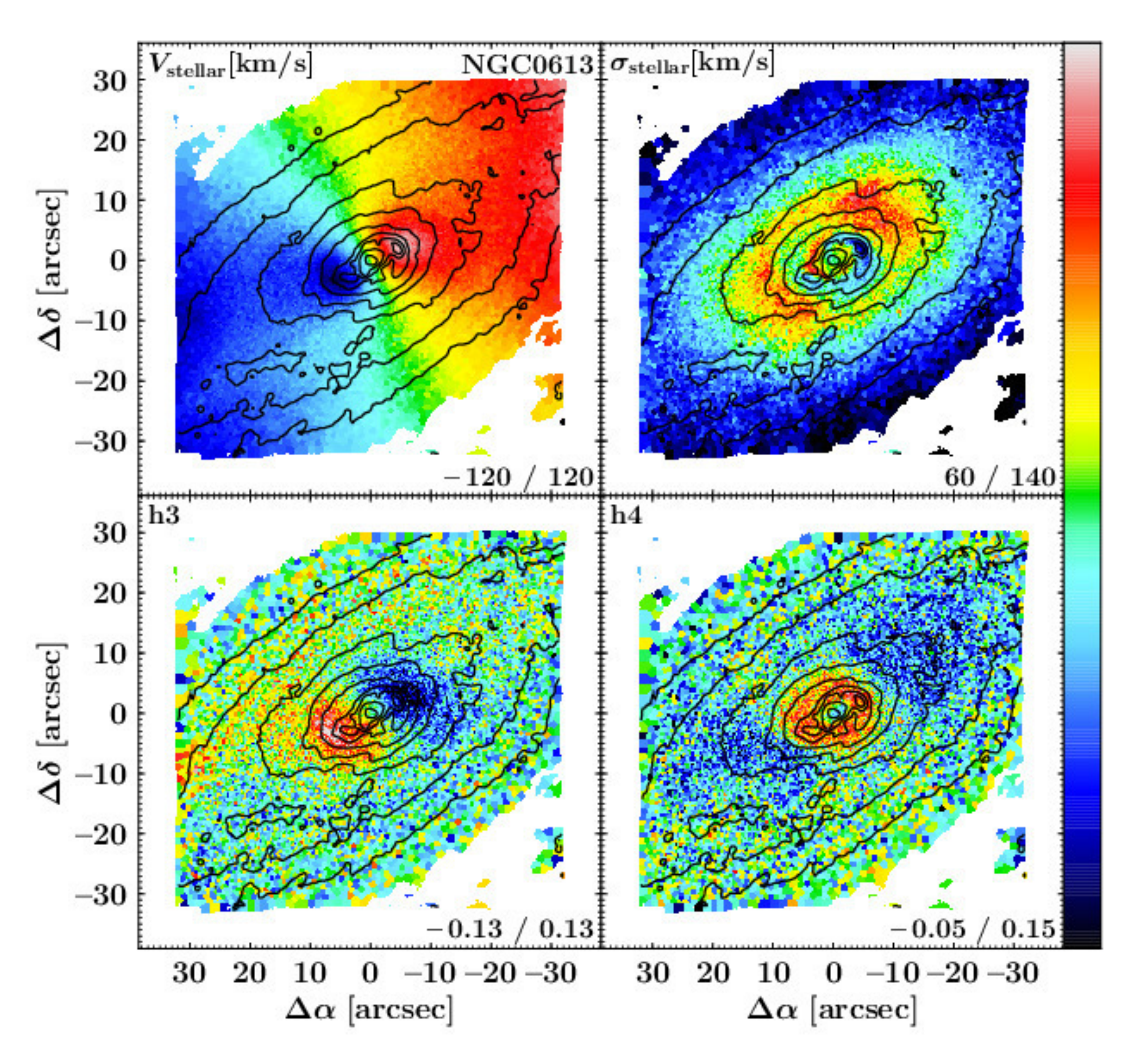}\vskip0.2cm
	\includegraphics[clip=true, trim=20 20 20 20, width=0.85\columnwidth]{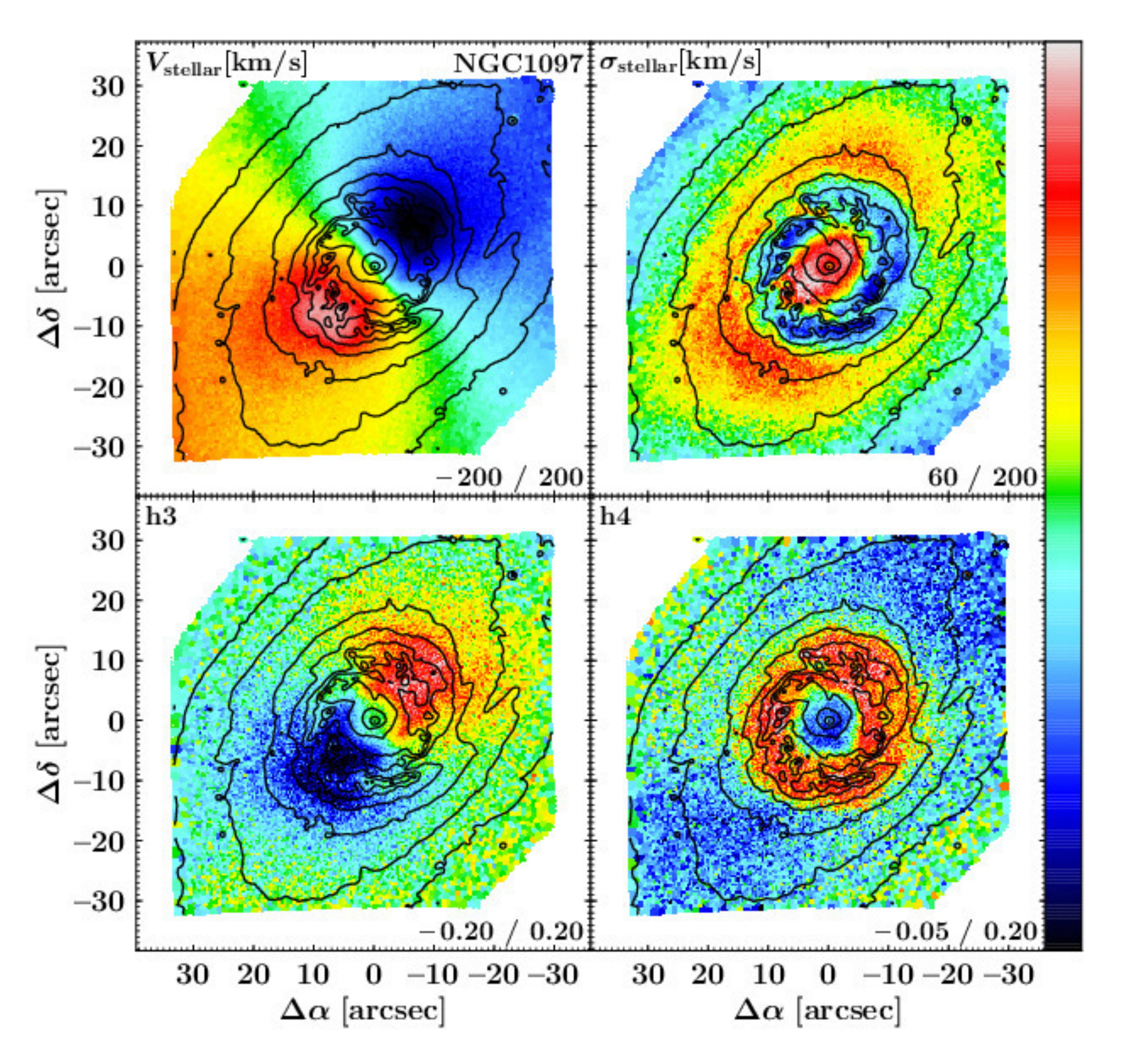}\hskip0.8cm
	\includegraphics[clip=true, trim=20 20 20 20, width=0.8\columnwidth]{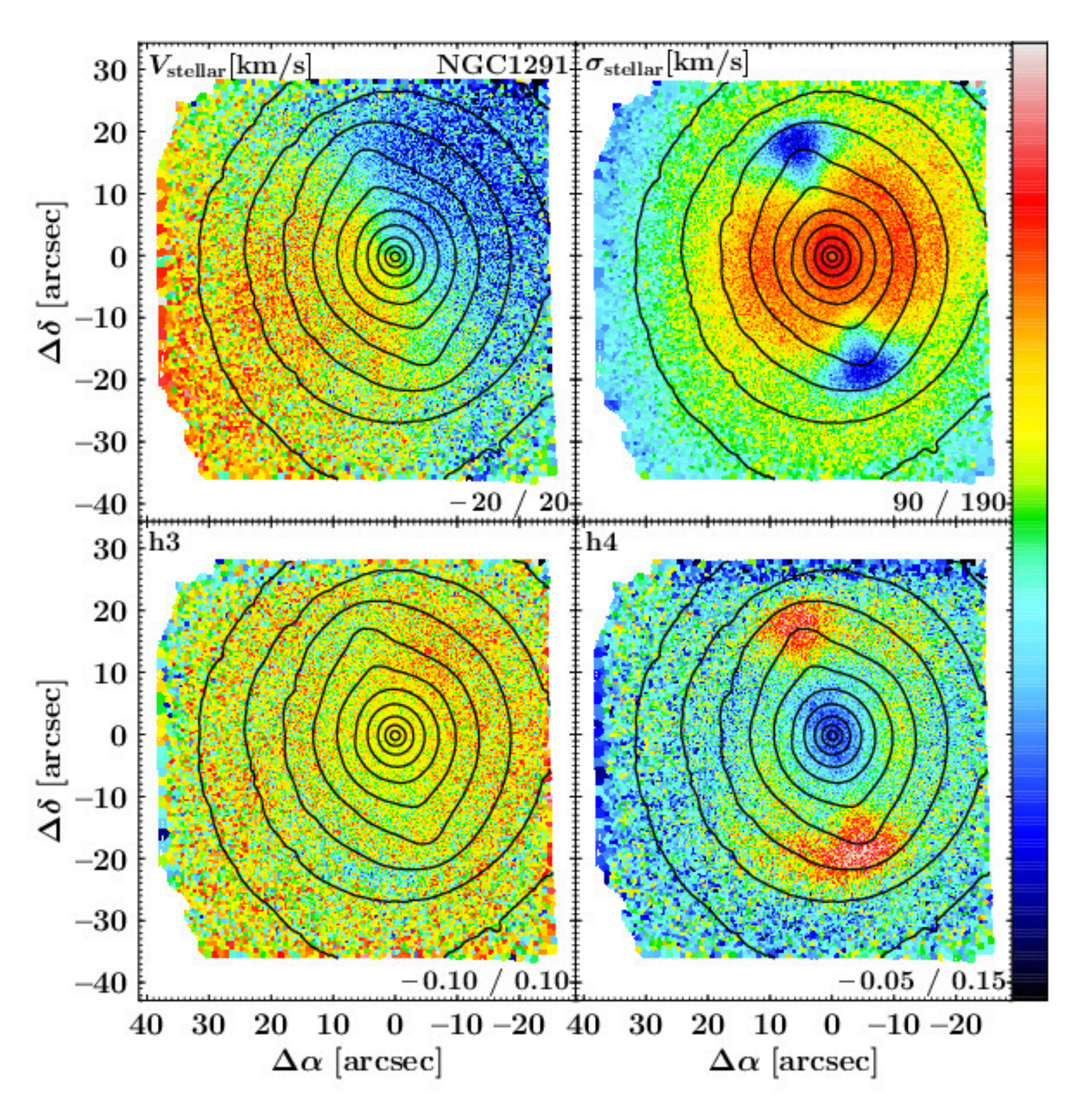}\vskip0.2cm
	\includegraphics[clip=true, trim=20 20 20 20, width=0.85\columnwidth]{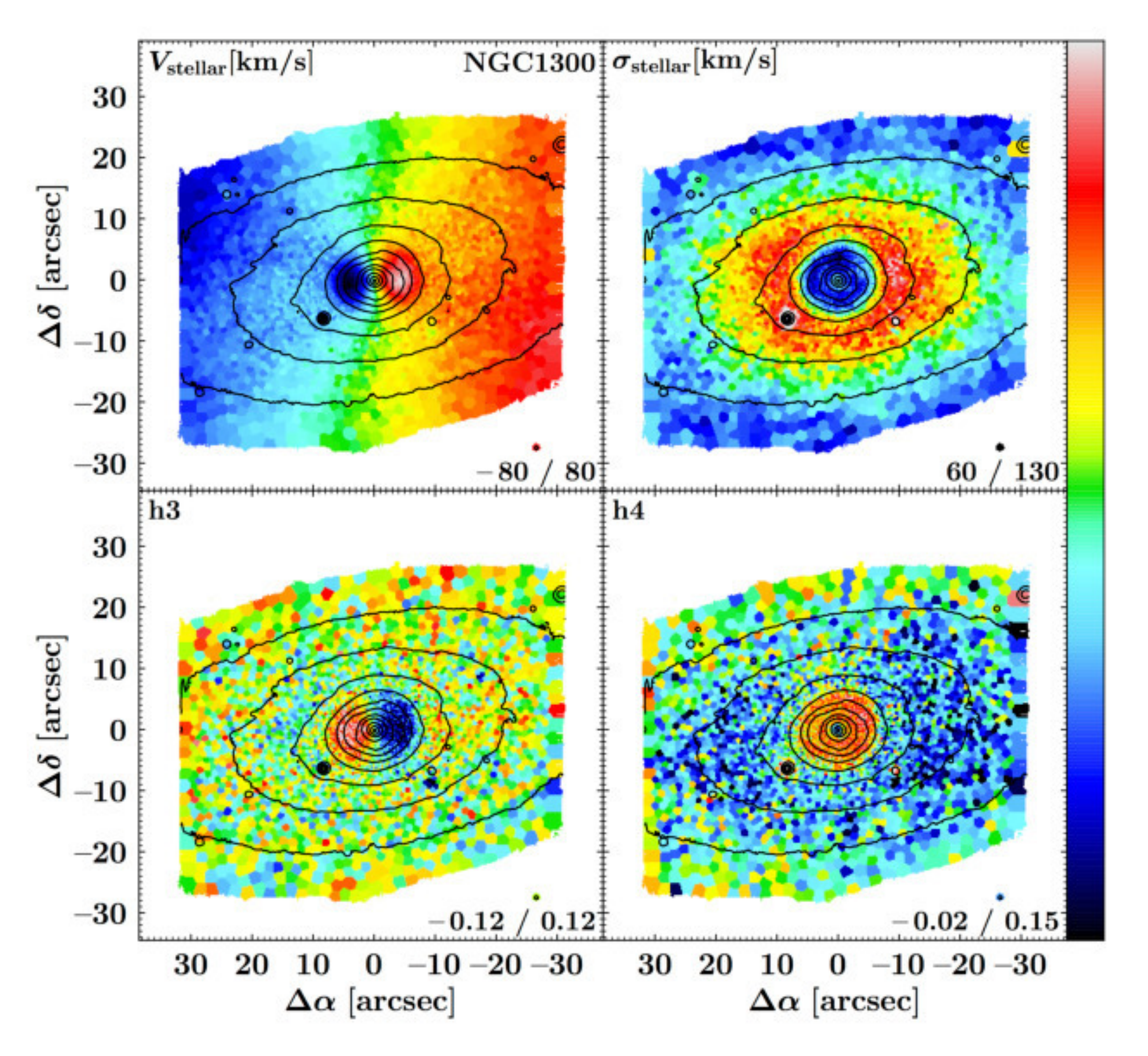}\hskip0.5cm
	\includegraphics[clip=true, trim=20 20 20 20, width=0.825\columnwidth]{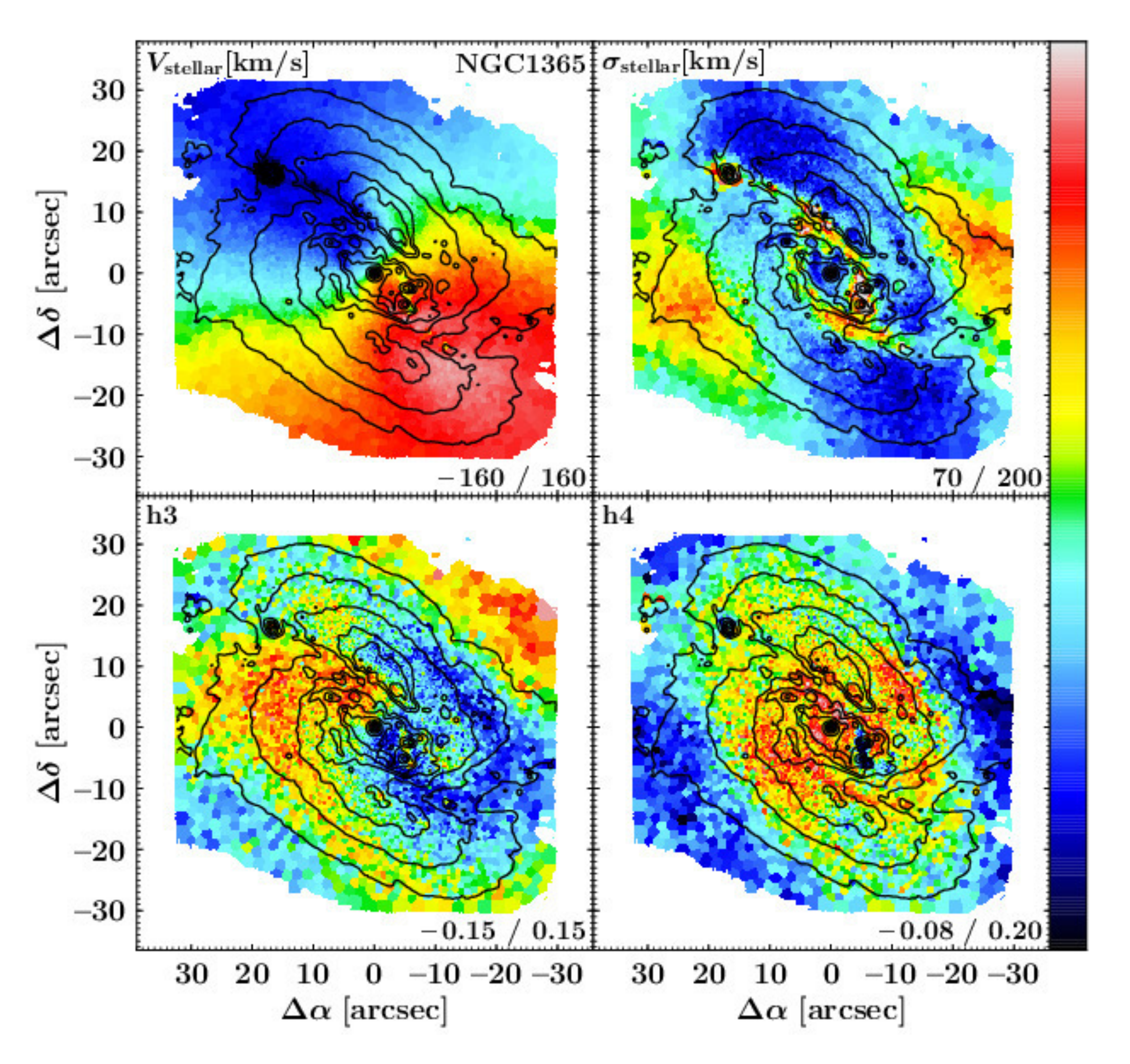}
\end{center}
\vskip -0.25cm
    \caption{Radial velocity, velocity dispersion, h$_3$ and h$_4$ maps for the stellar components in the TIMER galaxies, as indicated. The plotted range of the parameter measured is indicated at the bottom right corner of each panel. For radial velocity and velocity dispersion these are given in km\,s$^{-1}$. The isophotes shown are derived from the MUSE data cube reconstructed intensities and are equally spaced in steps of about 0.5 magnitudes. On average, $1\arcsec$ corresponds to $\approx100\,\rm{pc}$. North is up, east to the left. See Appendix\,\ref{app:maps} for the maps corresponding to the rest of the galaxies in TIMER.}
    \label{fig:maps}
\end{figure*}


In this section, we present and discuss signatures found in the kinematic maps derived with GIST that reveal the presence of different structural components. Here we focus on the general trends and leave to Appendix\,\ref{app:maps} a discussion on the more complex cases. The high-level data products derived to produce all kinematic maps discussed in this study are publicly available at \url{https://www.muse-timer.org/}.

\begin{figure*}
\begin{center}
	\includegraphics[clip=true, trim=20 20 20 20, width=0.65\columnwidth]{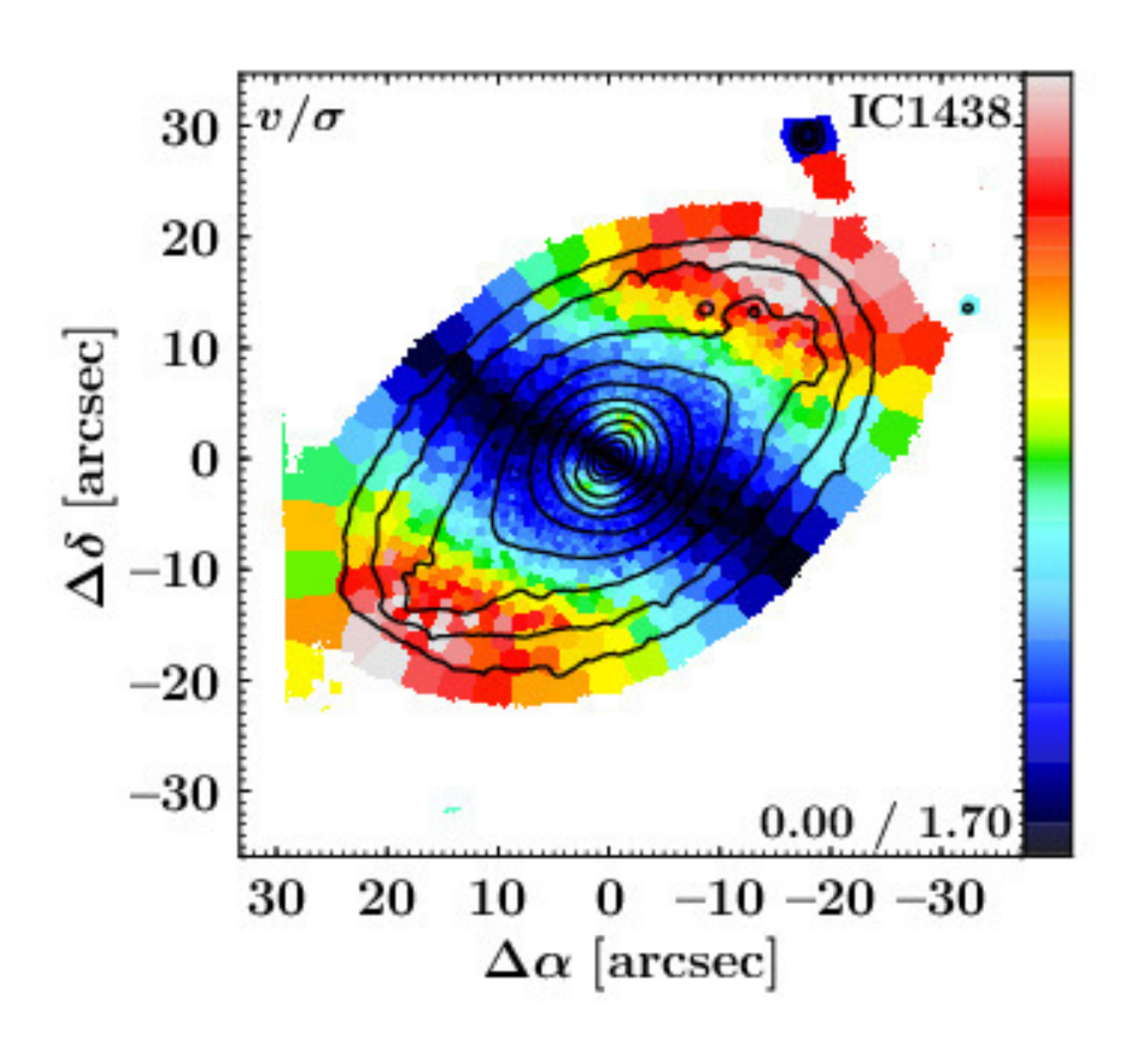}\hskip0.3cm
	\includegraphics[clip=true, trim=20 20 20 20, width=0.65\columnwidth]{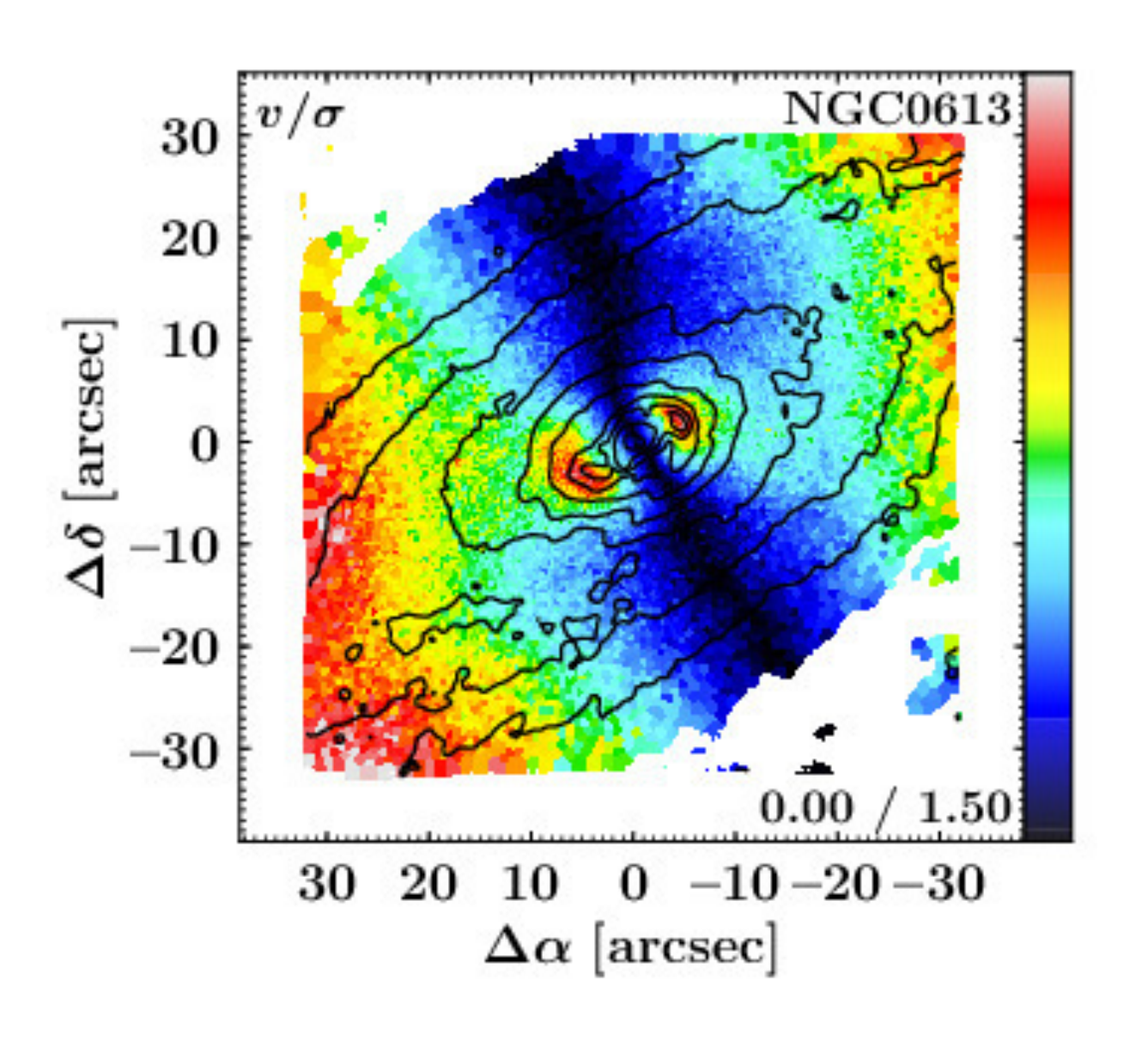}\hskip0.3cm
	\includegraphics[clip=true, trim=20 20 20 20, width=0.65\columnwidth]{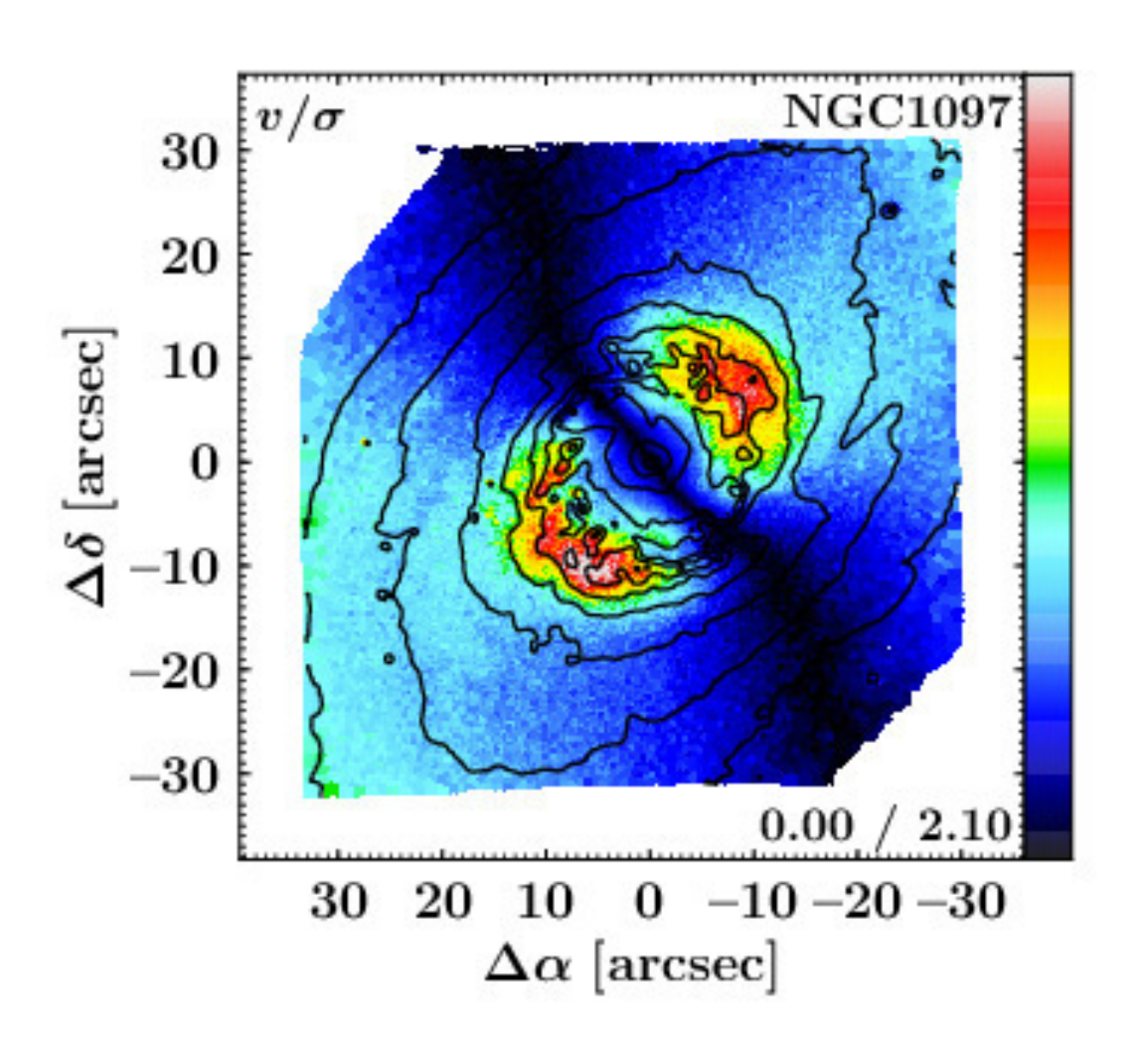}\vskip0.3cm
	\includegraphics[clip=true, trim=20 20 20 20, width=0.65\columnwidth]{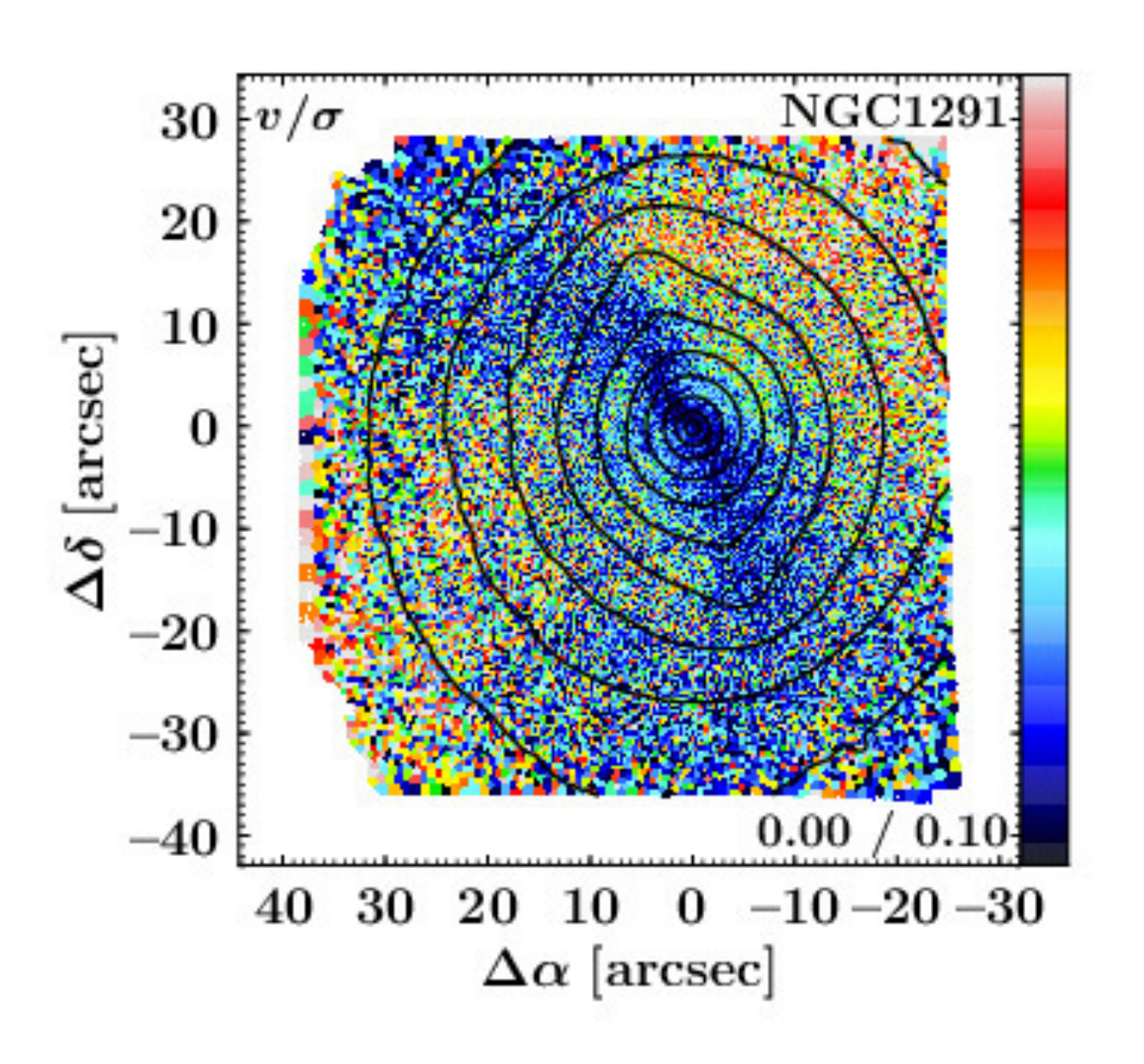}\hskip0.3cm
	\includegraphics[clip=true, trim=20 20 20 20, width=0.65\columnwidth]{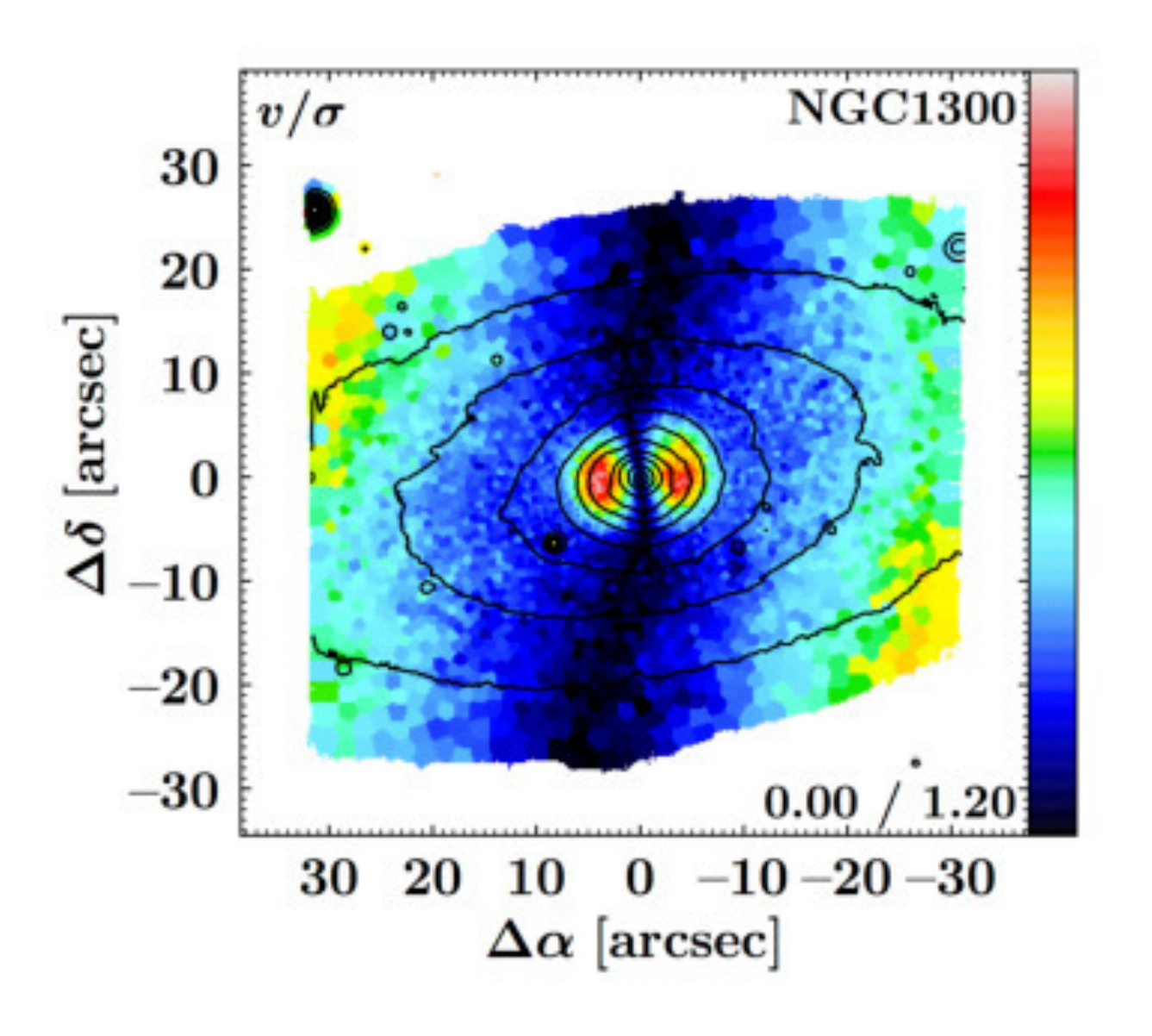}\hskip0.3cm
	\includegraphics[clip=true, trim=20 20 20 20, width=0.65\columnwidth]{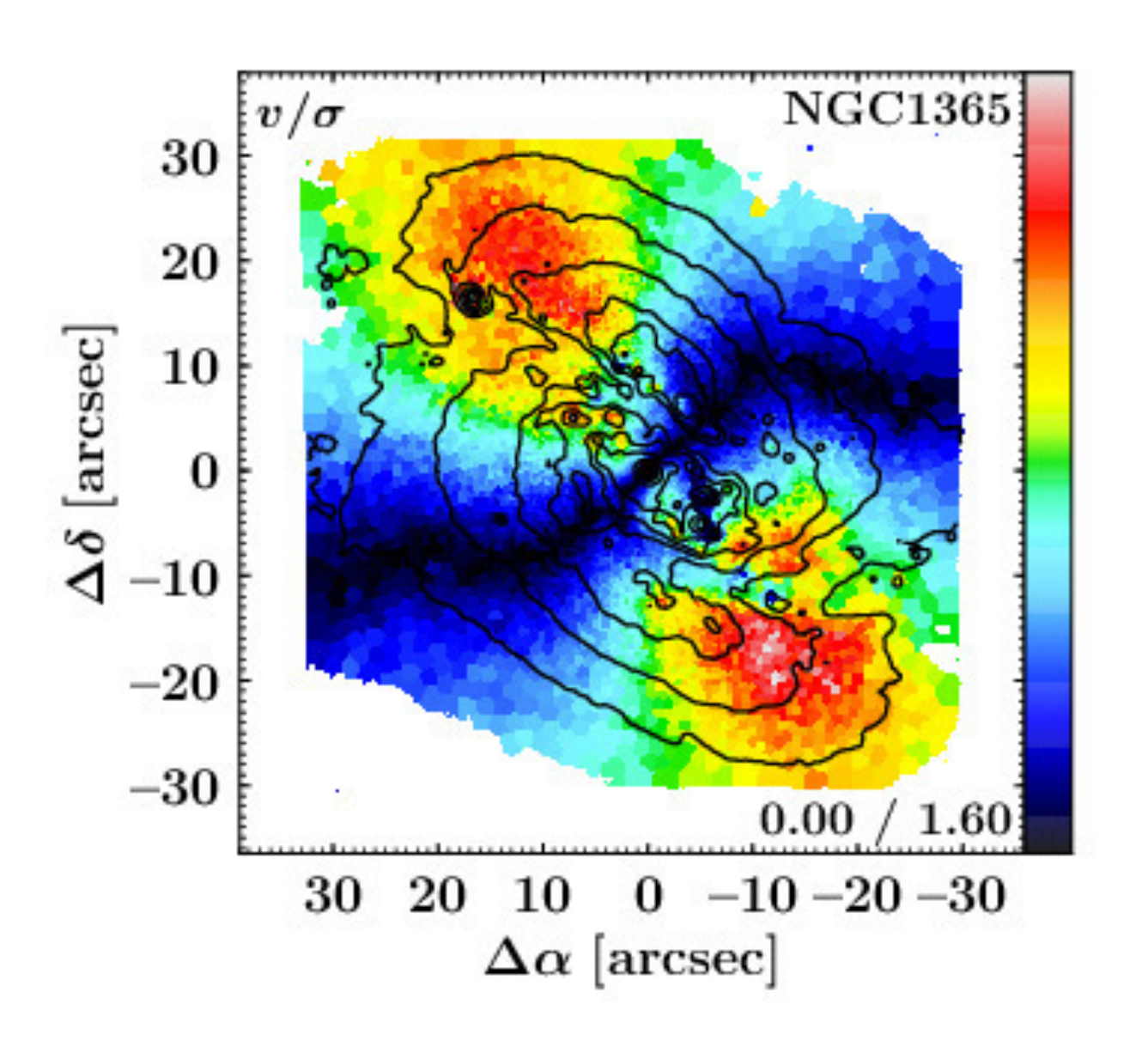}\vskip0.3cm
	\includegraphics[clip=true, trim=20 20 20 20, width=0.65\columnwidth]{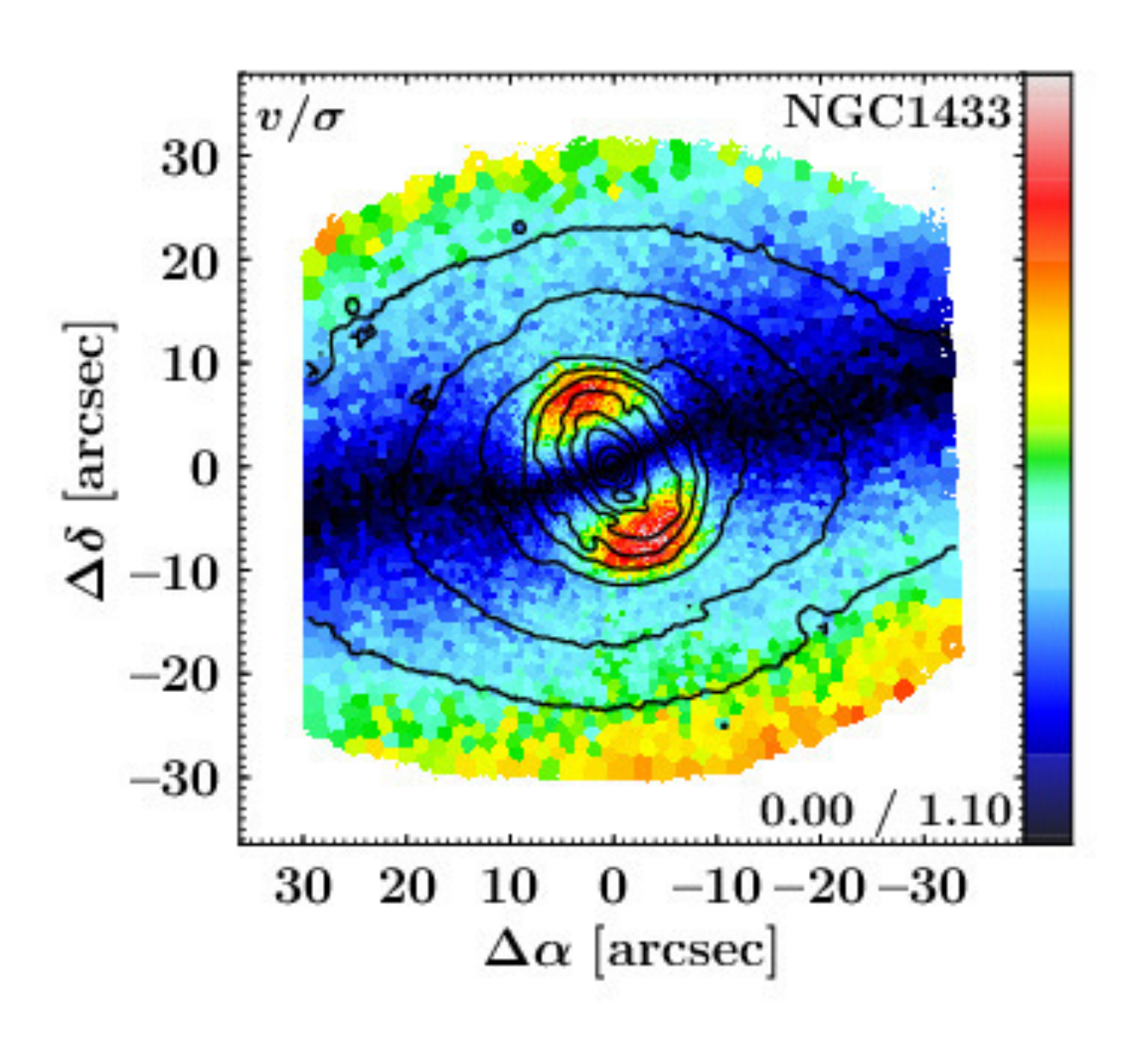}\hskip0.3cm
	\includegraphics[clip=true, trim=20 20 20 20, width=0.65\columnwidth]{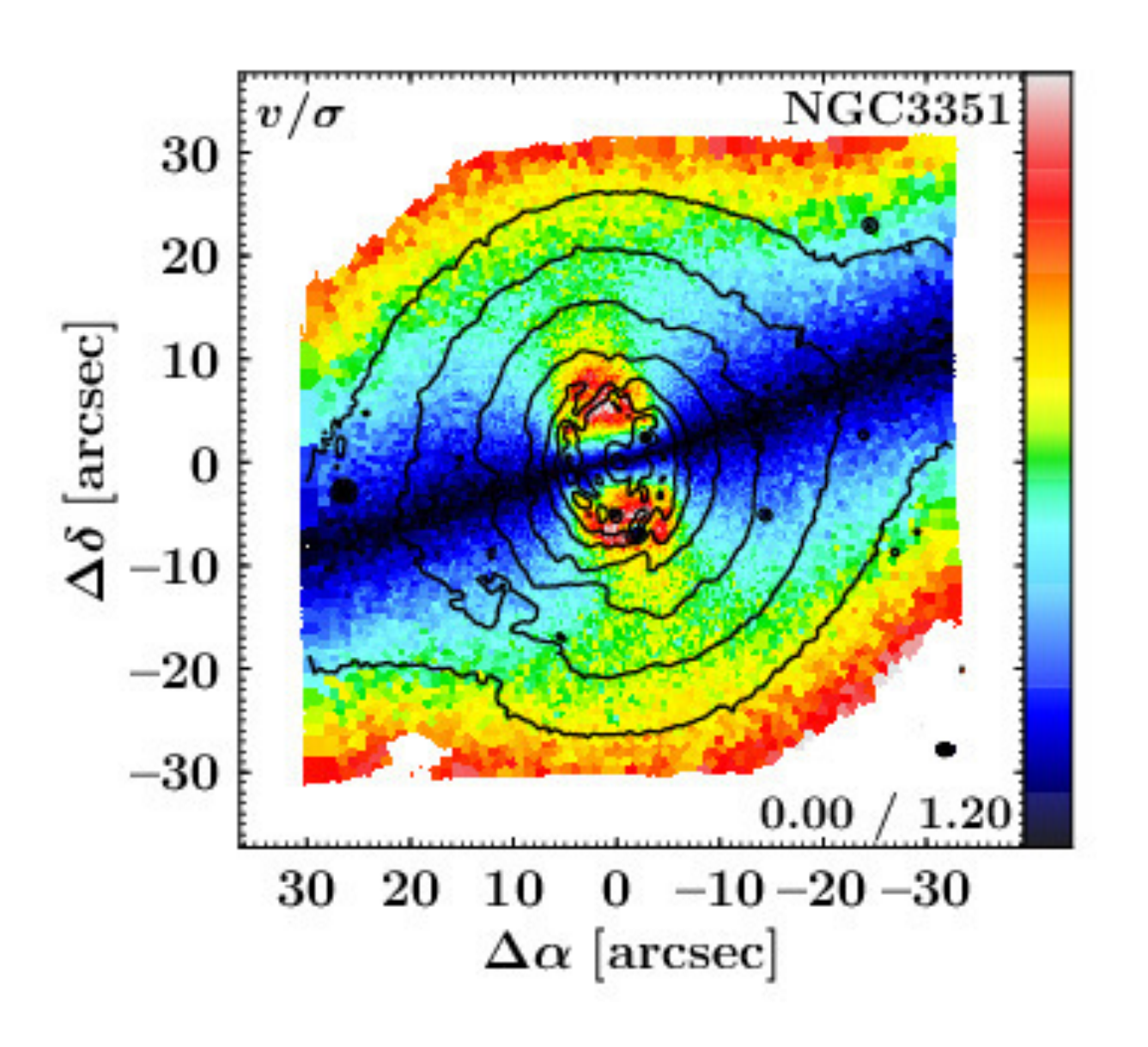}\hskip0.3cm
	\includegraphics[clip=true, trim=20 20 20 20, width=0.65\columnwidth]{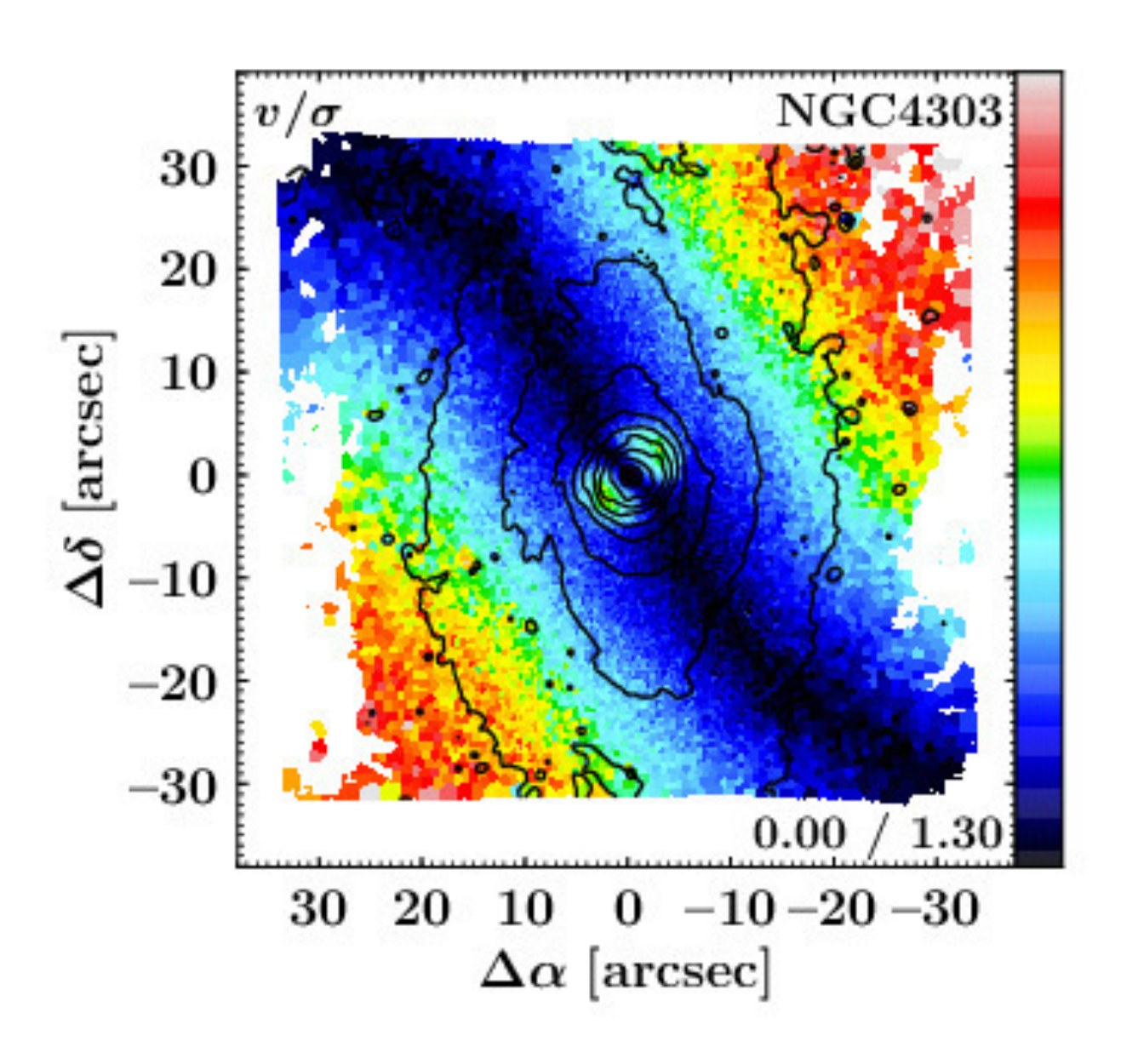}\vskip0.3cm
	\includegraphics[clip=true, trim=20 20 20 20, width=0.65\columnwidth]{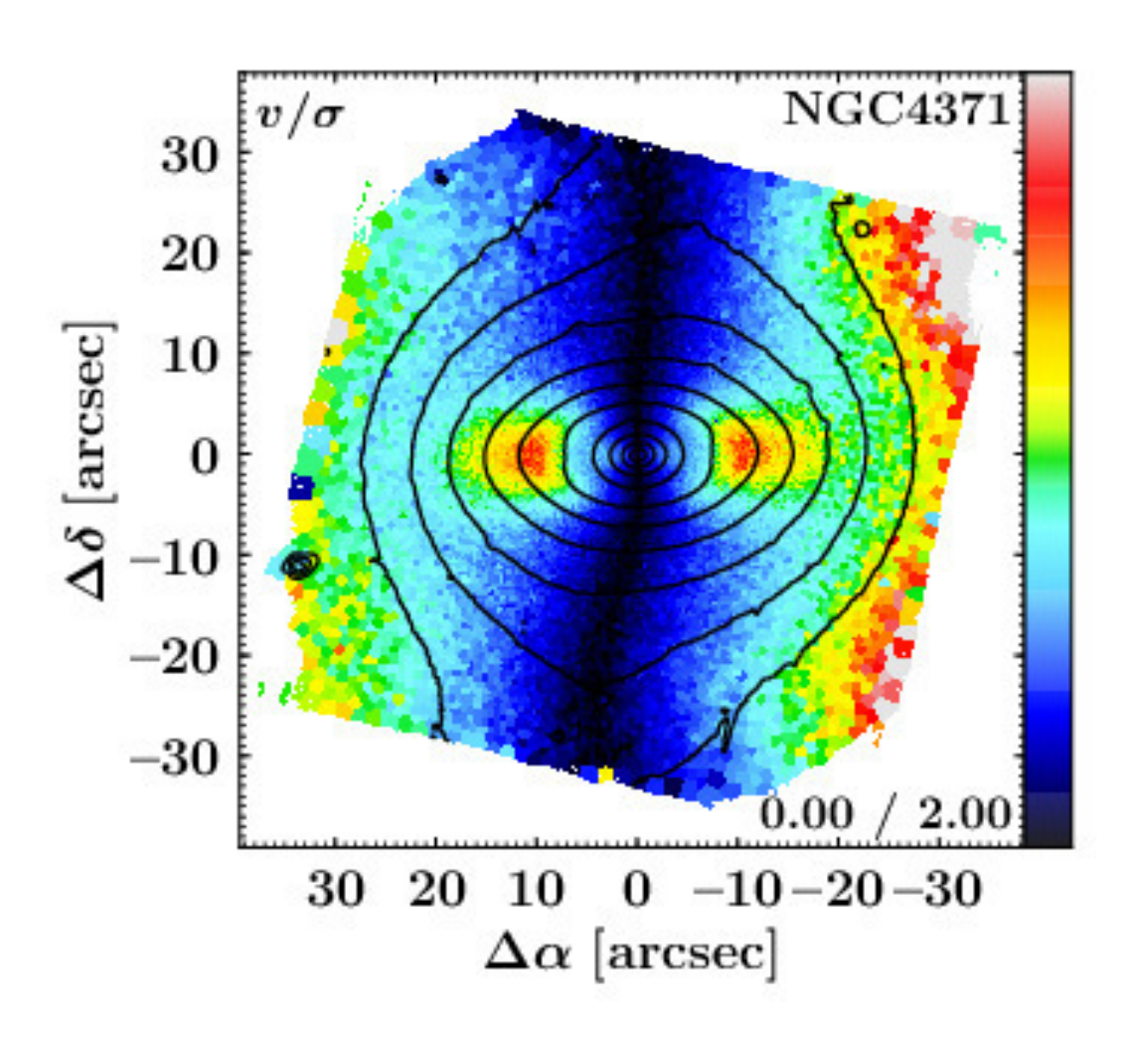}\hskip0.3cm
	\includegraphics[clip=true, trim=20 20 20 20, width=0.65\columnwidth]{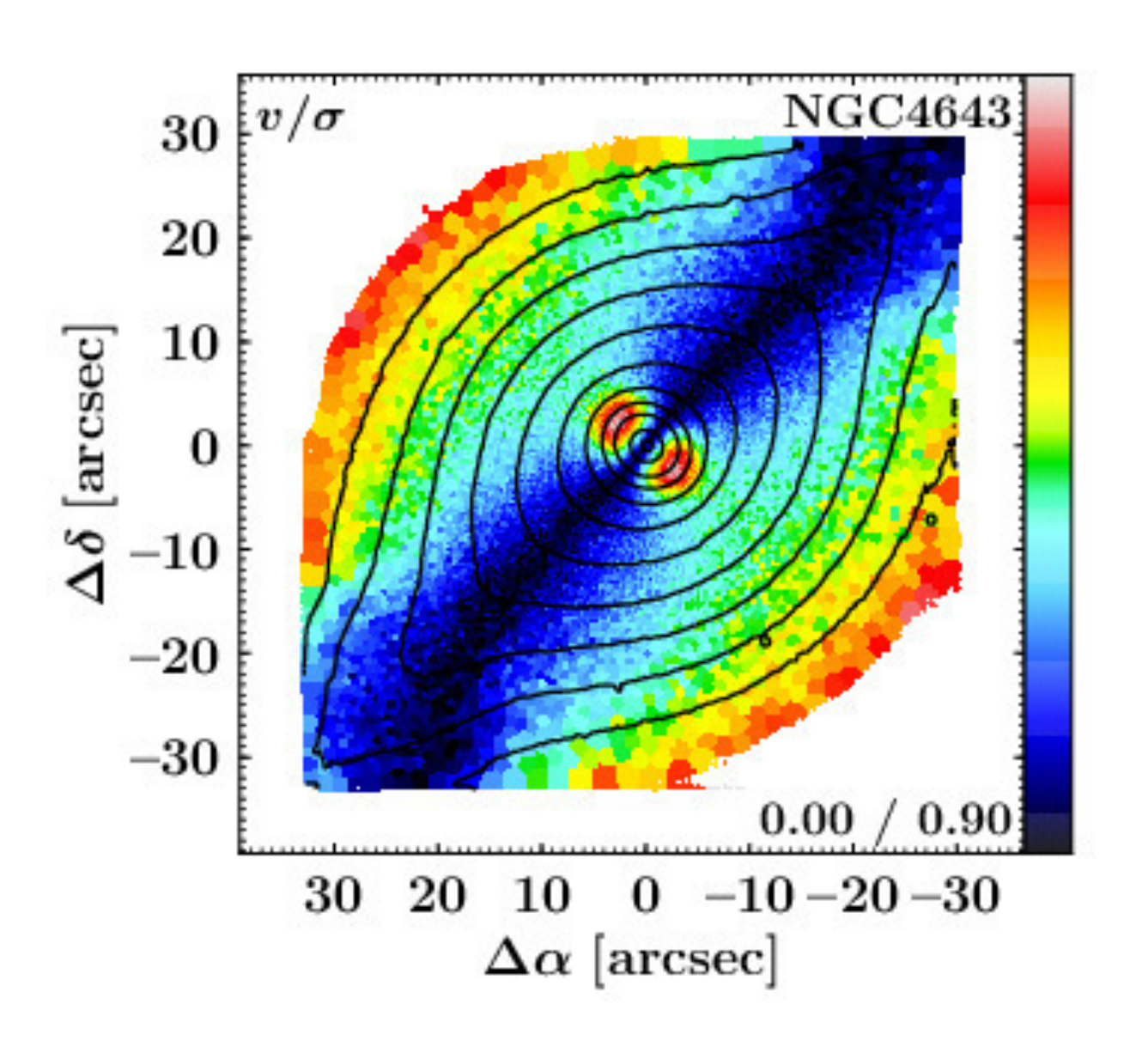}\hskip0.3cm
	\includegraphics[clip=true, trim=20 20 20 20, width=0.65\columnwidth]{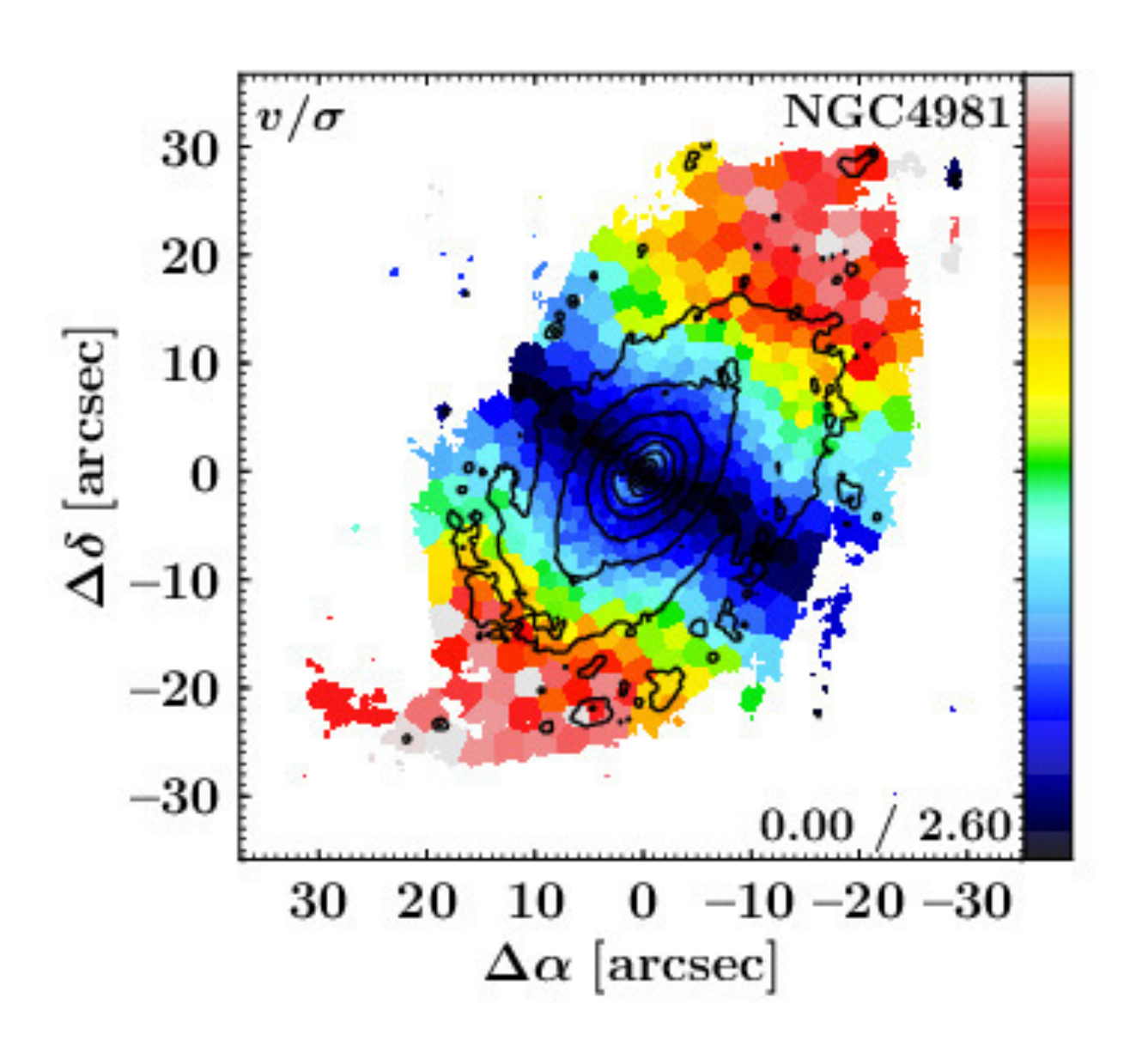}
\end{center}
    \caption{Maps of $v/\sigma$ for the stellar components in the TIMER galaxies, as indicated. The plotted range is indicated on the bottom right corner. The isophotes shown are derived from the MUSE data cube reconstructed intensities and are equally spaced in steps of about 0.5 magnitudes. On average, $1\arcsec$ corresponds to $\approx100\,\rm{pc}$. North is up, east to the left.}
    \label{fig:vos}
\end{figure*}

\begin{figure*}
\begin{center}
	\includegraphics[clip=true, trim=20 20 20 20, width=0.65\columnwidth]{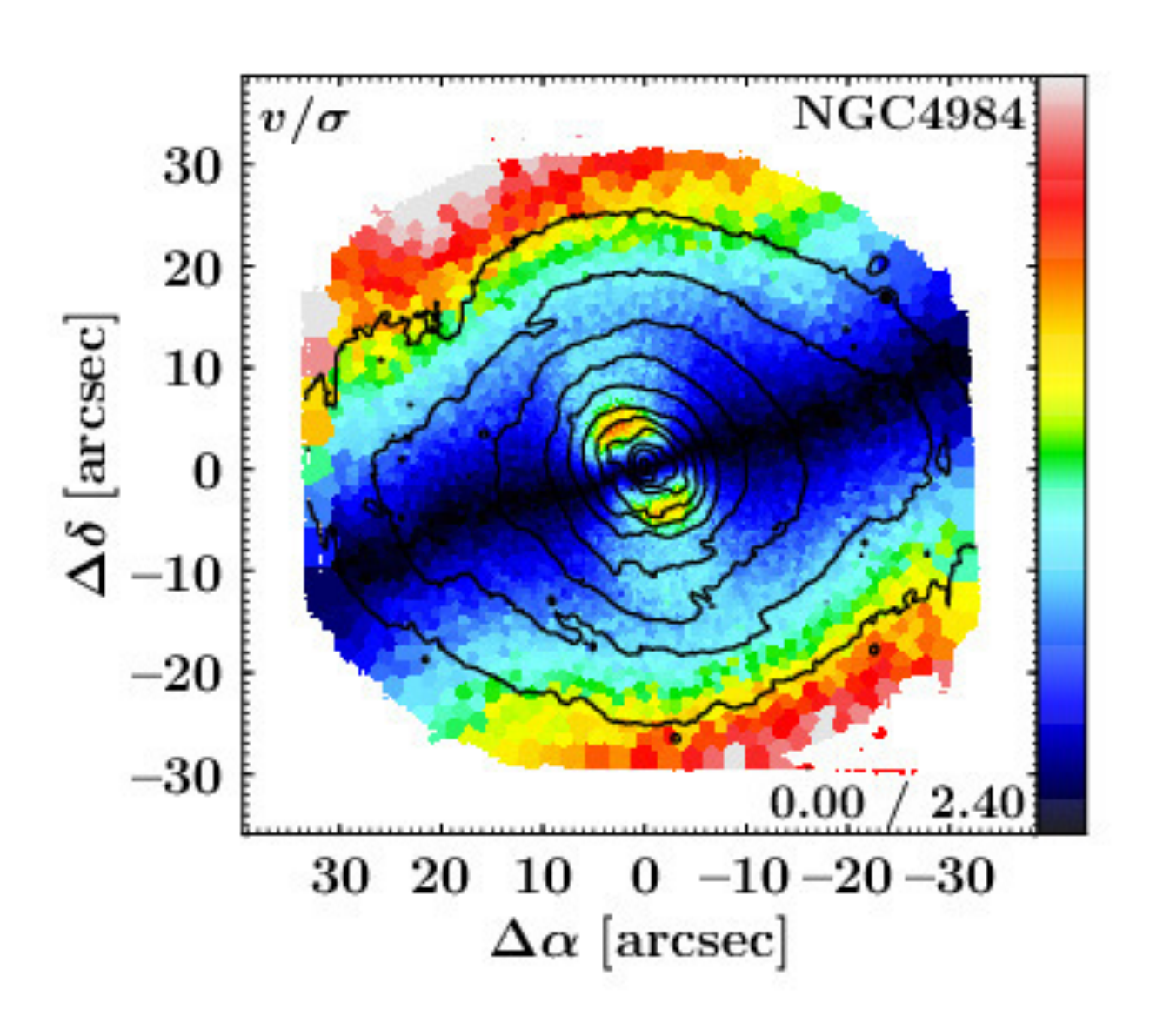}\hskip0.3cm
	\includegraphics[clip=true, trim=20 20 20 20, width=0.65\columnwidth]{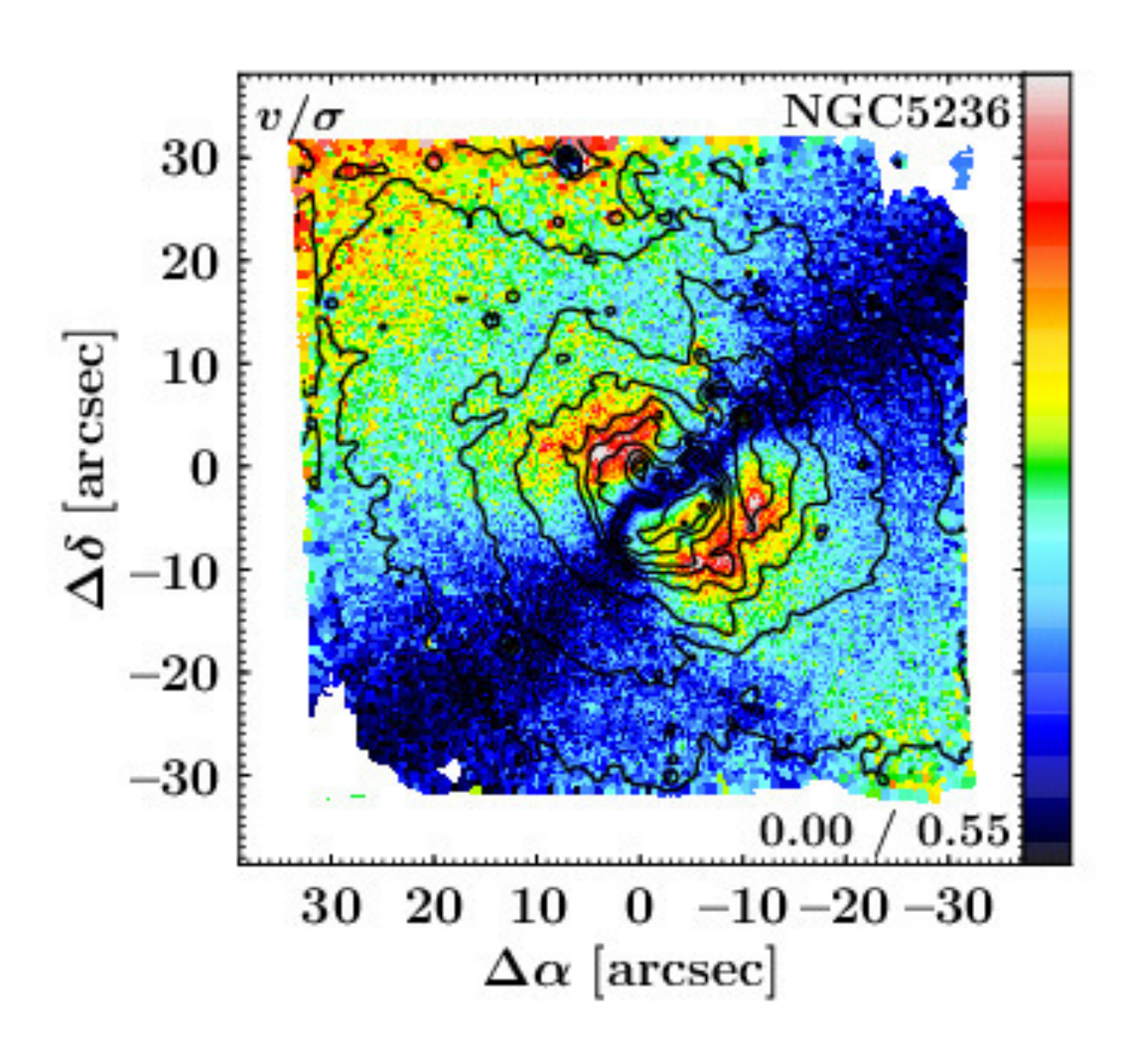}\hskip0.3cm
	\includegraphics[clip=true, trim=20 20 20 20, width=0.65\columnwidth]{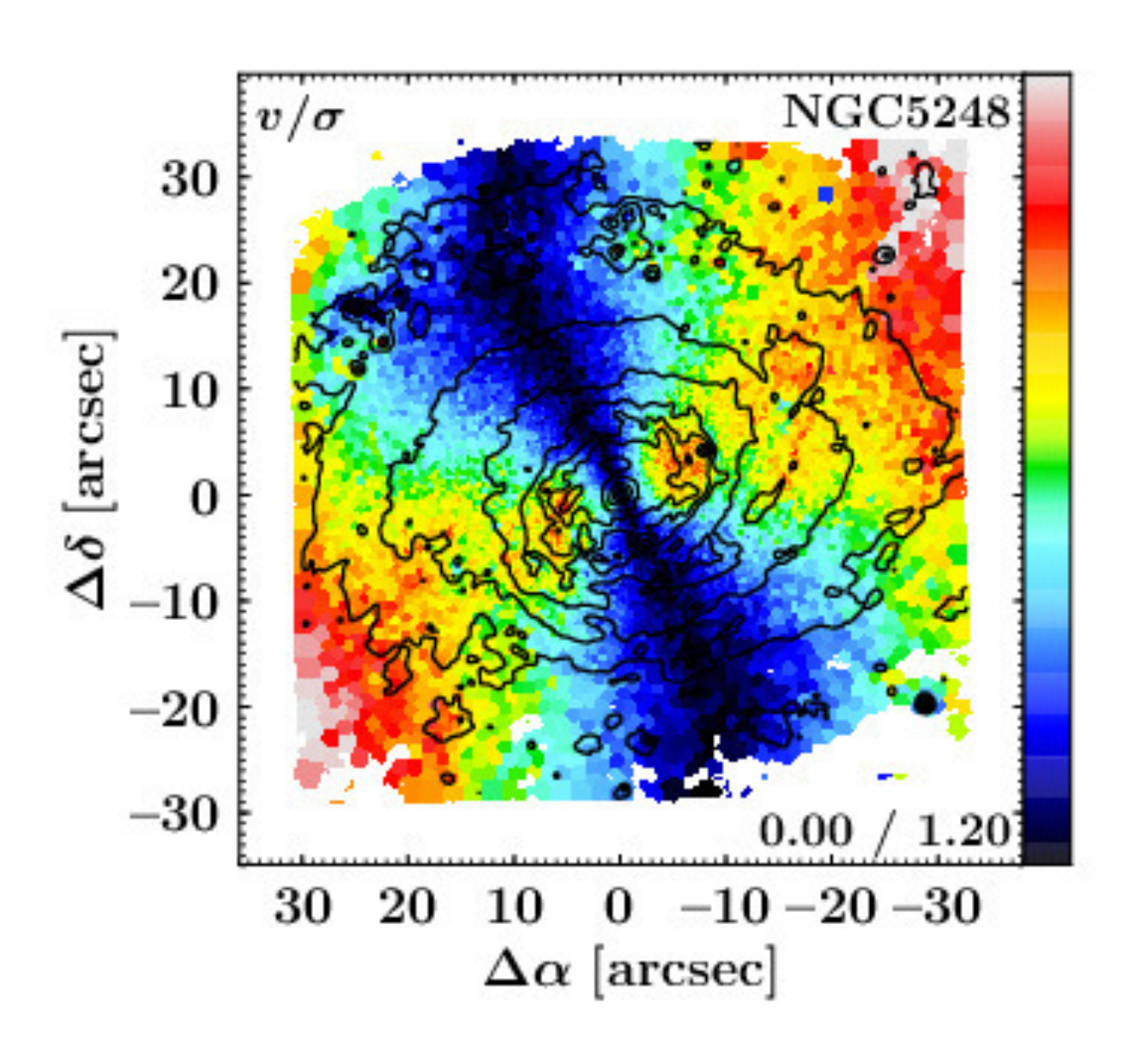}\vskip0.3cm
	\includegraphics[clip=true, trim=20 20 20 20, width=0.65\columnwidth]{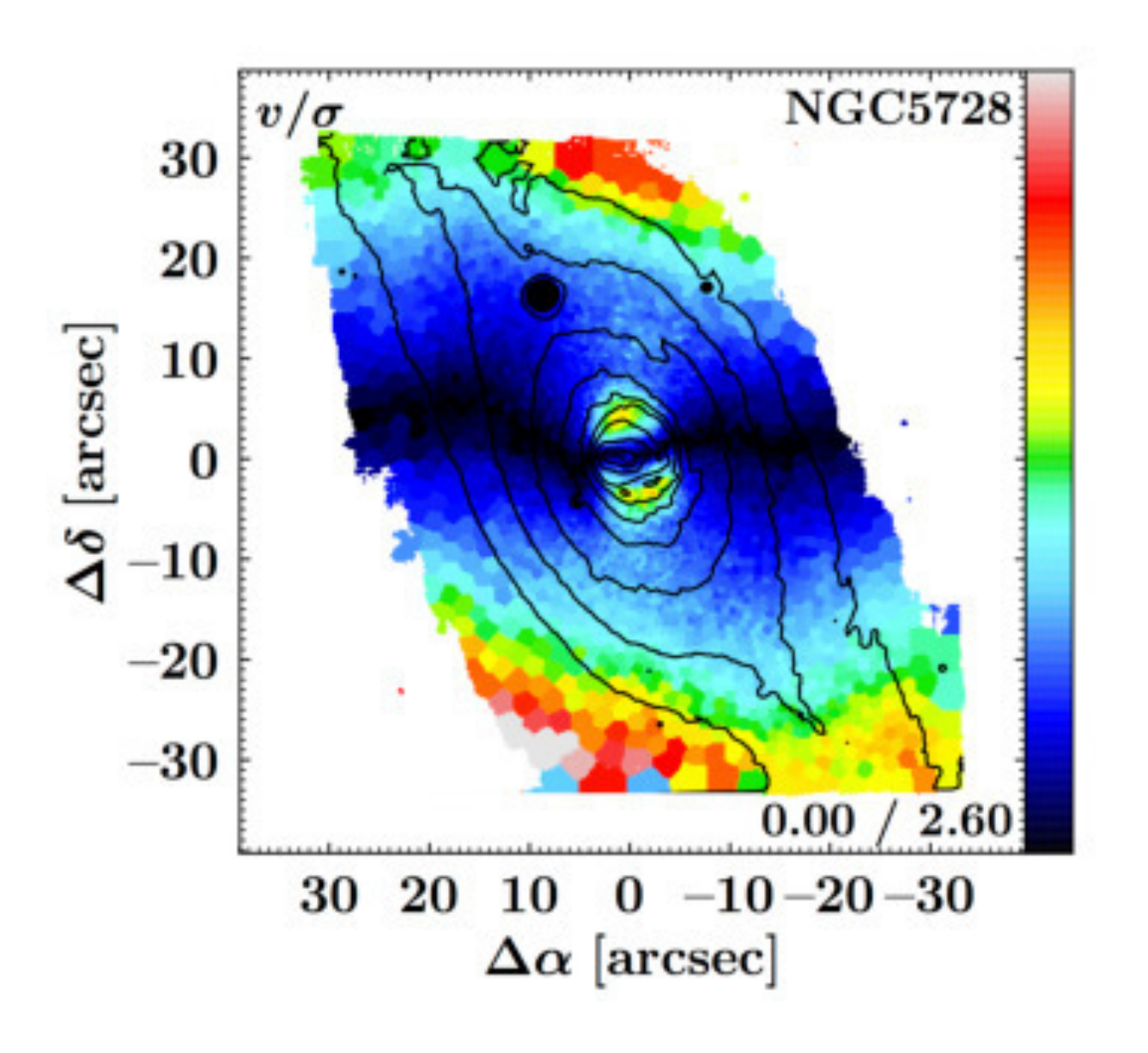}\hskip0.3cm
	\includegraphics[clip=true, trim=20 20 20 20, width=0.65\columnwidth]{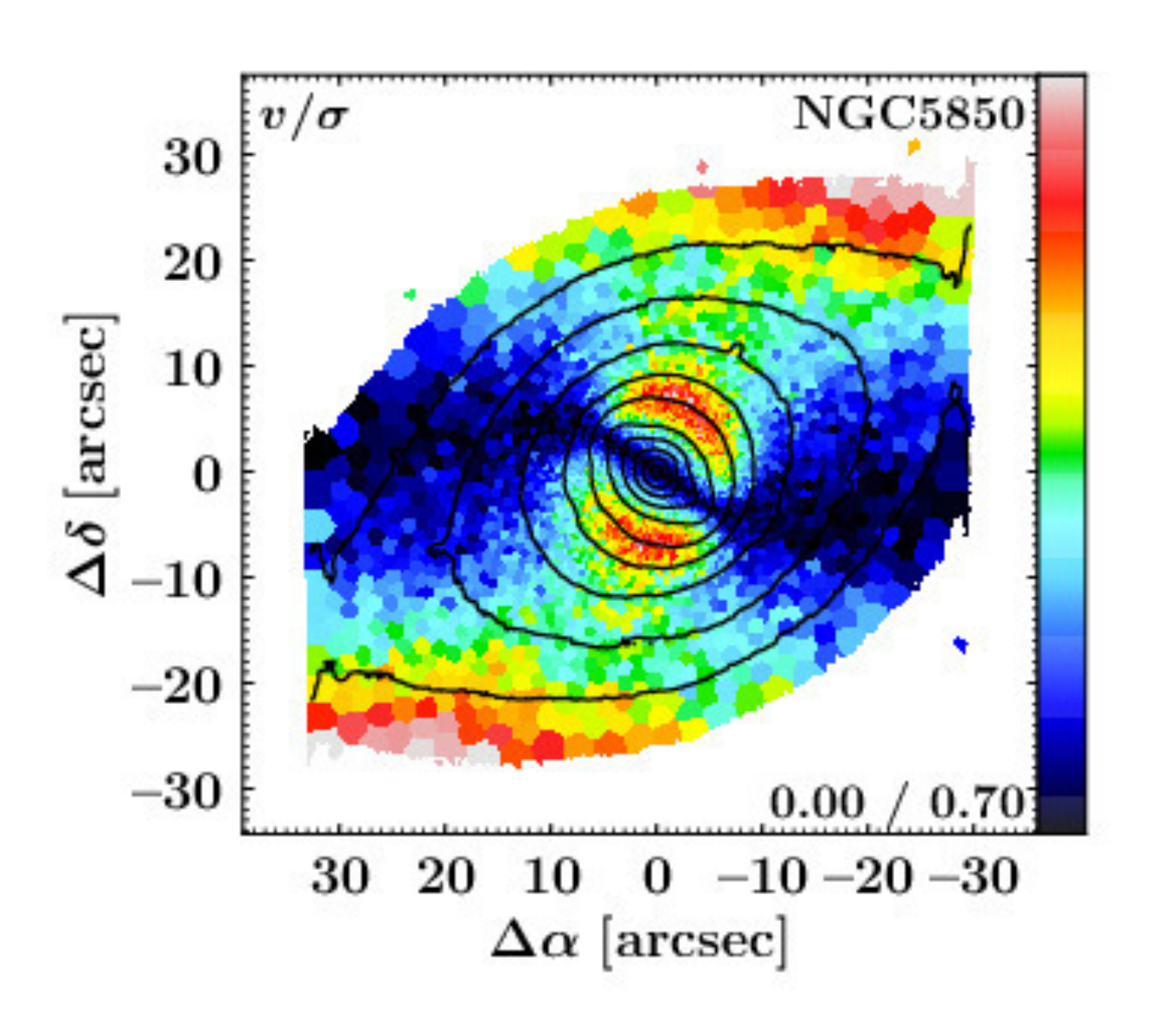}\hskip0.3cm
	\includegraphics[clip=true, trim=20 20 20 20, width=0.65\columnwidth]{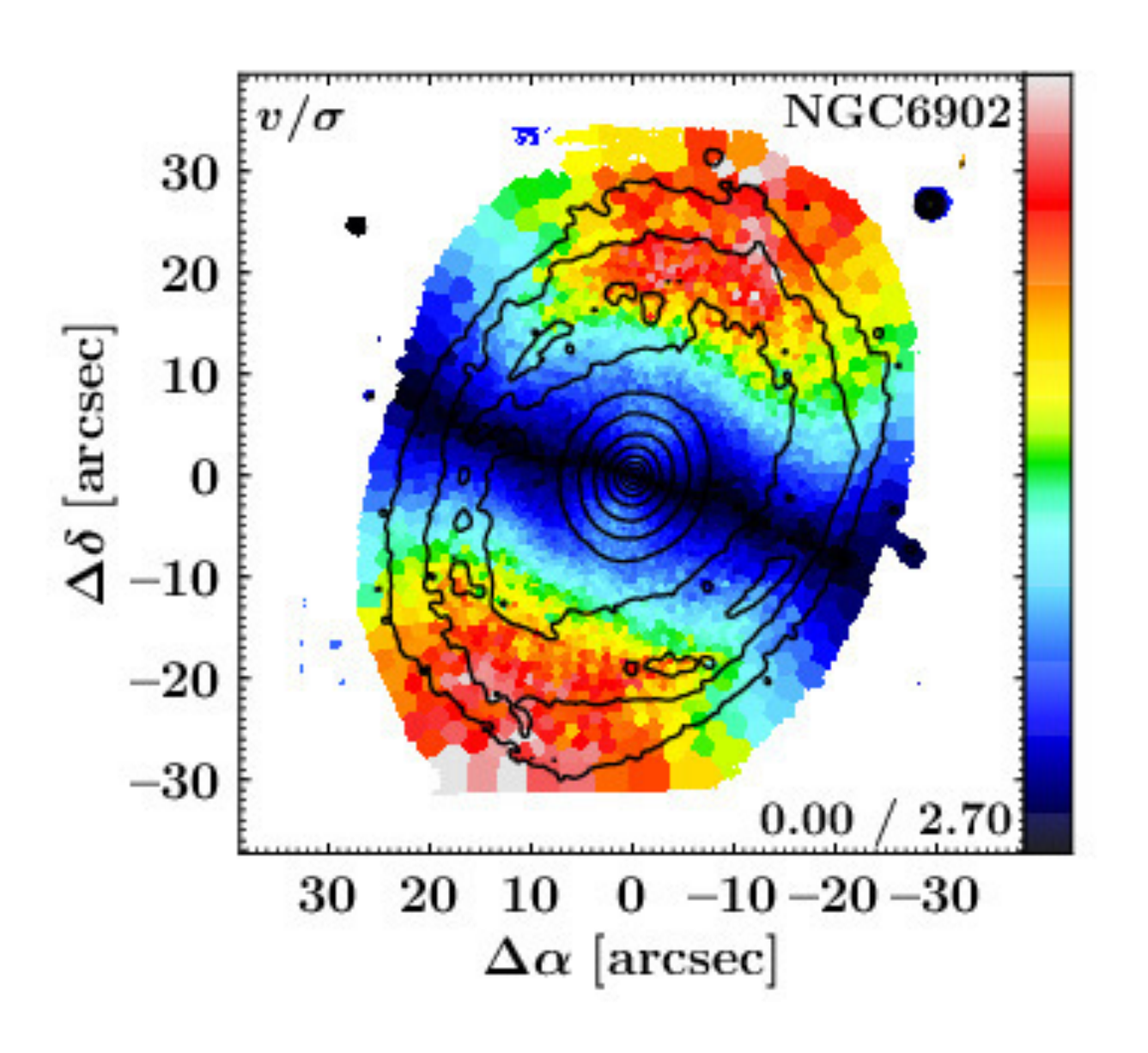}\vskip0.3cm
	\includegraphics[clip=true, trim=20 20 20 20, width=0.65\columnwidth]{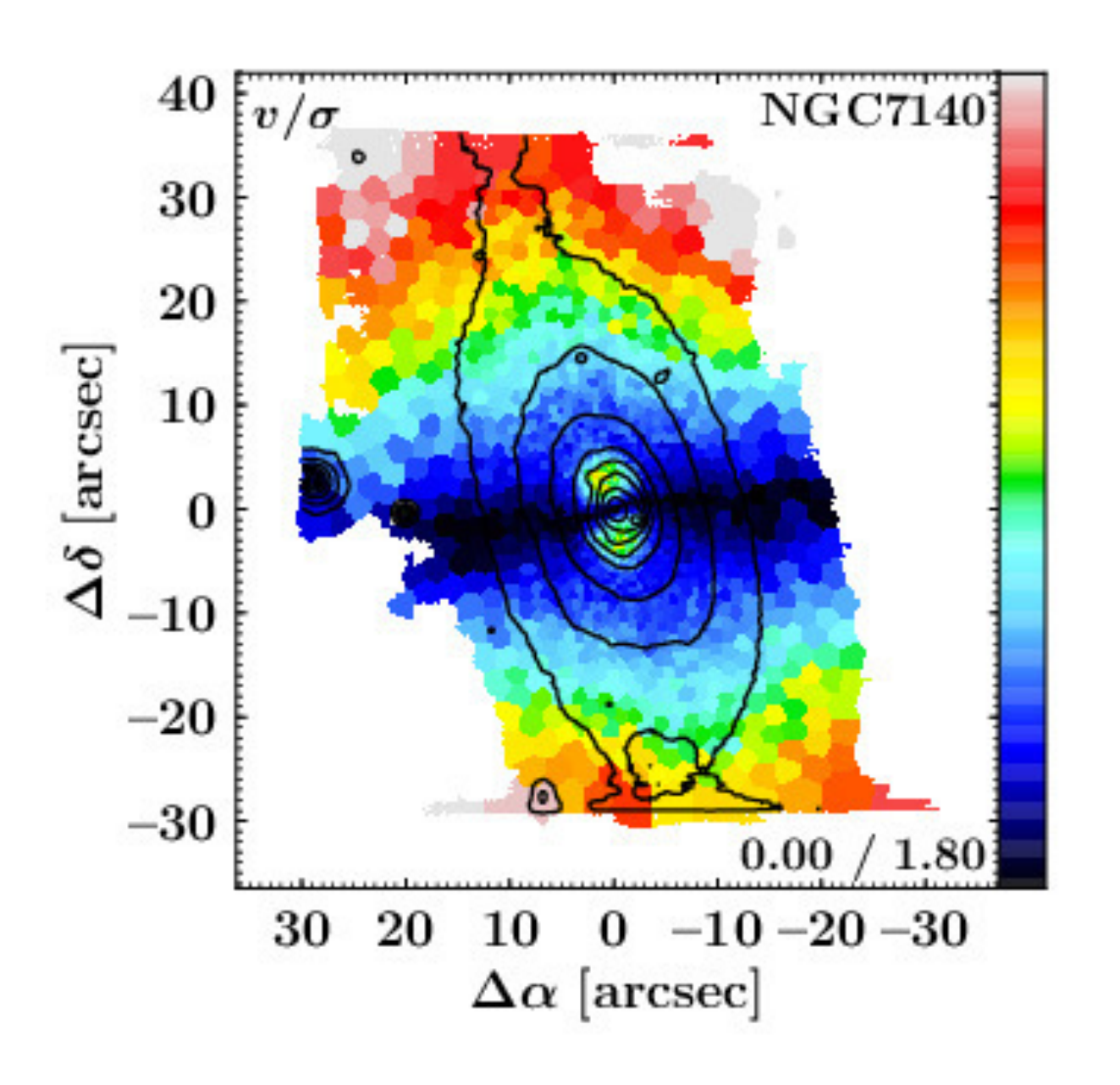}\hskip0.3cm
	\includegraphics[clip=true, trim=20 20 20 20, width=0.65\columnwidth]{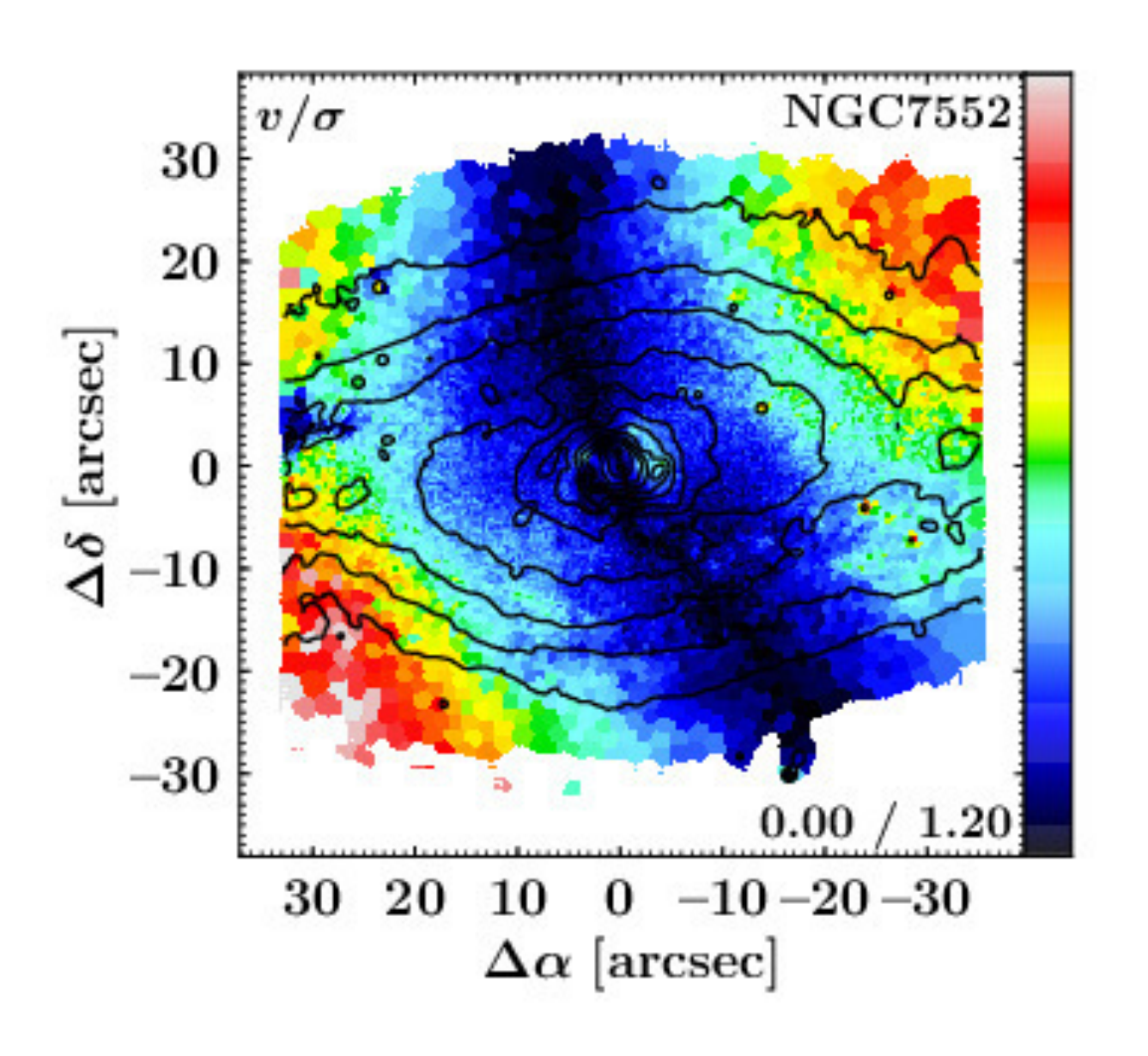}\hskip0.3cm
	\includegraphics[clip=true, trim=20 20 20 20, width=0.65\columnwidth]{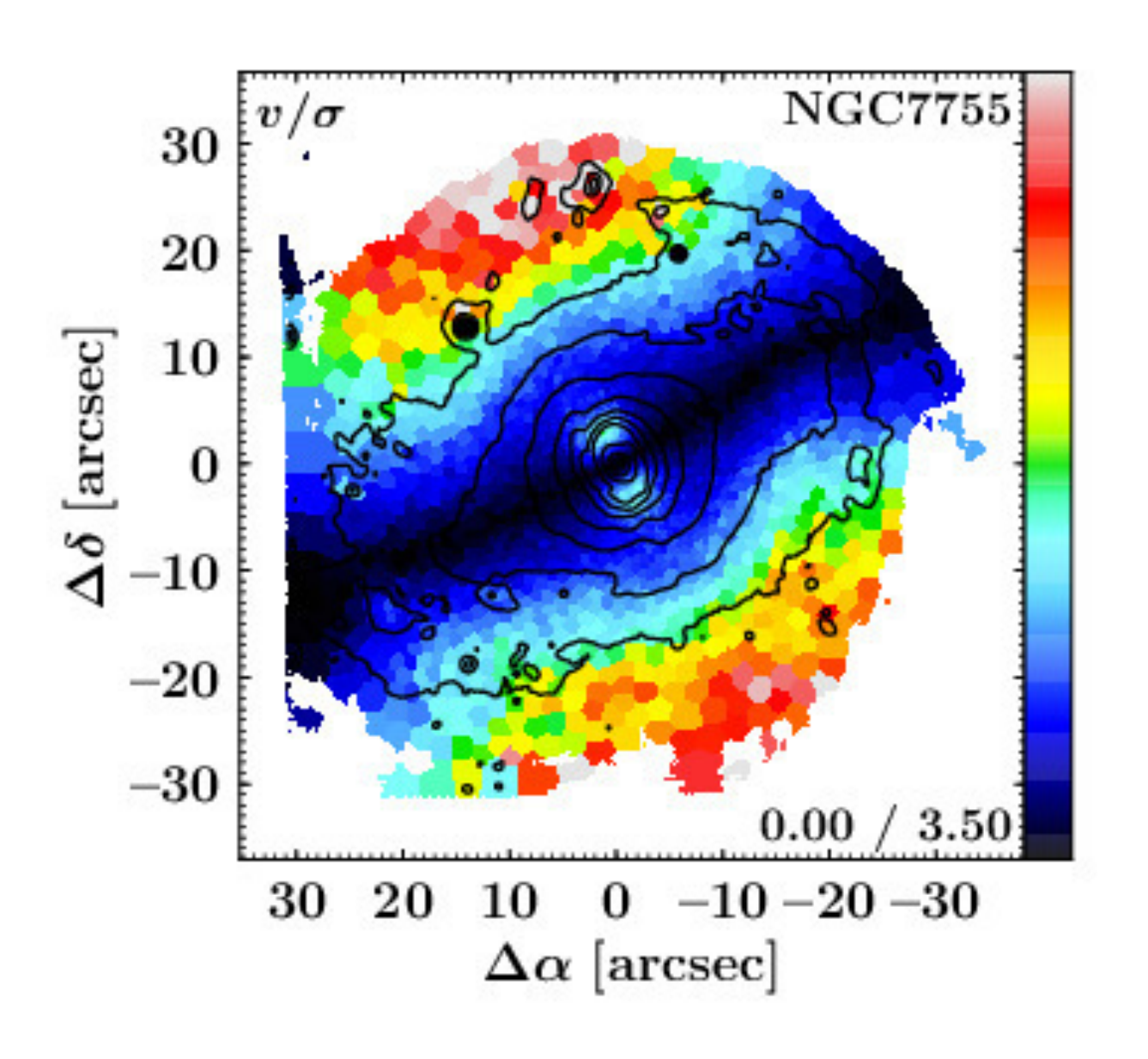}
\end{center}
\addtocounter{figure}{-1}
\caption{continued.}
\end{figure*}

\subsubsection{Kinematic signatures of nuclear discs}
\label{sec:discs}

\begin{figure*}
\begin{center}
	\includegraphics[width=2\columnwidth]{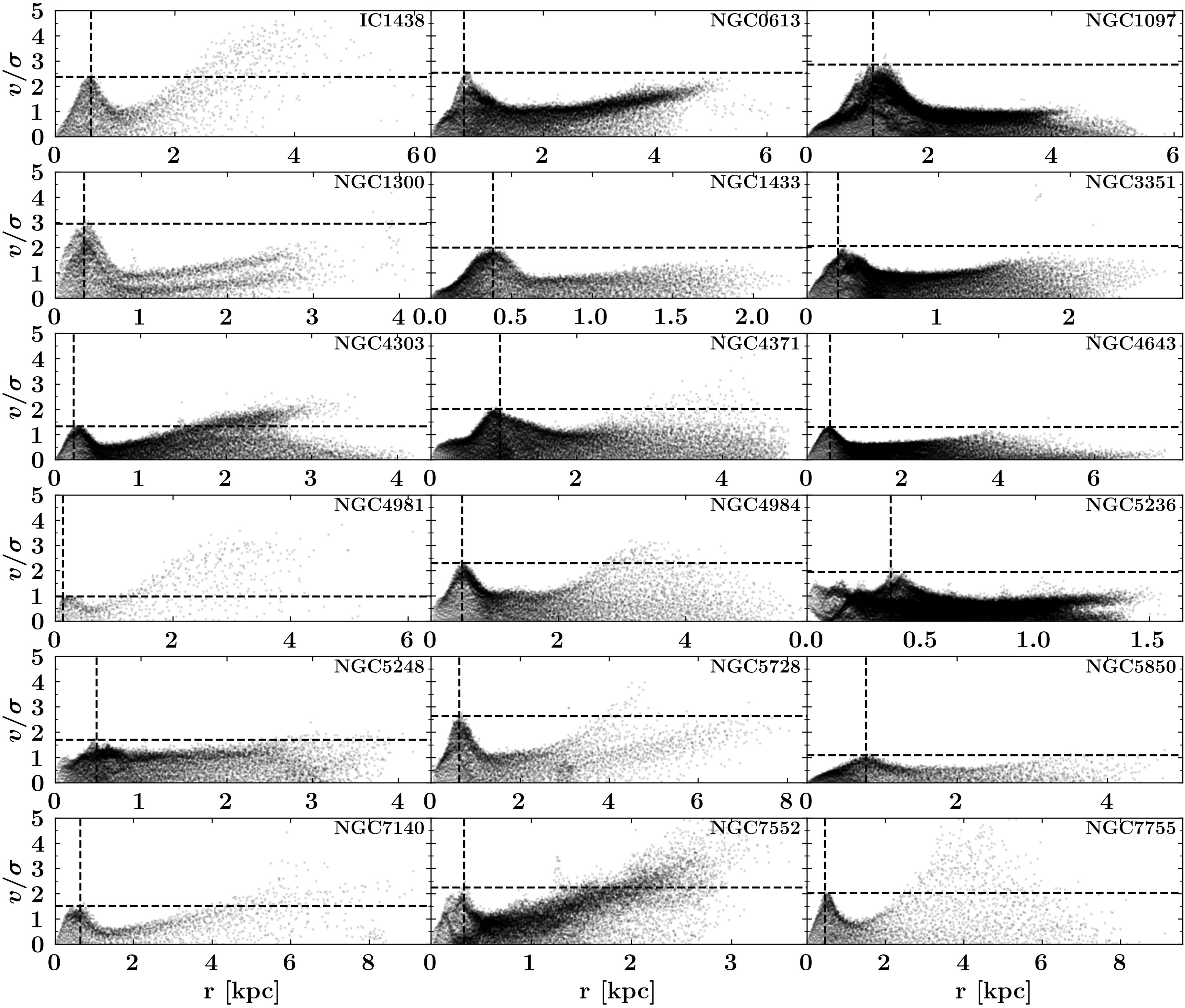}
	\end{center}
    \caption{Deprojected radial profiles of $v/\sigma$ at each Voronoi bin, with $v$ corrected for inclination. NGC\,1291, 1365 and NGC\,6902 are not included. The values of maximum $v/\sigma$ and r$_{\rm k}$ are shown with the horizontal and vertical dashed lines, respectively.}
    \label{fig:vsprofs}
\end{figure*}

\begin{figure}
\begin{center}
	\includegraphics[width=0.9\columnwidth]{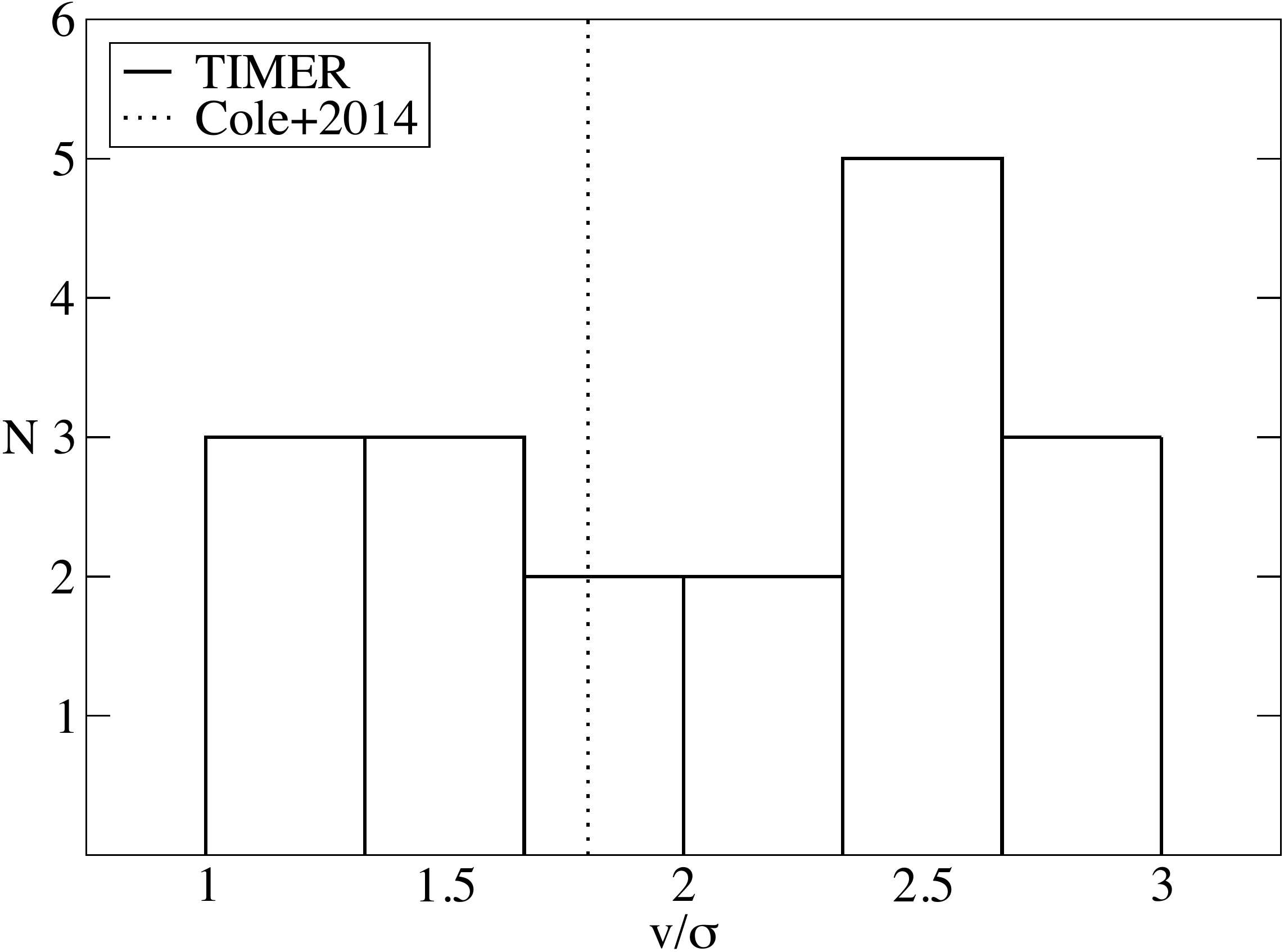}
	\includegraphics[width=0.9\columnwidth]{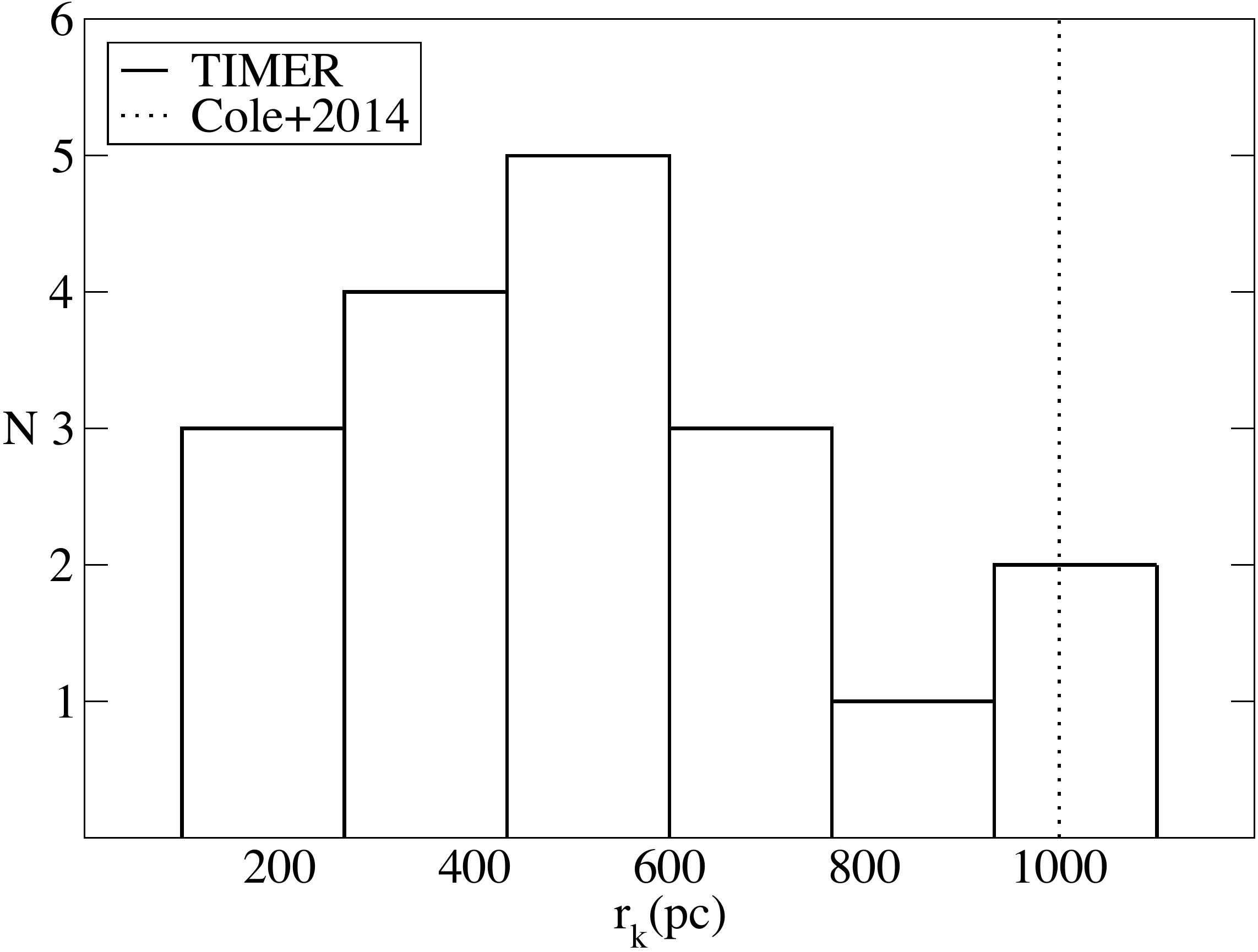}
	\end{center}
    \caption{Distribution of the maximum value of $v/\sigma$ (top) and of the kinematic radius (bottom) of the observed nuclear discs. NGC\,1291, 1365 and NGC\,6902 are not included. The vertical dotted lines mark the values corresponding to the bar-built nuclear disc in the model of \citet{ColDebErw14}.}
    \label{fig:vos_d}
\end{figure}

\begin{figure}[t]
\begin{center}
	\includegraphics[width=\columnwidth]{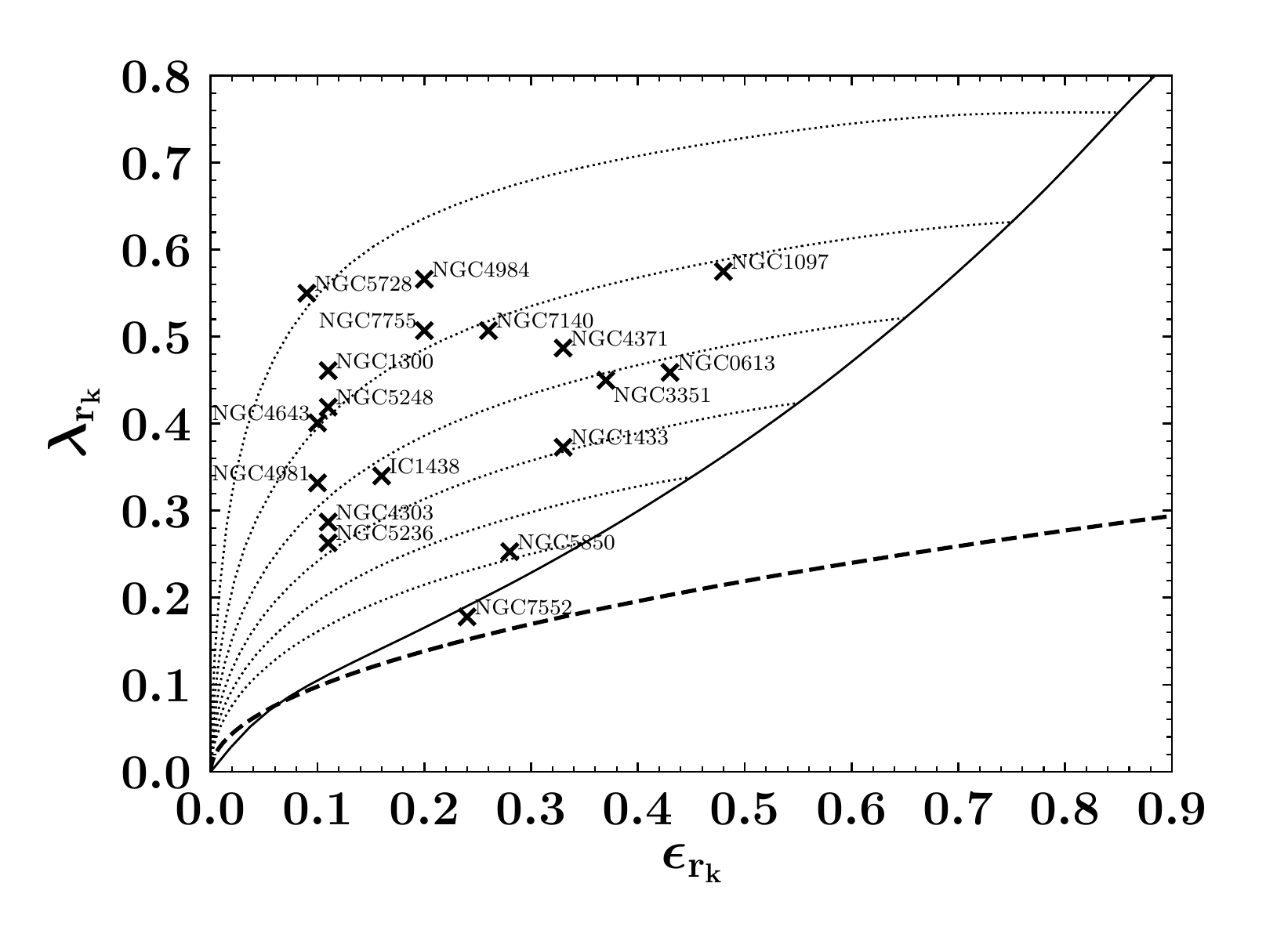}
	\end{center}
    \caption{$\lambda_{r_{\rm k}}$ plotted against the observed (projected) ellipticity at the radius of the nuclear disc. The dashed line is the upper envelope of slow rotators in \citet{EmsCapKra11}. The solid line shows the relation for a model of oblate stellar systems viewed edge-on \citep[see][]{Bin05,CapEmsBac07,EmsCapKra07}. The dotted lines correspond to the location of the same model for varying edge-on ellipticity (intrinsic flatness), from top (0.85) to bottom (0.35) in steps of 0.1, with edge-on systems on the relation and face-on systems towards the origin, as in \citet{EmsCapKra11}. NGC\,1291, 1365 and NGC\,6902 are not included.}
    \label{fig:lambda}
\end{figure}

The 2D kinematic maps provide a test to understand if the observed nuclear stellar structures indeed arise from bar-driven secular evolution processes. If they were built via such dissipative processes within the main disc, they are foreseen to have dynamical properties similar to main discs (which are also formed through dissipative processes, albeit not induced by bars), namely, elevated rotational support, low velocity dispersion and near-circular orbits\footnote{In the presence of an inner bar one expects to see signatures of elongated orbits where the inner bar dominates.} with an angular momentum vector aligned to that of the main disc. Further, since they should be dynamically distinct from the underlying main galaxy disc, the LOSVDs measured from spectra from such structures are likely to show elevated values of h$_4$ (and elevated absolute values of h$_3$), a signature of the overlapping of stellar structures with different velocity distributions: the nuclear structure that dominates the light in such regions and the underlying main galaxy disc \citep[see][]{ColDebErw14}. In principle, external gas accretion unrelated to the bar could also build stellar structures with these properties, but such structures are likely to show at some stage an angular momentum vector that is not aligned to that of the main disc. Moreover, in simulations, bars are the most common mechanism to remove angular momentum from the gas in a uniform fashion down to the central few hundred parsecs \citep[see, e.g.,][]{Ath92b}. On the other hand, if violent processes such as dry mergers build such nuclear structures, then they are expected in most merger configurations to have elongated orbits (showing no $v-$h$_3$ anti-correlation) with relatively low rotational support and high velocity dispersion. We discuss such scenarios further in Sect.\,\ref{sec:origins}.

A careful analysis of Figs.\,\ref{fig:maps} and \ref{fig:maps2} reveals that indeed the TIMER galaxies fit remarkably well the picture in which nuclear structures originate from bar-driven processes. For example, IC\,1438 clearly hosts a rapidly-rotating nuclear component within a radius of about $5''$, with a kinematic axis well aligned with that of the main disc, as seen in the corresponding velocity map. The velocity dispersion map shows that the rapidly-rotating component is characterised by low velocity dispersion, producing a well defined and well localised dip in the central region of the velocity dispersion map, which would, in the absence of this component, simply show an increasing trend in $\sigma$. The h$_3$ map shows also very clearly that this moment of the LOSVD is anti-correlated with $v$, a robust signature of near-circular orbits with a range of values for $v/\sigma$. This is also the case for the outer, main disc of the galaxy, e.g., at radii around $15''$. Finally, the h$_4$ map shows a sharp increase at the region dominated by the rapidly-rotating nuclear component, due to the superposition of the low-$\sigma$ nuclear component on top of the relatively high-$\sigma$ underlying component.

IC\,1438 thus satisfies {\em all} the kinematic criteria discussed above that need to be satisfied by nuclear stellar discs built from dissipative processes on a pre-existent underlying disc. In addition, it is important to stress that the kinematic maps show that the properties of the nuclear stellar disc are {\em not} part of a continuous distribution covering the whole MUSE field. For example, one sees clearly in IC\,1438 that $v$ does not vary monotonically from the outer radii inwards. The stellar velocity peaks at the outer parts of the field and steadily decreases until it increases sharply again at the region of the nuclear disc. This is a clear indication that the nuclear disc is a separate component from the main galaxy disc. In fact, at the same galactocentric radius, stars in the nuclear disc are rotating around the galaxy centre faster than the stars in the main disc (see Fig.\,\ref{fig:maps}). This implies that -- in the region where the nuclear disc resides -- the nuclear disc is dynamically colder than the main disc and its stars have orbits closer to near-circular orbits\footnote{Here we assume that both discs are in the same plane, which is justified by the alignment of their kinematic axes seen in Figs.\,\ref{fig:maps}, \ref{fig:vos} and \ref{fig:maps2}.}. As discussed above, this is also the reason behind the high values of h$_4$ in the nuclear disc.

To corroborate these findings we also produced maps of local $v/\sigma$ (see Fig.\,\ref{fig:vos}). The figure shows again and even more clearly that the nuclear structures morphologically identified by means of visual inspection by \citet{ButSheAth15} are stellar structures with very elevated dynamical rotational support and separate from the main galaxy disc. We note that Buta et al. have not necessarily identified these nuclear structures as nuclear discs but rather as nuclear disc features, such as rings, lenses, spiral arms and bars. However, Figs.\,\ref{fig:maps} and \ref{fig:vos} show that these nuclear structures are extended, and in our accompanying paper (Bittner et al. 2020, subm.) we present evidence from the stellar population properties that these nuclear structures appear to extend all the way to the galaxy centre. We thus have very strong evidence that the TIMER galaxies host nuclear discs with kinematic properties consistent with a bar-driven origin. In Sect.\,\ref{sec:origins} we will elaborate on alternative scenarios for the building of these nuclear discs, and also discuss further the connection between nuclear discs and nuclear rings.




Figure \ref{fig:vsprofs} shows deprojected radial profiles of the $v/\sigma$ measurements at each Voronoi bin, after correcting $v$ for inclination. We use the values determined with S$^4$G data by \citet{MunSheReg15} for the inclination and position angle of each galaxy. NGC\,1365 and NGC\,6902 are not included in this figure, since the signatures of a nuclear disc in these galaxies are not as clear as in the remaining of the sample (these cases are discussed in Appendix\,\ref{app:maps}). NGC\,1291 is also excluded, as determining $v/\sigma$ is difficult due to the low inclination of the galaxy (11$^\circ$). These profiles allow us to obtain the maximum value of $v/\sigma$ in the nuclear discs, assuming that the nuclear disc is in the same plane as the main disc. The region within which the maximum is searched for is delimited by a circumference centred at the galaxy centre and crossing the position of the first minimum in $v/\sigma$ along the disc major axis, beyond the centre. The measurements of $v/\sigma$ considered in this region are those for each Voronoi bin. We also define the kinematic radius (r$_{\rm k}$) of the nuclear disc as the deprojected distance from the galaxy centre of the Voronoi bin showing the maximum value of $v/\sigma$ in the region dominated by the rapidly-rotating nuclear disc. The values of maximum $v/\sigma$ and r$_{\rm k}$ are shown in Fig.\,\ref{fig:vsprofs} with the horizontal and vertical dashed lines, respectively.

In Fig.\,\ref{fig:vos_d} we present the distributions of the values of maximum $v/\sigma$ and r$_{\rm k}$. Interestingly, although the $v/\sigma$ distribution is skewed towards higher values, it also shows a broad range of $v/\sigma$ values, as low as unity. We note that the bar-built nuclear disc in the hydrodynamical simulations of \citet{ColDebErw14} has a peak $v/\sigma$ of $\approx1.8$ (the vertical dotted line in Fig.\,\ref{fig:vos_d}), fitting well within the distribution of observed values. The same is true for the size of the simulated nuclear disc, although it lies on one of the extremes of the observed r$_{\rm k}$ distribution. The simulated nuclear disc is as large as the largest observed nuclear discs. We note that the values of $v/\sigma$ and r$_{\rm k}$ taken from \citet{ColDebErw14} correspond to the end of their simulation, at 10\,Gyr, when the nuclear disc has reached its maximum size.

Finally, further insights can be gained by employing the $\lambda_R$ parameter introduced by \citet{EmsCapKra07}, which quantifies the projected stellar angular momentum per unit mass. We thus calculated $\lambda_{r_{\rm k}}$, i.e., $\lambda_R$ integrated within the radius $r_{\rm k}$, as in \citet{EmsCapKra07}, again using the position angle and ellipticity derived for the main disc by \citet{MunSheReg15} with S$^4$G data. In Fig.\,\ref{fig:lambda} we plot $\lambda_{r_{\rm k}}$ against the observed (projected) ellipticity at the radius of the nuclear disc, r$_{\rm k}$, derived with the S$^4$G radial profiles of ellipticity. The positions of our measurements in this diagram show clearly that the region within r$_{\rm k}$ in the TIMER galaxies has the same angular momentum support as the fast rotators of \citet{EmsCapKra11}. In addition, most systems are consistent with the elevated edge-on ellipticities (or intrinsic flatness) of disc systems. However, as discussed in \citet{CapEmsBac07}, this diagram is only rigorously valid for stellar systems with a density stratified on homologous oblate ellipsoids \citep[as in the model of][]{Bin05}, which is evidently not the case of our sample galaxies. The central region of the TIMER galaxies hosts not only the nuclear disc but at least also the main disc and bar, which means that the intrinsic flatness of the TIMER nuclear discs as seen in this diagram is at best a lower limit.

\subsubsection{Kinematic signatures of bars}
\label{sec:bs}

The kinematic maps in Figs.\,\ref{fig:maps} and \ref{fig:maps2} also show kinematic signatures of the presence of bars. Take IC\,1438 again as an example. Between the main disc and the rapidly-rotating nuclear disc, both showing an anti-correlation between $v$ and h$_3$, one sees a region along the bar in which $v$ and h$_3$ are actually {\em correlated}. From the bottom-left corner of the h$_3$ map, towards the top-right, one first sees blue/green bins, then red/yellow bins, then blue bins again and the inverse patterns after crossing the centre. The correlation between $v$ and h$_3$ was shown by \citet{BurAth05} and \citet{IanAth15} to result from the superposition of a bar and a disc, and is seen where both components contribute more or less equally \citep[see also][]{LiSheBur18}.

A similar signature is also seen in NGC\,1300, 1433, 3351, 4643, 5850, 7140 and NGC\,7755. Possibly due to projection effects, dust extinction and/or lower physical spatial resolution, this signature is somewhat less clear in NGC\,613, 4303, 4981, 4984, 5248, 5728 and NGC\,7552. Since all these galaxies are known to have bars from studies of their morphology, this result is not terribly surprising, but it lends support to our theoretical understanding of the stellar dynamics in barred galaxies.

Interestingly, NGC\,1433 shows a correlation between $v$ and h$_3$ also within $\sim4$ arcsec from the centre, which suggests the presence of an inner bar \citep[see also][for a zoom-in view of the kinematics in this region]{BitFalNed19}, although archival HST images inspected by \citet{deLSanMen19} show no clear morphological signature of an inner bar \citep[but see][]{Erw04,ButSheAth15}.

We also point out the conspicuous drops in velocity dispersion at the ends of the inner bars in NGC\,1291 and NGC\,5850. These $\sigma$-hollows were found by \citet{deLFalVaz08} to be a characteristic signature of inner bars, and these two galaxies were studied in detail already in \citet{MendeLGad19} and \citet{deLSanMen19}.

\subsubsection{Kinematic signatures of box/peanuts}
\label{sec:bps}

Our maps also reveal kinematic signatures of the presence of box/peanuts. \citet{DebCarMay05} employed numerical simulations to show that box/peanuts can be detected in face-on galaxies by examining the spatial distribution of h$_4$: box/peanuts imprint two significant minima along the bar major axis at the positions of the box/peanut vertices \citep[see also][]{IanAth15,LiSheBur18}. \citet{MenCorDeb08} showed that this diagnostic works observationally, detecting box/peanuts in primary bars, and, more recently, \citet{MendeLGad19} were able to find for the first time a box/peanut in an inner bar, that of NGC\,1291.

IC\,1438 provides again a good example. The very central region, i.e., within a radius of about $2''$, shows values of h$_4$ close to zero. This region is surrounded by elevated values of h$_4$ where the rapidly-rotating nuclear disc dominates, as discussed above. Just outside this region -- at radii between about 5 and 10 arcsec -- h$_4$ takes negative values, becoming close to zero again at larger radii. This is the signature just mentioned of the presence of a box/peanut.

We find the same signature clearly in NGC\,613, the inner bar of NGC\,1291 \citep[as shown already in][]{MendeLGad19}, NGC\,1300, 4303, 4643, 4981, 4984, 5728, 5850, 7140, 7552, and NGC\,7755. Simulations show that this signature is weaker for galaxies with inclination angles larger than 30$^\circ$ \citep[see, e.g., Fig.\,30 in ][]{IanAth15}. Most of the aforementioned galaxies are below or close to this threshold, with the exception of NGC\,4981, 4984, 7140 and NGC\,7755, where this interpretation should be taken carefully. 

NGC\,5728 is an interesting case. The h$_4$ map shows clear minima along the bar major axis, and the isophotal contours show regions that are slightly offset from the bar major axis on opposite sides of the bar at each side from the centre. This S-shape configuration is produced by what \citet{ErwDeb13} called spurs. Such spurs are photometric signatures of box/peanuts seen at an angle \citep[][see also \citealt{BeaMajGuh07}]{AthBea06}, and thus this galaxy is a beautiful illustration where both photometric and kinematic signatures of box/peanuts coincide.

Considering the TIMER sample altogether, we find a clear signature of the presence of a box/peanut in 13 of the 21 galaxies. As discussed in Appendix\,\ref{app:maps}, at least in some cases, the signature absence may be due to the restricted fields studied here. Therefore, our results indicate a lower limit for the fraction of box/peanuts in our sample of massive barred galaxies of at least 62\%. This is consistent with the estimate presented by \citet{ErwDeb17} also for massive barred galaxies, namely, 79\%, even though our analysis relies entirely on kinematics and theirs on photometry. On the other hand, in a study employing the kinematic diagnostic on a sample of 10 galaxies, \citet{MenDebCor14} found a box/peanut fraction of 50\%.


\subsubsection{Kinematic signatures of barlenses}
\label{sec:bls}

As mentioned in the Introduction, barlenses are thought to be the face-on projection of box/peanuts, and thus also part of the bar, even though they extend further from the bar major axis than the remaining of the bar. \citep{LauSalBut05,LauSalBut07,LauSalBut11,AthLauSal15,LauSal16}. If that is the case, one expects to see, in the region dominated by the barlens, both the h$_4$ minima that characterise box/peanuts and the $v-$h$_3$ correlation that characterise bars \citep[see][although one should be careful to disregard regions dominated by a nuclear disc]{IanAth15}. In the S$^4$G images, we see a clear barlens morphology in five of the galaxies studied here: NGC\,1300, 3351, 4643, 4984 and NGC\,7755. Except for the latter, all are classified by \citet{ButSheAth15} as indeed showing a barlens. In these five galaxies, but more strongly in NGC\,4984 and NGC\,7755, the expected signatures are seen. In NGC\,4984, the h$_4$ minima form a thick ring with inner and outer radii of about $10''$ and $20''$, respectively. In the same region one sees the $v-$h$_3$ correlation. Likewise, NGC\,7755 shows the same signatures in radii between $5''$ and $10''$. These features are very similar to what is seen in the simulations of \citet[their fig. 29]{IanAth15}.

While this is consistent with our current understanding of barlenses and box/peanuts, we point out that seven other galaxies in this sample are classified by \citet{ButSheAth15} as having barlenses, but the expected signatures are not clear. Evidently, part of the reason could be attributed to the relatively small spatial coverage in some of these galaxies. On the other hand, the visual classification of barlenses is prone to ambiguities and should be considered carefully, as nuclear discs, and even classical bulges, can be confused as barlenses. As discussed in \citet{GadSeiSan15}, the case of NGC\,4371 is very illustrative. Morphologically, the signature of a barlens appears very clear. However, once kinematic information is included in the analysis, one concludes that the concerned structure is actually a rapidly-rotating -- and rather large -- nuclear disc. This demonstrates the importance of assessing the nature of morphological components through a detailed analysis of the corresponding stellar kinematics.

Apart from NGC\,4371, Buta et al. also include NGC\,613, 1097, 1291, 5728, 5850 and NGC\,7552 as having barlenses. In the case of NGC\,1097 and NGC\,1291 the TIMER fields are too restricted to properly study the corresponding regions. Recently, \citet{deLSanMen19} found that the component visually classified by Buta et al. in NGC\,5850 as a barlens is actually the nuclear disc, and that this is possibly the case for NGC\,1291 also but it is difficult to ascertain that in the case of this galaxy given its low inclination. For the remaining three galaxies, the signatures are weak: the h$_4$ minima, clearly seen along the bar major axis are not prominent in the perpendicular direction. As discussed in Sect.\,\ref{sec:bs}, the presence of a $v-$h$_3$ correlation is not very significant in NGC\,613, 5728 and NGC\,7552, possibly due to projection effects, dust extinction and/or lower physical spatial resolution.

\subsection{Previous observations with integral-field spectrographs}

Some galaxies in the TIMER sample have been observed before with different integral-field spectrographs. It is particularly instructive to qualitatively compare our observations with those performed with SAURON \citep{BacCopMon01,deZBurEms02} and WiFeS \citep{DopHarMcG07,DopRheFar10}, which as MUSE operate in the optical wavelength range. Four TIMER galaxies were studied previously with SAURON: NGC\,4371 \citep{CapEmsKra11}, NGC\,4643 \citep{CapEmsKra11,SeiFalMar15}, NGC\,5248 \citep{DumMunEms07} and NGC\,5850 \citep{deLFalVaz08,deLFalVaz13}. NGC\,7552 was studied with WiFeS by \citet{SeiCacRui15}. The spectral resolution obtained with SAURON is about a factor two lower than in MUSE observations, which in turn is a factor $\gtrsim2$ lower than that provided by WiFeS. However, MUSE excels in sensitivity and spatial sampling. Part of the gains in sensitivity comes from the hosting telescopes. The collecting area of the VLT is roughly  about four times larger than that of the William Herschel Telescope, which hosts SAURON, and roughly 16 times larger than the 2.3-m telescope at Siding Spring Observatory, which hosts WiFeS. The spatial sampling of MUSE corresponds to 0.2\arcsec\ per spatial element, whereas in both SAURON and WiFeS this is about 1\arcsec.

The remarkable spatial sampling and sensitivity of MUSE weigh significantly in studies of central structures in disc galaxies, such as this study. This can be appreciated when comparing the TIMER observations with those mentioned above. The MUSE kinematic maps reveal a multitude of details that remain hidden in the SAURON and WiFeS datasets \citep[see, e.g., the analysis concerning NGC\,5850 in][]{deLSanMen19}. Surely, indications of a kinematically colder nuclear component are seen in the SAURON and WiFeS data, but the lower spatial sampling makes it more difficult to ascertain that the component is a nuclear disc, separate from the main galaxy disc. Furthermore, a more detailed analysis of the kinematic maps, e.g., by searching for the kinematic signatures of bars, box/peanuts and barlenses, as performed above, is certainly precluded by the relatively low spatial resolution of the SAURON and WiFeS data.

We can also compare our measurements of velocity and velocity dispersion with those presented by \citet[][for NGC\,1365]{VenNarMar18},  \citet[][for NGC\,5728]{ShiDavLut19}, and \citet[][for NGC\,1291]{BosGaddeB10}. The first two studies employed the exact same datasets as employed here, but performed independent analyses, and, reassuringly, the maps presented agree both qualitatively and quantitatively with those we present in Figs.\,\ref{fig:maps} and \ref{fig:maps2}. The study presented in \citet{BosGaddeB10} is based on spectra taken with the deployable integral-field units of the FLAMES/GIRAFFE spectrograph at the VLT. The spectral resolution is about five times better than that of MUSE but the spectra are restricted to the calcium triplet wavelength range. The authors reported a central velocity dispersion of $\approx195$\,km\,s$^{-1}$, which is slightly above the value we derive, namely 168\,km\,s$^{-1}$, but given the typical uncertainty of $\sim10$\,km\,s$^{-1}$, the two measurements are statistically equivalent. Bosma et al. also reported measurements off the main bar in NGC\,1291 at about 20\arcsec\ from the centre: one can see a drop in velocity dispersion to $\approx110$\,km\,s$^{-1}$ and a radial velocity of only a few km\,s$^{-1}$ (the galaxy is very close to face-on), and both features match our own measurements very well (see Fig.\,\ref{fig:maps}).

\section{A Comparison with Structural Analysis from Photometry}
\label{sec:decomps}

As shown above, detailed maps of $v$, $\sigma$, h$_3$ and h$_4$ are a powerful tool to understand the nature of the different stellar structures in a galaxy. These maps show that in all galaxies studied here (with no more than two possible exceptions) there clearly is a fast rotating nuclear disc, separate from the main galaxy disc. 

In photometric decompositions, the central component in disc galaxies is often fitted using the \citet{Ser68} function, where the S\'ersic index $n$ regulates how centrally concentrated stars are, with large values of $n$ produced by high concentrations of stars. Typically, one would expect that nuclear discs have an exponential surface brightness radial profile (i.e., with $n\approx1$), just as main discs. In practice, it is common to associate disc-like bulges to values of $n\leq2$, whereas classical bulges are associated to higher values.  


However, establishing the physical nature of bulges via photometric structural analysis is not straightforward, since a stellar component with $n\approx1$ is, in principle, not necessarily dynamically supported by rotation. The latter can only be directly probed with measurements of the stellar kinematics. Conversely, a stellar structure with $n>2$ may not necessarily be dispersion-dominated. Given the large body of ancillary data for the TIMER galaxies, we are in an excellent position to directly compare our stellar kinematic measurements with results from studies on the photometric properties of these galaxies. This allows us to test whether the nuclear discs we find in the TIMER sample are correctly identified via photometric structural analysis, and this is the main goal of this section.

It is important to stress that an accurate image decomposition depends on the physical spatial resolution, depth and nature of the data \citep[dust effects at short wavelengths are known to bias the results, see, e.g.,][]{deJ96b,GadBaeFal10,PasPopTuf13}. It also depends on the procedures employed to account for the exceedingly complex stellar structure in disc galaxies. In fact, many authors have argued that the distinction between disc-like and classical bulges using the S\'ersic index alone is prone to uncertainties, and that this analysis is only more robust when several criteria are used together, including criteria based on stellar kinematics and intrinsic shape \citep[see, e.g.,][]{FisDro16,NeuWisCho17,CosCorMen18,CosMenCor18}.

Despite these caveats, to keep the comparison straightforward, we will simply verify what values of the S\'ersic index $n$ are obtained for the photometric bulges in decompositions of the TIMER galaxies, and use the criterion $n\leq2$ for a successful identification of a nuclear disc, as opposed to a classical bulge. We will base this comparison on two recent studies separately in the following subsections. These studies were chosen here given the large overlap with the TIMER sample and the effort put in producing accurate results, with sophisticated models and careful procedures (which typically include individual inspection of the fits and numerous fits per galaxy to understand the uncertainties and avoid local $\chi^2$ minima).

\subsection{Kim et al. (2014)}


This work made use of S$^4$G 3.6$\mu$m images of 144 barred galaxies to derive structural parameters of bulges, bars and discs using BUDDA \citep{deSGaddos04,Gad08}, taking advantage of the exquisite depth of the S$^4$G data and the minimised impact of dust at these wavelengths. Nuclear point sources and outer disc breaks were accounted for to avoid biasing the bulge parameters. The point spread function (PSF) was determined for each image separately and modelled as a circular \citet{Mof69} function. The surface brightness radial profiles of bars were modelled using a S\'ersic function, and a bar was included in the models of all 13 TIMER galaxies that also belong to their sample. 


For eight of the 13 galaxies, the photometric bulge S\'ersic index is less than two, so in the majority of the cases the photometric bulge is identified as a disc-like bulge, in agreement with our kinematic analysis. For the remaining five galaxies, the bulges are found to have $n>2$, which would in principle indicate that in these cases the S\'ersic index fails to correctly identify the nature of the central component. However, one important caveat is that some galaxies may host composite bulges, i.e., a small classical bulge at the very centre surrounded by a nuclear disc \citep[see, e.g.,][who reported effective radii for small classical bulges ranging from 25 to 430\,pc]{ErwSagFab15}. For  example, NGC\,1291 clearly has a nuclear disc and an inner bar, and other studies have shown that it has also a small classical bulge with an effective radius of 416\,pc within the nuclear disc \citep[see][]{MendeLGad19,deLSanMen19}. Thus, the bulge component in the fits of \citet{KimGadShe14} accounts for both the nuclear disc and the small classical bulge, which pushes the S\'ersic index of the central component to $n=2.7$. The other bulges with $n>2$ are those in IC\,1438, NGC\,1433, NGC\,4303 and NGC\,7140. In IC\,1438 there is a clear and substantial peak in the stellar velocity dispersion in the inner $1-2''$, combined with a drop in h$_4$, that suggests that this galaxy too has a small dispersion-dominated component within the nuclear disc. Nevertheless, for the remaining three galaxies we see no evidence of a small classical bulge in the kinematic maps.

In this context, it is important to note that the S$^4$G images, with a typical PSF full width half maximum (FWHM) of about $2''$, are inadequate to separate central components on those spatial scales. Therefore, the issue of composite bulges highlights the importance of high physical spatial resolution in imaging data, and the accounting of the different structural components in the models used to fit galaxy images.

We also point out that no bulge was found to have $n\approx4$, which is expected for central regions dominated by classical bulges and for massive elliptical galaxies \citep[e.g.,][and references therein]{Gad09b}. In fact, the largest value of $n$ found is below 3, which indicates that there is no galaxy in the TIMER sample with a dominant classical bulge, a conclusion that is corroborated in our accompanying paper (Bittner et al. 2020, subm.) through an analysis of stellar populations properties. This may result from the selection of the TIMER sample, which favours galaxies hosting nuclear components with a disc-like morphology. However, it is not clear if this criterion rejects galaxies hosting large classical bulges.

\subsection{Salo et al. (2015)}


All galaxies in the TIMER sample were studied in \citet{SalLauLai15}, which also employed S$^4$G 3.6$\mu$m images but did not account for disc breaks. A single oversampled PSF image that accurately reproduces the particular PSF 2D shape in the S$^4$G images was used in all decompositions, and bars were modelled using a modified Ferrers profile. The decompositions were performed using GALFIT \citep{PenHoImp02,PenHoImp10}. 
Interestingly, bars were not included in the models for NGC\,4303, 4981, 5248 and NGC\,6902.


Only two of the photometric bulges were modelled with $n>2$. One is again in NGC\,1291, and the other is in NGC\,7552. As discussed above, NGC\,1291 hosts a small classical bulge, but we find no evidence for a similar component in NGC\,7552. As in the study by Kim et al. no bulge was found with $n>3$.

A comparison between the values obtained by the two studies for the S\'ersic index of the photometric bulges shows some noticeable discrepancies, which again highlight how large is the uncertainty in the measurement of the S\'ersic index. While for many galaxies the agreement between the two studies is remarkable, with an absolute difference in $n$ of only $0.1-0.2$, in some cases the difference reaches values above unity, and the measurements of Kim et al. are systematically above those of Salo et al. The median and mean values of the absolute difference in the S\'ersic index measurements are 0.9 and 0.7, respectively. These discrepancies also show that even using the same dataset, a different outcome may result if models and/or techniques employed are different. 

\subsection{Exponential photometric bulges are nuclear discs}

In the previous two subsections we have seen that photometric decompositions can retrieve reasonably well the nature of photometric bulges, and do so in the majority of the cases studied here. It is important to stress, however (and in addition to the caveats on photometric decompositions mentioned above), that the TIMER sample consists of nearby galaxies, and thus the physical spatial resolution of the images employed in these studies is relatively high. Decompositions employing images of more distant galaxies, with lower physical spatial resolution, will presumably not perform as well, but to quantify this is beyond the scope of this paper.

\begin{figure}
\begin{center}
	\includegraphics[width=.98\columnwidth]{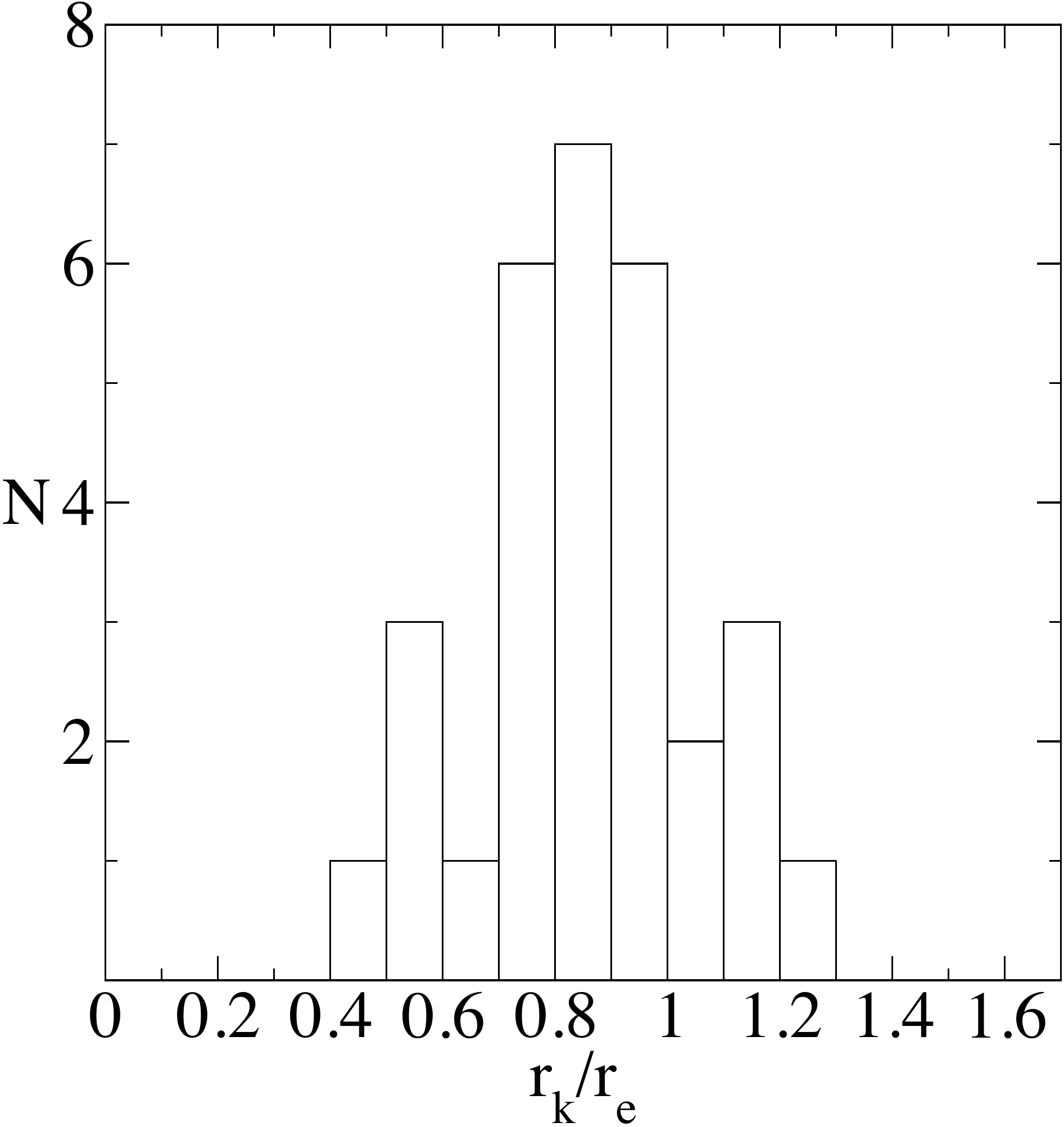}
\end{center}
       \caption{Distribution of the ratio between nuclear disc size, defined kinematically as the radius with maximum $v/\sigma$, and the effective radius of the photometric bulge component fitted through careful 2D multicomponent photometric decompositions. The values are clustered around a single value, following a normal distribution (${\rm r}_{\rm k}/{\rm r}_{\rm e}=0.86\pm0.19$), which shows that the photometric component defined as the `bulge' is (or is dominated by) the rapidly-rotating nuclear disc.}
    \label{fig:sizes}
\end{figure}

A powerful comparison to further test whether the photometric bulges modelled by Kim et al. and Salo et al. are indeed the rapidly-rotating nuclear discs we identify with the MUSE TIMER data cubes -- or are at least dominated by them, in the case of composite bulges -- concerns comparing their sizes, as measured by the different techniques. The photometric decompositions readily provide the effective radii (r$_{\rm e}$) of the fitted component, but a different approach is needed to derive the dimensions of the nuclear discs from the kinematic measurements. To this end, we use the kinematic radius (r$_{\rm k}$) defined above as the distance from the galaxy centre of the spatial bin showing the maximum value of  $v/\sigma$ in the region dominated by the rapidly-rotating nuclear disc. If the photometric and kinematic components are in fact the same physical entity then the ratio ${\rm r}_{\rm k}/{\rm r}_{\rm e}$ must be clustered around a single value, following a normal distribution. Figure \ref{fig:sizes} shows that this is indeed the case: ${\rm r}_{\rm k}/{\rm r}_{\rm e}$ is clustered around 0.86. Fitting a normal distribution to our measured values of ${\rm r}_{\rm k}/{\rm r}_{\rm e}$ yields a mean value of 0.86 and $\sigma=0.19$. In addition, we applied to our data the statistical tests presented in \citet{DAgBelDAg90} and \citet{AnsGly83} to verify that the distribution in Fig.\,\ref{fig:sizes} has indeed both skewness and kurtosis that are statistically compatible with a normal distribution. Furthermore, the value of $\sigma$ derived above is compatible with the typical uncertainty in the measurement of ${\rm r}_{\rm e}$, which is $\approx20\%$.

One important aspect to consider is that correlations between ${\rm r}_{\rm k}$ and ${\rm r}_{\rm e}$ with other parameters could produce the results presented in Fig.\,\ref{fig:sizes}, even if there is no physical connection between the nuclear discs and photometric bulges in our sample. For example, one could argue that a galaxy with a large disc would naturally have a large photometric bulge and a large nuclear disc, and thus the correlation between ${\rm r}_{\rm k}$ and ${\rm r}_{\rm e}$ would be trivial. We test this possibility using as a proxy for disc size (R$_{25.5}$) the 25.5\,mag\,arcsec$^{-2}$ isophotal radius at 3.6$\mu$m, as derived by \citet{MunSheReg15} with S$^4$G images (see Table\,\ref{tab:ids}). We find that the Pearson correlation coefficient R for the three relations, ${\rm r}_{\rm k}$ vs ${\rm r}_{\rm e}$,  ${\rm r}_{\rm k}$ vs R$_{25.5}$, and ${\rm r}_{\rm e}$ vs R$_{25.5}$, is, respectively: 0.87, 0.63 and 0.75. Thus, the correlation between ${\rm r}_{\rm k}$ and ${\rm r}_{\rm e}$ is  stronger, and therefore not trivial. Considering these results altogether, we can safely state that exponential `bulges' in photometric decompositions are nuclear discs. It is also important to bear in mind that ${\rm r}_{\rm k}$ and ${\rm r}_{\rm e}$ are measured through completely different techniques, which lends support to this conclusion.

\section{The origin of nuclear discs}
\label{sec:origins}

\begin{table*}
	\centering
	\caption{Properties of nuclear discs, bars and galaxies in the TIMER sample. Column (1) gives the galaxy designation, while columns (2) and (3) show, respectively, the peak value of $v/\sigma$ in the nuclear disc, and the radius at which this peak is located, r$_{\rm k}$. In columns (4) and (5) we show the effective radius of the photometric bulge derived by \citet{KimGadShe14} and \citet{SalLauLai15}, respectively. Columns (6), (7) and (8) present the bar radius, ellipticity and bar-to-total luminosity ratio as derived by \citet{KimGadShe14}. In columns (9) and (10) we present the values of Q$_{\rm B}$ and A$_2$ calculated by \citet{DiaSalLau16}. Finally, column (11) shows the values of R$_{25.5}$ obtained by \citet{MunSheReg15}. All radii are in kpc, using the distances tabulated in Paper I, and the bar radii and ellipticities derived by Kim et al. are deprojected following the 2D approach described in \citet{GadAthCar07}.}
	\label{tab:ids}
	\begin{tabular}{ccccccccccc}
		\hline
Galaxy & $v/\sigma$ & r$_{\rm k}$ & r$_{\rm e, K}$ & r$_{\rm e, S}$ & R$_{\rm bar}$ & $\epsilon_{\rm bar}$ & Bar/T & Q$_{\rm B}$ & A$_2$ & R$_{25.5}$ \\
(1) & (2) & (3) & (4) & (5) & (6) & (7) & (8) & (9) & (10) & (11)\\
		\hline
IC\,1438   &  2.57  &  0.60  & 0.77  & 0.56 &  4.43  & 0.53  & 0.11   &  0.178  & 0.838 & 12.76  \\
NGC\,613  &  2.35  &  0.59  & 0.71  & 0.68 &  9.72  & 0.62  & 0.26   &  0.489  & 0.903 & 23.60  \\
NGC\,1097  &  2.82  &  1.07  & 0.95  & 1.24 & 10.40  & 0.45  & 0.26   &  0.254  & 0.709 & 36.24  \\
NGC\,1300  &  2.98  &  0.33  & 0.44  & 0.38 &  6.70  & 0.75  & 0.08   &  0.58   & 0.603 & 17.88  \\
NGC\,1433  &  2.00  &  0.38  & 0.42  & 0.42 &  3.63  & 0.68  & 0.08   &  0.366  & 0.560 & 12.82  \\
NGC\,3351  &  2.57  &  0.24  & 0.41  & 0.33 &  4.02  & 0.70  & 0.09   &  0.227  & 0.513 & 12.41  \\
NGC\,4303  &  1.36  &  0.21  & 0.39  & 0.29 &  3.52  & 0.56  & 0.06   &  0.535  & 0.550 & 18.14  \\
NGC\,4371  &  2.02  &  0.95  & \omit & 0.82 &  \omit & \omit & \omit  &  0.234  & 0.618 & 16.38  \\
NGC\,4643  &  1.31  &  0.50  & \omit & 0.86 &  \omit & \omit & \omit  &  0.272  & 0.813 & 24.45  \\
NGC\,4981  &  0.99  &  0.14  & \omit & 0.29 &  \omit & \omit & \omit  &  0.093  & 0.172 & 13.56  \\
NGC\,4984  &  2.49  &  0.49  & 0.53  & 0.55 &  6.18  & 0.48  & 0.14   &  0.176  & 0.836 & 18.98  \\
NGC\,5236  &  1.89  &  0.37  & \omit & 0.32 &  \omit & \omit & \omit  &  0.472  & 0.467 & 19.35  \\
NGC\,5248  &  1.66  &  0.49  & \omit & 0.58 &  \omit & \omit & \omit  &  0.138  & 0.324 & 17.28  \\
NGC\,5728  &  2.84  &  0.63  & 0.83  & 0.86 &  9.70  & 0.51  & 0.28   &  0.387  & 1.149 & 20.58  \\
NGC\,5850  &  1.13  &  0.80  & 0.62  & 0.74 &  7.84  & 0.64  & 0.16   &  0.327  & 0.742 & 20.80  \\
NGC\,7140  &  1.45  &  0.63  & 0.99  & 0.66 & 11.16  & 0.36  & 0.23   &  0.399  & 0.805 & 24.86  \\
NGC\,7552  &  2.61  &  0.33  & 0.34  & 0.36 &  4.90  & 0.64  & 0.32   &  0.358  & 1.060 & 9.79   \\
NGC\,7755  &  2.20  &  0.47  & \omit & 0.53 &  \omit & \omit & \omit  &  0.401  & 0.841 & 20.29  \\
		\hline
	\end{tabular}
\end{table*}

\begin{figure*}
\begin{center}
	\includegraphics[width=2\columnwidth]{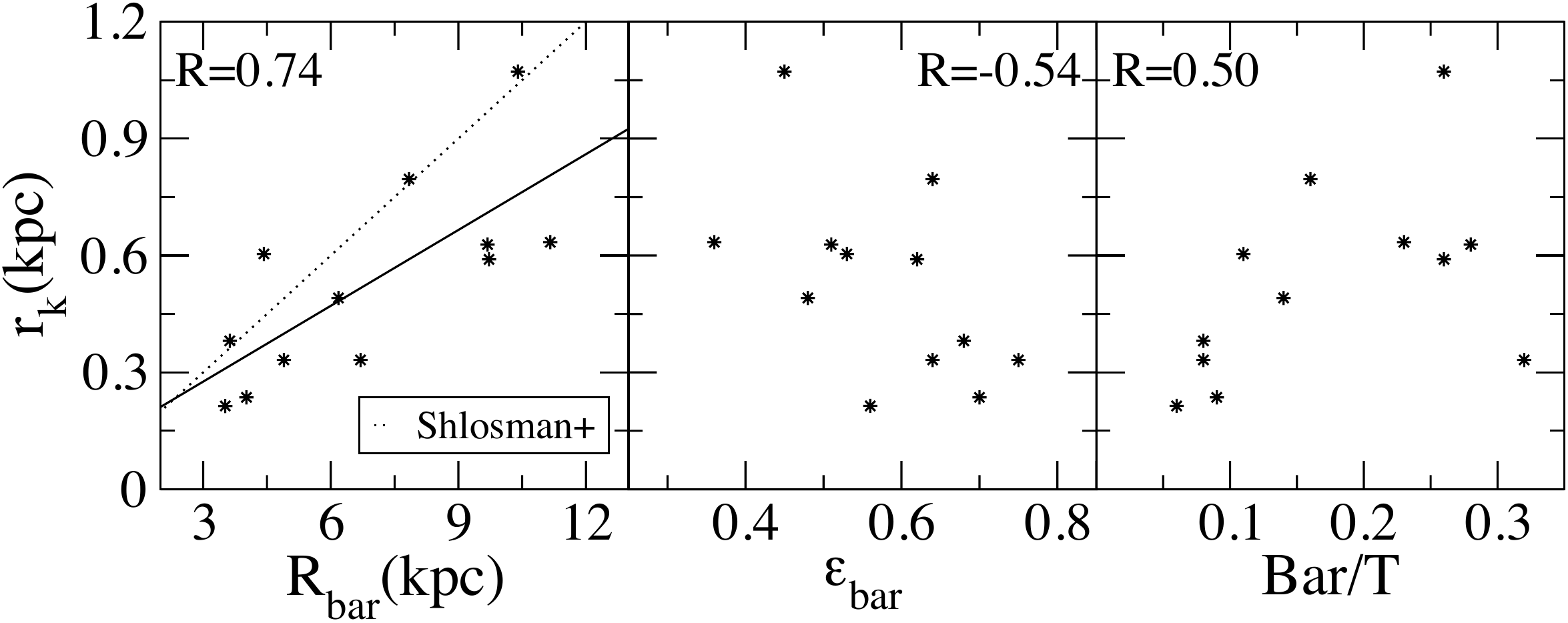}\\
	\vspace{0.2cm}
	\includegraphics[width=1.33\columnwidth]{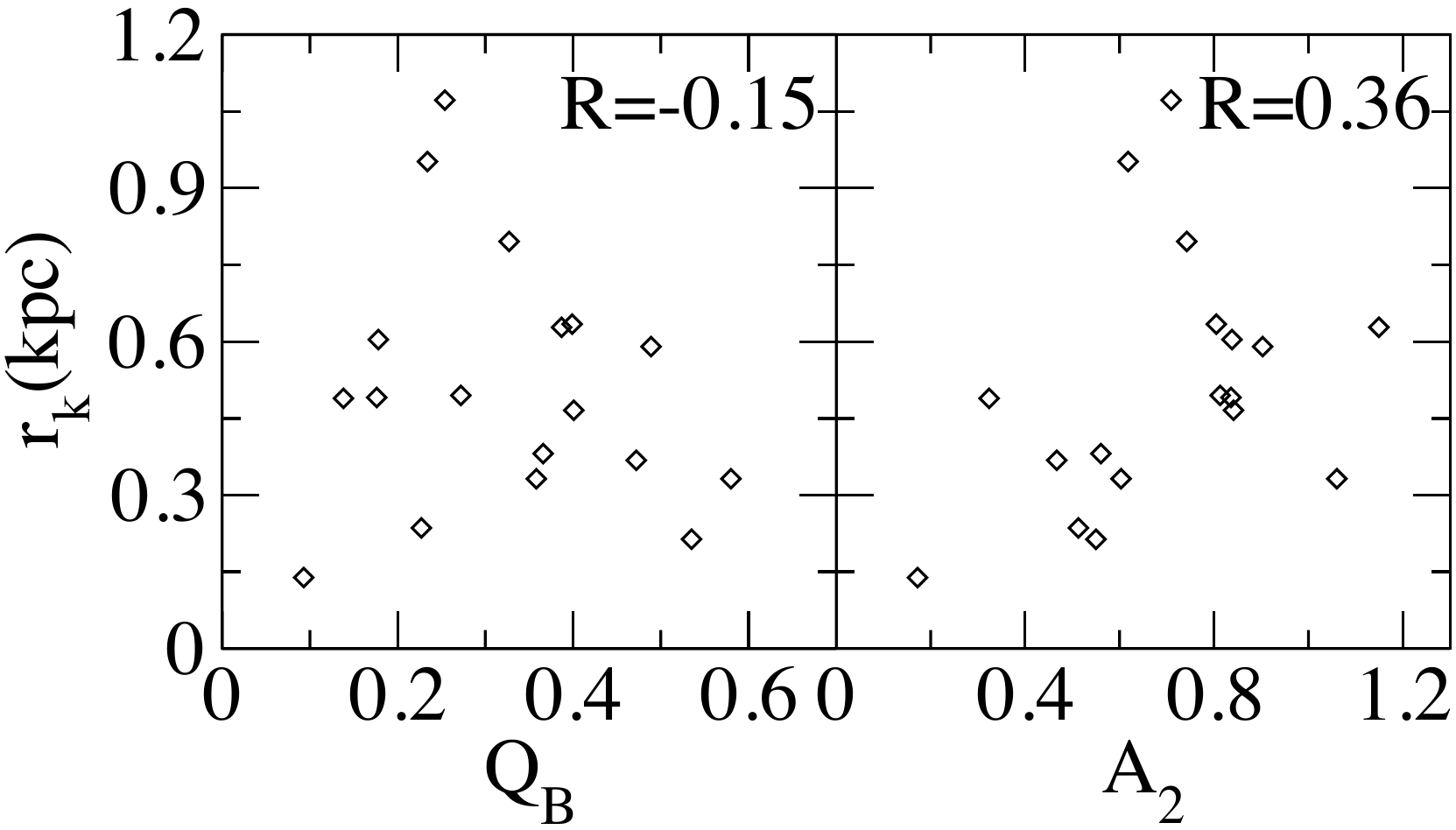}
\end{center}
    \caption{Relations between the kinematic sizes of nuclear discs and selected bar properties. The top left panel shows that r$_{\rm k}$ is correlated with bar size (semi-major axis). The solid line is a simple linear regression fit to the data, whereas the dotted line corresponds to ${\rm r}_{\rm k}=0.1{\rm R}_{\rm bar}$, the relation suggested in \citet{ShlFraBeg89}. In the top middle and right panels we present how the kinematic size of nuclear discs relates to bar ellipticity and bar-to-total luminosity ratio, respectively. The bottom panels show the relations between r$_{\rm k}$ and different measures of bar strength \citep[from][see text for details]{DiaSalLau16}. Bar radii and ellipticities are from \citet{KimGadShe14} and are deprojected. The Pearson correlation coefficient R is indicated in each panel. In Appendix\,\ref{app:stats} we present an analysis of the statistical properties of these relations.}
    \label{fig:barsQb}
\end{figure*}

While the results above are consistent with the scenario in which bars drive gas to the central region and ultimately build the observed nuclear discs, they do not rule out that, at least in some cases, accretion of gas could have been promoted by, e.g., an interaction, {\em before} the formation of the bar, and therefore be unrelated to it. Thus, to shed more light in this discussion (and given that bars are expected to grow longer and stronger with time; e.g., \citealt{Ath03}), in this section, we explore the properties of nuclear discs as derived from the TIMER kinematic maps in conjunction with other relevant bar properties. Table \ref{tab:ids} presents the physical parameters we explore. In particular, we use the values determined by \citet{KimGadShe14} from their photometric decompositions for the bar radius (R$_{\rm bar}$), ellipticity ($\epsilon$) and bar-to-total luminosity ratio (Bar/T). Further, \citet{DiaSalLau16}, presented a number of measurements of bar strength, from which we use Q$_{\rm B}$ and A$_2$. Q$_{\rm B}$ indicates how strong the non-axisymmetric potential of the bar is as compared to the axisymmetric component of the total stellar mass distribution. It thus provides an indication of the impact of the bar on the dynamics of gas and stars in the host galaxy, by taking into consideration the effects of any central spheroid and the main disc itself. On the other hand, A$_2$ is a measure of the $m=2$ Fourier component that is directly connected to the bar non-axisymmetry. More specifically, A$_2$ is calculated radially, normalised by the value of A$_0$ (the axisymmetric Fourier component) {\em at each radius}, and the value we employ here is the peak value of A$_2$/A$_0$ in the bar. Therefore, Q$_{\rm B}$ and A$_2$ provide complementary information on the bar and associated secular evolution processes. 

Figure \ref{fig:barsQb} shows that the radii of nuclear discs correlate significantly with bar radii, which is expected in the theoretical framework of the orbital structure in bars and how it evolves. In fact, the gas brought to the central region by bars is expected to accumulate in the region where the x$_2$ orbits of the bar dominate over the bar x$_1$ orbits\footnote{The x$_1$ orbits are eccentric and parallel to the bar major axis, and are present throughout the bar. They are thus considered the `backbone' of bars. The x$_2$ orbits are less eccentric and perpendicular to the bar major axis, and are only found in the bar inner region.}, where a nuclear stellar structure is thus formed \citep{Ath92b,KimSeoSto12,LiSheKim15,SorSobFra18,SeoKimKwa19}. As the bar evolves and grows, the region where the x$_2$ orbits dominate grows as well, hence the expected correlation\footnote{We point out that, with all things being equal, more eccentric bars will have smaller nuclear discs, since the extent of the x$_2$ orbits is then shorter \citep[see][]{Ath92b}. A nuclear disc will not grow in a bar that gets longer but keeps its semi-minor axis constant. In fact, Fig.\,\ref{fig:barsQb} shows that more eccentric bars tend to have smaller nuclear discs.}. This correlation between r$_{\rm k}$ and R$_{\rm bar}$ is therefore consistent with the picture in which the nuclear stellar discs studied here were formed from gas brought to the central region by the bar, and not built before the formation of the bar. In addition, in this picture, gas falling onto the outer boundary of the region dominated by x$_2$ orbits is prone to form stars and produce star-forming nuclear rings. Indeed, \citet{SeoKimKwa19} found in their simulations  that nuclear rings are larger when bars evolve for a longer period of time. As expected, \citet{ComKnaBec10} found that the relation between the sizes of bars and nuclear rings is such that the upper limit of the distribution of nuclear ring sizes correlates with bar size. It is unclear if the larger scatter seen in their work is due to uncertainties in the measurements, or whether it is real and brought out by the larger sample. Nevertheless, the connection between nuclear ring size and bar radius in that work is also clear.

Interestingly, in the pioneering work of \citet{ShlFraBeg89}, a model is put forward in which the bar sweeps gas from the region of the main disc within R$_{\rm bar}$, building a nuclear disc that is limited by the ILR radius. In their model, this radius is of the order of $0.1{\rm R}_{\rm bar}$, and the relation between r$_{\rm k}$ and R$_{\rm bar}$ we show in Figure \ref{fig:barsQb} is close to that. This is an indication of the correctness of their model.

Figure \ref{fig:barsQb} also shows that r$_{\rm k}$ tends to be larger for lower values of bar ellipticity and higher values of Bar/T. Again this can be naturally understood considering the connection between nuclear discs and the extent of the bar x$_2$ orbits. As x$_2$ orbits cannot extend past the bar edges, more elongated bars will tend to have less extended x$_2$ orbits, and therefore smaller nuclear discs, as we observe. As bars evolve, they capture stars from the disc, grow longer and become more massive, which increases Bar/T. In addition, with the buckling of the inner parts of the bar, and the formation of the box/peanut and barlens structure, the inner region of the bar becomes less elongated, decreasing the overall bar ellipticity. The less elongated inner region promotes the expansion of the x$_2$ orbits, creating the trends between r$_{\rm k}$ with bar ellipticity and Bar/T. Nevertheless, it is important to point out that in this study some of the observed trends are not strong correlations, and therefore more work is needed to confirm these conjectures (see Appendix\,\ref{app:stats} for an analysis of the statistical properties of these relations).

It is interesting to note that the range of bar radii in Fig.\,\ref{fig:barsQb} spans a factor of $\approx4$, although in simulations bars grow by factors no larger than $\approx2$ \citep[see, e.g.,][]{MarShlHel06}. This implies that the trends discussed here are not only the result of the way bar properties evolve, but also of the initial bar properties. On the other hand, the bar radius has been shown to be an increasing function of the {\em bulge}-to-total luminosity ratio (B/T) when normalised by R$_{25.5}$ \citep{KimGadShe14}. Moreover, the function appears to be the same whether the photometric bulge component is a classical bulge or a nuclear disc. \citet{KimGadShe14} suggested that this relation is expected since more prominent classical bulges can absorb substantial angular momentum from the bar, which leads to more significant bar growth \citep[see][]{AthMis02,Ath03}. On the other hand, Kim et al. argue that a similar relation is expected also in the case of nuclear discs, as longer bars will push more gas inwards to the central region, leading to more prominent nuclear discs \citep{Ath92b}. However, understanding why the relation is {\em the same} for classical bulges and nuclear discs remains a puzzle.

One also sees in Fig.\,\ref{fig:barsQb} that the relation between r$_{\rm k}$ and Q$_{\rm B}$ is such that for low values of Q$_{\rm B}$ nuclear discs show a wide range of sizes, but only small nuclear discs are seen when Q$_{\rm B}$ is large. This can in part be explained by the reduced extent of x$_2$ orbits in more eccentric bars. Again, a similar result was found by \citet{ComKnaBec10} for the sizes of nuclear rings. Interestingly, we find a trend (albeit with some scatter) between r$_{\rm k}$ and A$_2$ that to the best of our knowledge has not been reported elsewhere. While A$_2$ is a purely photometric parameter, r$_{\rm k}$ is a purely kinematic one, and therefore this trend too is consistent with a picture in which the formation of the nuclear disc is connected to the bar.


Interestingly, the relation between r$_{\rm k}$ and A$_2$ does not show the same properties as the relation between r$_{\rm k}$ and Q$_{\rm B}$. While the connection between r$_{\rm k}$ and Q$_{\rm B}$ is similar to the connection between r$_{\rm k}$ and bar ellipticity, the correlation between r$_{\rm k}$ and A$_2$ resembles that between r$_{\rm k}$ and Bar/T.  This can be understood in the way A$_2$ and Q$_{\rm B}$ are defined. Although both A$_2$ and Q$_{\rm B}$ are measurements of bar strength, Q$_{\rm B}$ takes into account the overall axisymmetric galactic potential, whereas A$_2$ accounts for it only at the radius where A$_2$ is measured, typically close to the end of the bar. As described above, this is done with the normalisation of A$_2$ by the value of A$_0$ at that radius. Higher values of A$_2$ are produced if the density of stars in the bar increases while it decreases in the region of the disc outside the bar but within the bar radius. \citet{KimGadAth16} showed that, as bars evolve, they capture stars in the disc region within the bar radius, reducing the density of stars in the region. Further, they showed that evolved bars thus tend to have elevated values of A$_2$, a result that was confirmed in \citet{But17}. This is corroborated by the finding from \citet{DiaSalLau16} that longer bars are stronger and that this correlation is particularly tight if bar strength is measured as A$_2$, since bars are expected to grow longer as they evolve. Therefore, the trend between r$_{\rm k}$ and A$_2$ suggests that galaxies with more evolved bars tend to have larger nuclear discs. We already discussed above that more evolved bars should naturally host larger nuclear discs due to the increase in the extension of the bar x$_2$ orbits as bars evolve. Furthermore, we speculate that this is also a result of the bar being able to induce gas inflows for longer, thus producing more massive nuclear discs. In fact, \citet{ColDebErw14} found that their simulations produce smaller nuclear discs if the inflow of gas is halted. Nevertheless, we stress that there is substantial scatter in the relation between r$_{\rm k}$ and A$_2$ seen in Fig.\,\ref{fig:barsQb}, and thus further studies are needed to confirm this relation (see Appendix\,\ref{app:stats}). In addition, more work is necessary to understand if the scatter is caused not only by difficulties in the measurements, but if other physical properties or processes play a role in the evolution of r$_{\rm k}$ and A$_2$.

Further work with larger samples would also be helpful to rule out that some of these correlations could result from underlying correlations, e.g., between bar size and disc size \citep[see][]{Erw05,Gad11}. However, we highlight that the anti-correlation we observe between r$_{\rm k}$ and bar ellipticity cannot be explained by the underlying {\em correlation} between bar size and ellipticity, simply because these correlations have opposite signs. Since the anti-correlation between r$_{\rm k}$ and bar ellipticity is predicted by the bar-driven model for the formation of nuclear discs, it suggests that the nuclear discs in the TIMER sample are indeed built by the bar.

We point out that while all galaxies plotted in Fig.\,\ref{fig:barsQb} have nuclear discs, not all of them have nuclear rings, according to the evaluation by \citet{ComKnaBec10}. These galaxies are NGC\,4643, 4981, 4984, 5850, 7140 and NGC\,7755. Nuclear rings are found at the edge of nuclear discs when both structures are present, close to or at the radius where $v/\sigma$ peaks, i.e., at r$_k$. We verified that these galaxies are not systematically off in any of the relations shown in this section. Therefore, both nuclear rings and nuclear discs follow the same relations, suggesting that the formation of both structures stems from the same fundamental process. The results discussed above suggest that the main processes providing the gaseous content for the formation of both nuclear rings and nuclear discs are bar-driven, and the connection between nuclear rings and nuclear discs suggests that a single set of joint mechanisms is in charge of transforming the gas component into these stellar structures. In our accompanying paper (Bittner et al. 2020, subm.) we study this connection in more detail through the analysis of the stellar population properties of nuclear discs and nuclear rings.

Processes unrelated to bars have also been studied as possible drivers of the building of nuclear discs. Recent simulations have shown that nuclear discs can be produced at the late stage of gas-rich mergers \citep[see][]{QueEliTap15,WanHamPue15,SauHamPue18}. However, we note that most of these nuclear discs are substantially larger than the largest nuclear disc we find. Also, it is important to point out that even in this case bar-driven secular evolution processes can still play a major role in the building of the nuclear disc. The merger may simply serve as a mechanism to bring gas to the disc region within the bar radius, and from this point the bar is responsible to stream the gas further inwards. The merger can also trigger the bar instability if it was not spontaneously triggered before the interaction. Bars can be short-lived in major mergers, so it is unclear whether bars have played no role in the aforementioned simulations. While we present above suggestive evidence that in the TIMER sample of barred galaxies the nuclear discs are built by bars, such mergers can result in unbarred galaxies hosting nuclear discs, even if a bar played a major role but is later dissolved due to the merger.

On the other hand, gas-poor mergers are not expected to produce such regular and rapidly rotating structures. However, as we described in the Introduction, \citet{EliGonBal11} argue that nuclear discs are built in their collisionless simulations when specific conditions are met. Nevertheless, an inspection of the corresponding kinematical maps (see their figures 12 to 15) reveals that these simulations do not reproduce the signatures of nuclear discs we discussed in Sect.\,\ref{sec:discs}. Specifically, our nuclear discs are characterised by conspicuous drops in $\sigma$ and an anti-correlation between $v$ and h$_3$ (indicating near-circular orbits). In contrast, in the aforementioned simulations, the central region shows an inward increase in $\sigma$, and, in most cases, a {\em correlation} between $v$ and h$_3$. Therefore, we conclude that, to date, there is no  evidence that the nuclear discs in TIMER could have been built in gas-poor mergers.

This is not to say that the building of nuclear discs through bar-driven processes is perfectly well understood in simulations. In fact, the bar-built nuclear disc in the model of \cite{ColDebErw14} is too large compared to its bar and does not fit in the observed relation we present in Fig.\,\ref{fig:barsQb}. While the observed ratio between nuclear disc size and bar size is of the order of 10\%, the nuclear disc size in the model of Cole et al. is about 30\% of the bar size. More theory work is necessary, and it would be particularly helpful to understand how simulations can reproduce the observed trends in Fig.\,\ref{fig:barsQb}.

\section{General discussion and conclusions}
\label{sec:conc}

The results presented in Sects.\,\ref{sec:discs} and \ref{sec:origins} show a good agreement with the bar-driven secular evolution scenario for the building of nuclear stellar components in disc galaxies, at least for massive galaxies with conspicuous bars and nuclear structures. The nuclear discs found with the TIMER data have all the kinematic properties expected in a stellar structure built via the collapse of molecular clouds brought to the central region along the leading edges of the bar and put into near-circular orbits in the region where the bar x$_2$ orbits dominate: large rotational support, low velocity dispersion and an anti-correlation between $v$ and h$_3$ consistent with near-circular orbits (see Figs.\,\ref{fig:maps} and \ref{fig:maps2}). Further, these nuclear discs are clearly an additional component on top of the original main galaxy disc. This can be seen from the elevated values of h$_4$, resulting from the fact that the stars in the nuclear discs have orbits closer to circular than those in the main discs at the radii where the nuclear disc dominates (Figs. \ref{fig:maps}, \ref{fig:maps2} and \ref{fig:vos}). These properties are also consistent with a formation scenario with gas accretion that is unrelated to bars. However, Fig.\,\ref{fig:barsQb} shows how the size of the nuclear disc depends on the bar radius, ellipticity, bar-to-total ratio and bar strength, which is qualitatively understood in the theoretical framework of bar evolution and the impact of bars on the gas component. These connections between the nuclear disc size and bar properties are only expected in a scenario where the accretion of gas that builds the nuclear disc is due to the bar. Figure \ref{fig:barsQb} also suggests that more evolved bars tend to build larger nuclear discs, which may help putting constraints on the ages of bars, although more theoretical work on the formation and evolution of nuclear discs is necessary. Finally, in our accompanying paper (Bittner et al. 2020, subm.), we present stellar population properties of the TIMER nuclear discs, and discuss how these properties too are consistent with the picture in which such nuclear discs are built via bar-driven processes.

As mentioned above, it is still unclear how rare are nuclear discs in unbarred galaxies \citep[but see][]{ComKnaBec10}. Weak bars and oval distortions in the disc can be difficult to identify morphologically but may as well produce nuclear discs in the same way as prominent bars. Nevertheless, a similar study as we present here but with a sample of unbarred galaxies showing nuclear discs would be very beneficial. A comparison between the physical properties of nuclear discs in barred and unbarred galaxies (such as size and angular momentum) would shed light on their formation process.


As we have seen in Sect.\,\ref{sec:decomps}, in photometric decompositions of galaxy images, one often finds an exponential or near-exponential central component, i.e., a photometric bulge with low S\'ersic index ($n\lesssim2$), which is thought to originate from bar-driven secular processes. In agreement with this expectation, the results presented in Sect.\,\ref{sec:decomps} show that the nuclear discs we find spectroscopically in TIMER, through an assessment of the stellar kinematics, are indeed often recognised as (near-)exponential bulges in the photometric studies we considered. This is encouraging, despite the caveats discussed above on using the S\'ersic index alone to separate nuclear discs from classical bulges. However, we stress again that the TIMER sample was built to include galaxies for which a visual assessment alone is already capable of identifying nuclear components that appear to have disc-like properties. This is a result of the relatively high physical spatial resolution of the images employed. Arguably, the recovery of nuclear discs via photometry is more prone to errors when the spatial resolution is not suitable, e.g., for more distant galaxies.

Even in these optimal circumstances, Sect.\,\ref{sec:decomps} shows that some difficulties are encountered, e.g., when performing photometric decompositions of galaxies with composite bulges. A small classical bulge within a nuclear disc can dominate the light emission in such a way that the presence of a bar-built nuclear component goes unnoticed in the photometry. This problem is of course exacerbated if the spatial resolution is not high enough to separate the two components, and/or a single model is used to fit the central region, erroneously encompassing all structures therein. It is as yet not clear how often disc galaxies host small classical bulges embedded in nuclear discs so the severity of this problem is still unknown. However, our results show that composite bulges may produce values for the S\'ersic index between 2 and 3, suggesting that pure classical bulges can be identified only by putting a more stringent threshold at $n>3$ (especially considering the corresponding uncertainties). Nevertheless, this is to be considered with caution and, particularly, in a statistical sense \citep[see][]{MenAguFal18}.

The TIMER maps shown in Figs.\,\ref{fig:maps}, \ref{fig:vos} and \ref{fig:maps2} show several kinematic signatures corroborating theoretical work on the dynamical properties of bars and inner bars. This is also the case for box/peanuts (Sect.\,\ref{sec:bps}). This allows us to put a lower limit in the fraction of massive barred galaxies with box/peanuts at 62\%, in broad agreement with previous results (with the caveat that our sample selection favours conspicuous bars and nuclear components). The case of NGC\,5728 is remarkable, in that it shows an agreement between kinematic and photometric considerations on the properties of box/peanuts in inclined disc galaxies. Concerning barlenses (Sect.\,\ref{sec:bls}), we find evidence corroborating previous studies that find that barlenses are simply the face-on projection of box/peanuts, i.e., of the inner parts of bars, since we see in the barlens region kinematic signatures of both a bar and a box/peanut, as expected from numerical simulations. This is the first time kinematic evidence is presented to support this picture. Barlenses are often difficult to identify photometrically and have morphologies similar to classical bulges, and therefore identifying them through their kinematic properties can help putting more accurate constraints on the impact of mergers in the evolution of disc galaxies. Altogether, these results show the power of spatially resolved kinematics in producing a straightforward understanding of the physical properties and nature of stellar structures in the inner regions of galaxies, and suggest that high-quality integral-field spectroscopy data is necessary to accurately decompose these complex inner regions.

\begin{acknowledgements}
We thank the anonymous referee for timely and helpful reports. Based on observations collected at the European Southern Observatory under ESO programmes 097.B-0640(A), 095.B-0532(A), 094.B-0321(A) and 060.A-9313(A). J. F-B, AdLC, and PSB acknowledge support through the RAVET project by the grants AYA2016-77237-C3-1-P, AYA2016-77237-C3-2-P and PID2019-107427GB-C31 from the Spanish Ministry of Science, Innovation and Universities (MCIU). J. F-B and AdLC acknowledge support through the IAC project TRACES which is partially supported through the state budget and the regional budget of the Consejer\'ia de Econom\'ia, Industria, Comercio y Conocimiento of the Canary Islands Autonomous Community. JMA acknowledges support from the Spanish Ministry of Economy and Competitiveness (MINECO) by grant AYA2017-83204-P. TK was supported by the Basic Science Research Program through the National Research Foundation of Korea (NRF) funded by the Ministry of Education (No. 2019R1A6A3A01092024). The Science, Technology and Facilities Council is acknowledged by JN for support through the Consolidated Grant Cosmology and Astrophysics at Portsmouth, ST/S000550/1. GvdV acknowledges funding from the European Research Council (ERC) under the European Union's Horizon 2020 research and innovation programme under grant agreement No 724857 (Consolidator Grant ArcheoDyn). This research has made use of NASA's Astrophysics Data System Bibliographic Services. We acknowledge the usage of the HyperLeda database (http://leda.univ-lyon1.fr). This research has also made use of the NASA/IPAC Extragalactic Database (NED), which is operated by the Jet Propulsion Laboratory, California Institute of Technology, under contract with the National Aeronautics and Space Administration.
\end{acknowledgements}

\bibliographystyle{aa} 
\bibliography{../../../gadotti_refs} 
%

\appendix
\section{Further kinematic maps and discussion on exceptional cases}
\label{app:maps}

\begin{figure*}
\begin{center}
	\includegraphics[clip=true, trim=20 20 20 20, width=0.85\columnwidth]{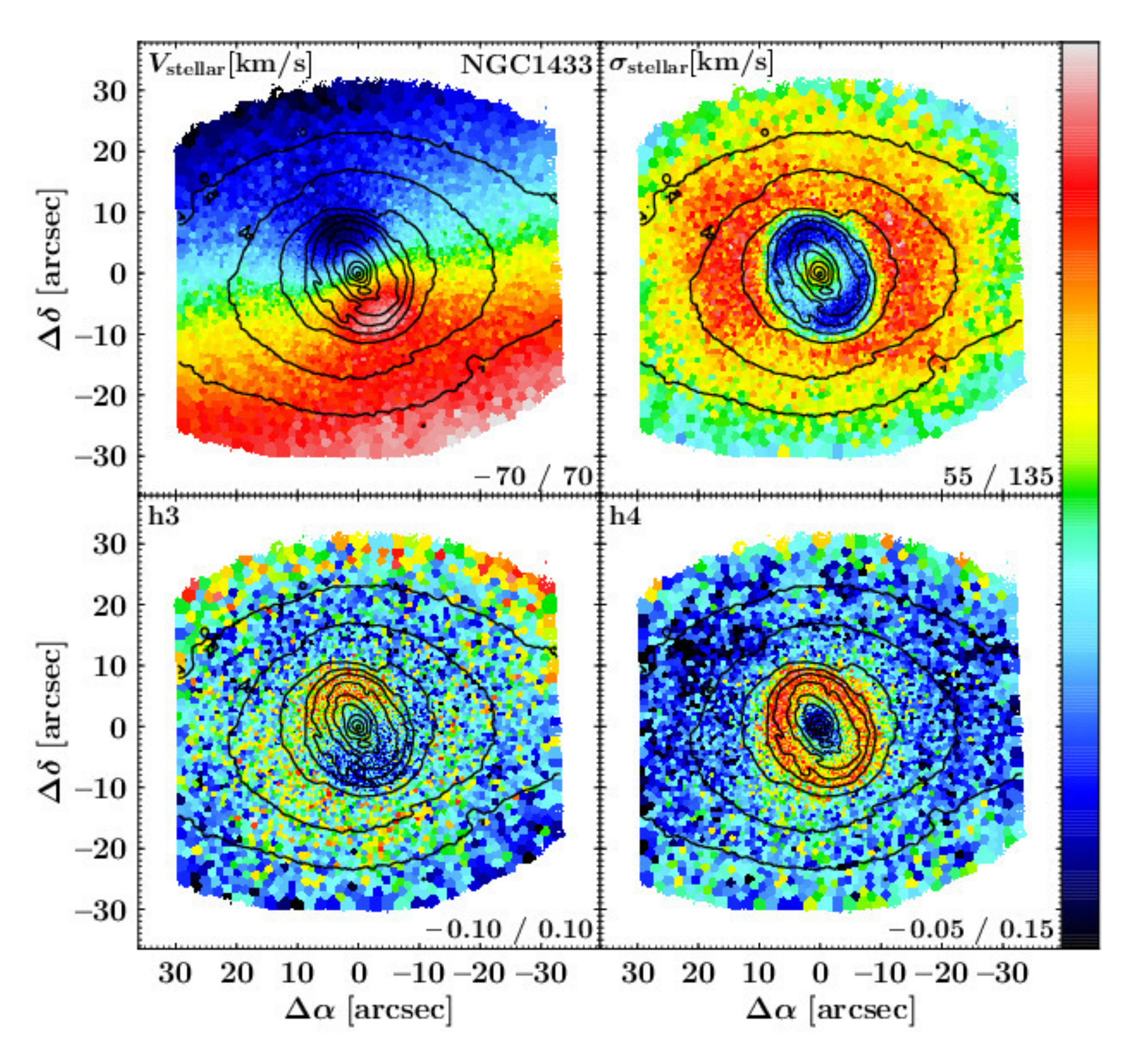}\hskip0.5cm
	\includegraphics[clip=true, trim=20 20 20 20, width=0.85\columnwidth]{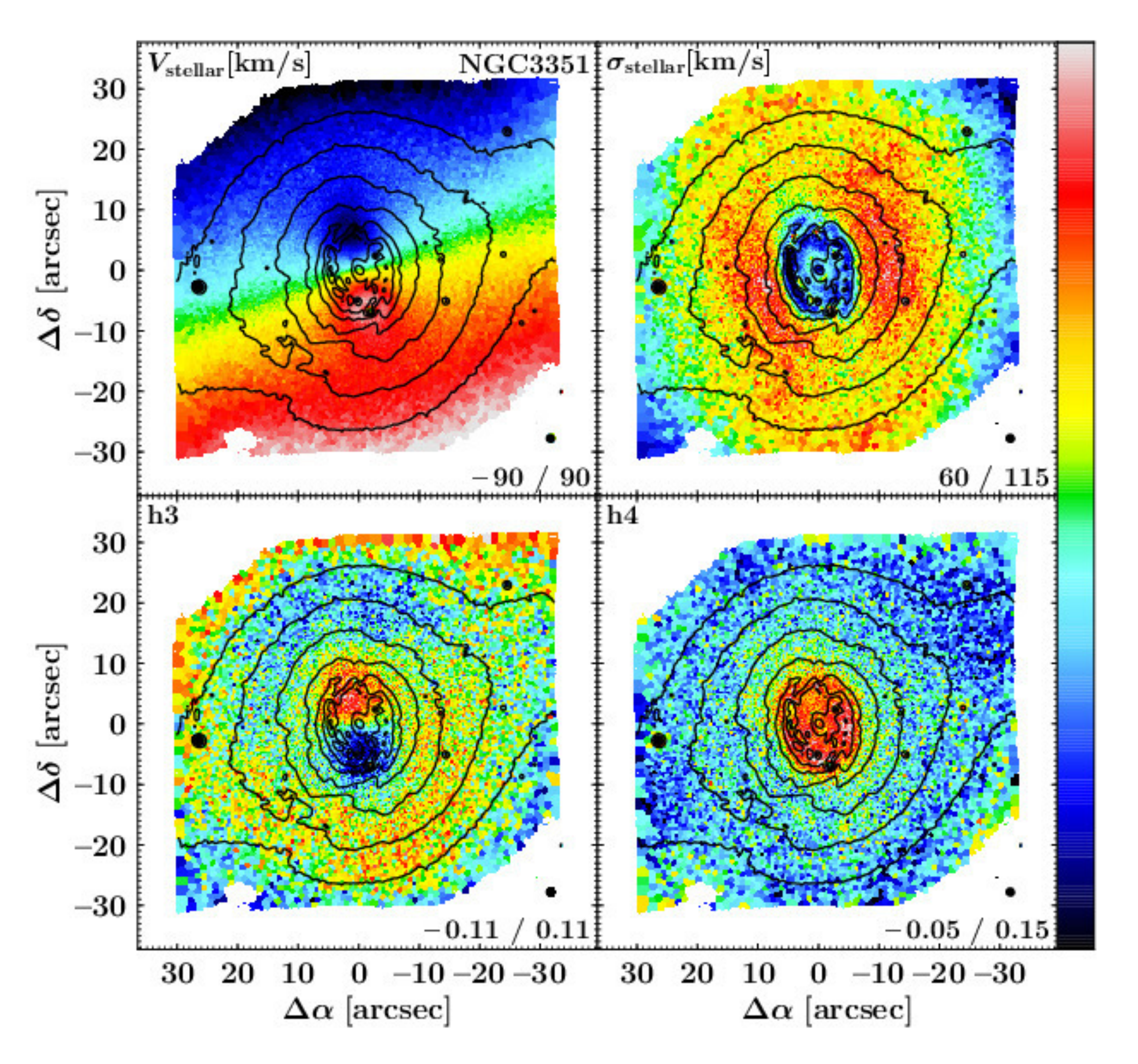}\vskip0.2cm
	\includegraphics[clip=true, trim=20 20 20 20, width=0.85\columnwidth]{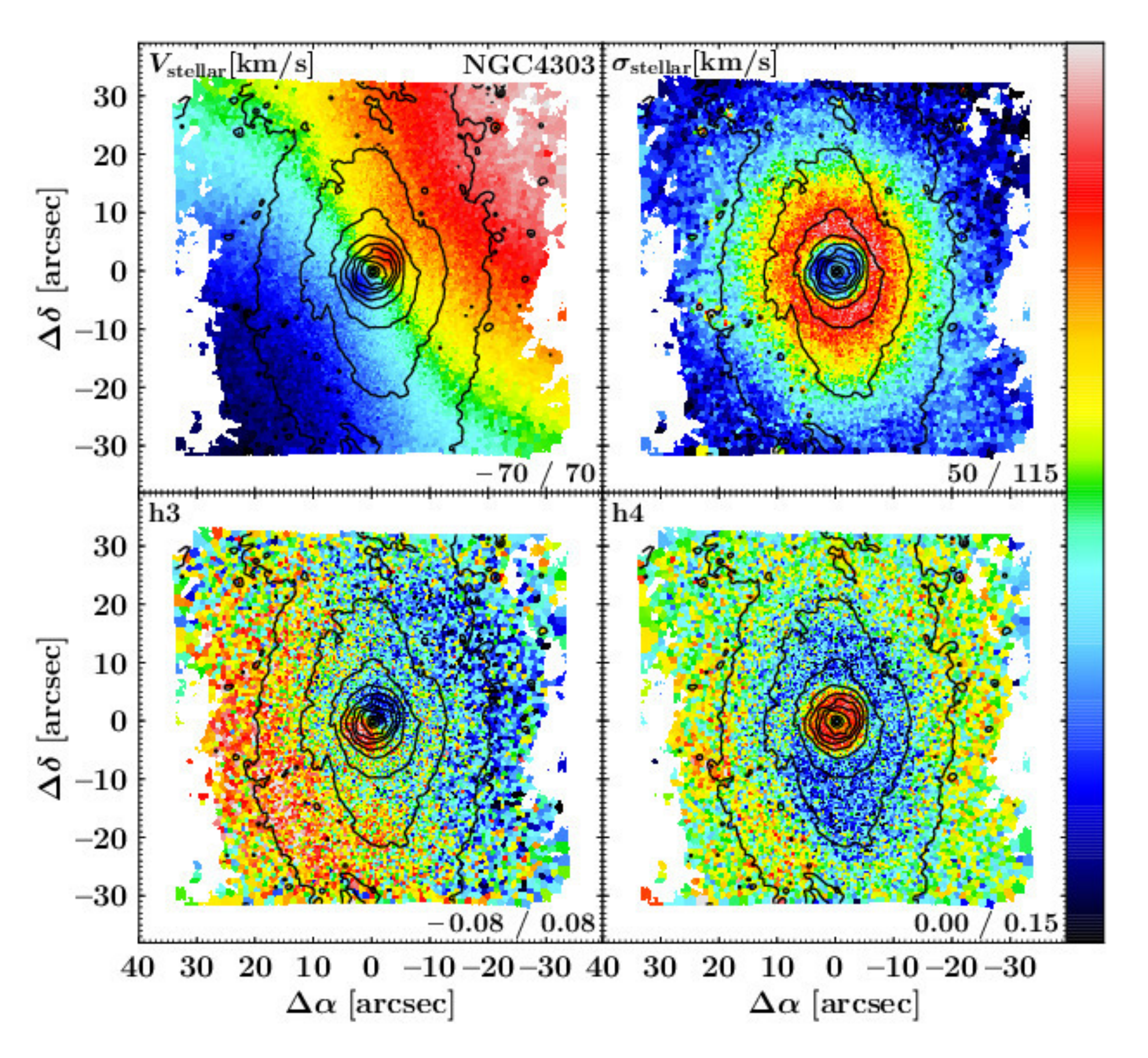}\hskip0.5cm
	\includegraphics[clip=true, trim=20 20 20 20, width=0.85\columnwidth]{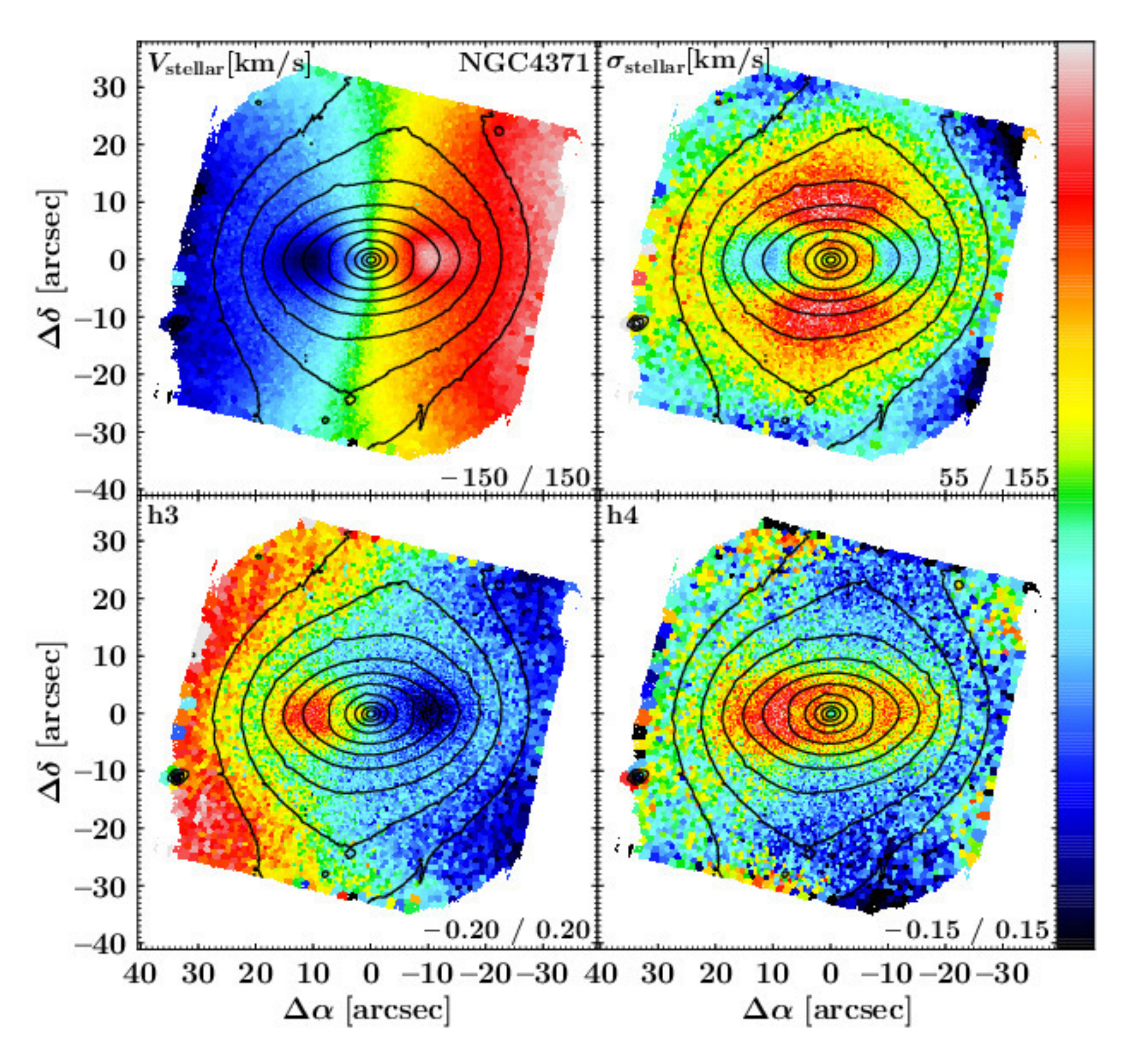}\vskip0.2cm
	\includegraphics[clip=true, trim=20 20 20 20, width=0.85\columnwidth]{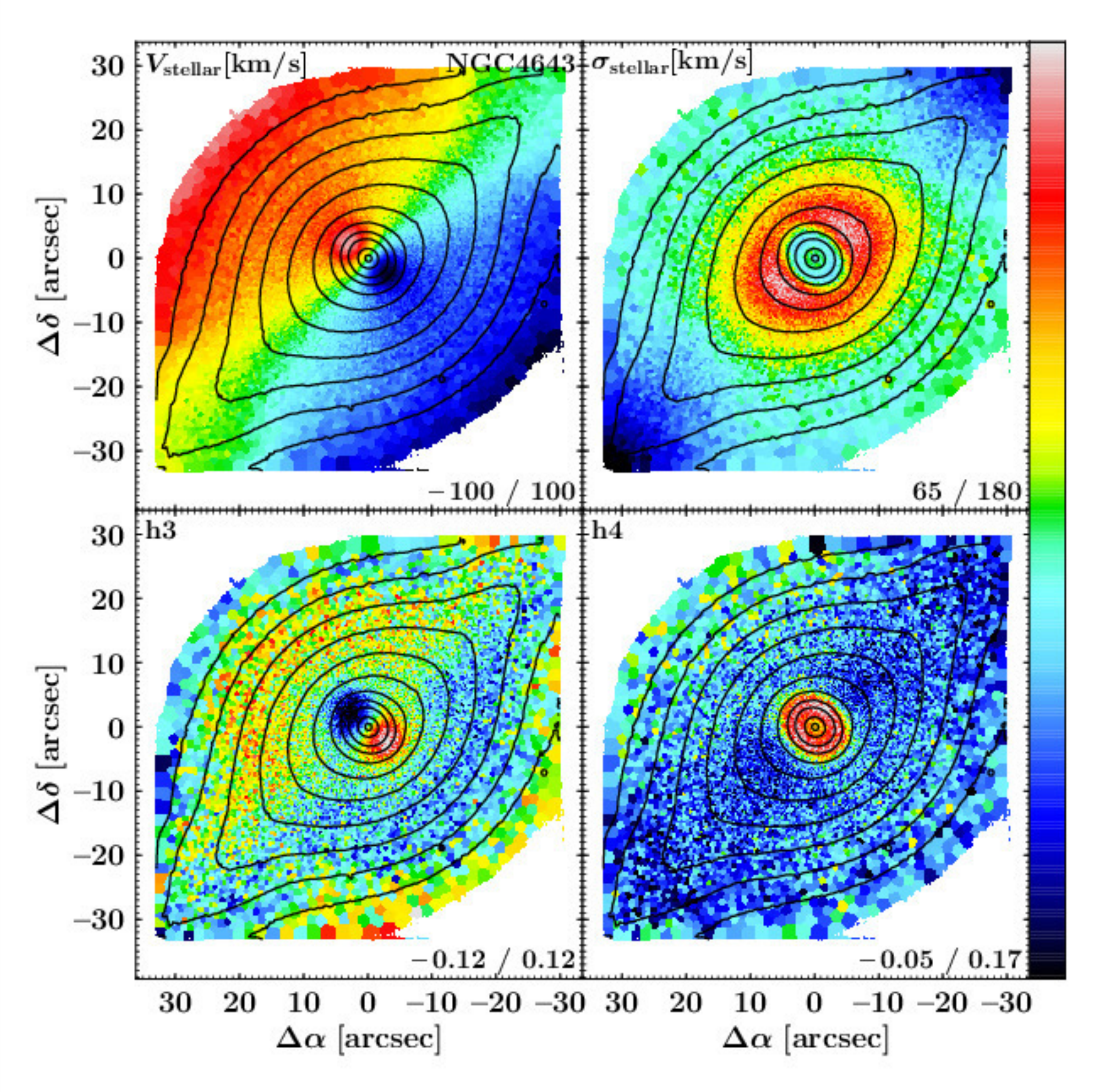}\hskip0.5cm
	\includegraphics[clip=true, trim=20 20 20 20, width=0.87\columnwidth]{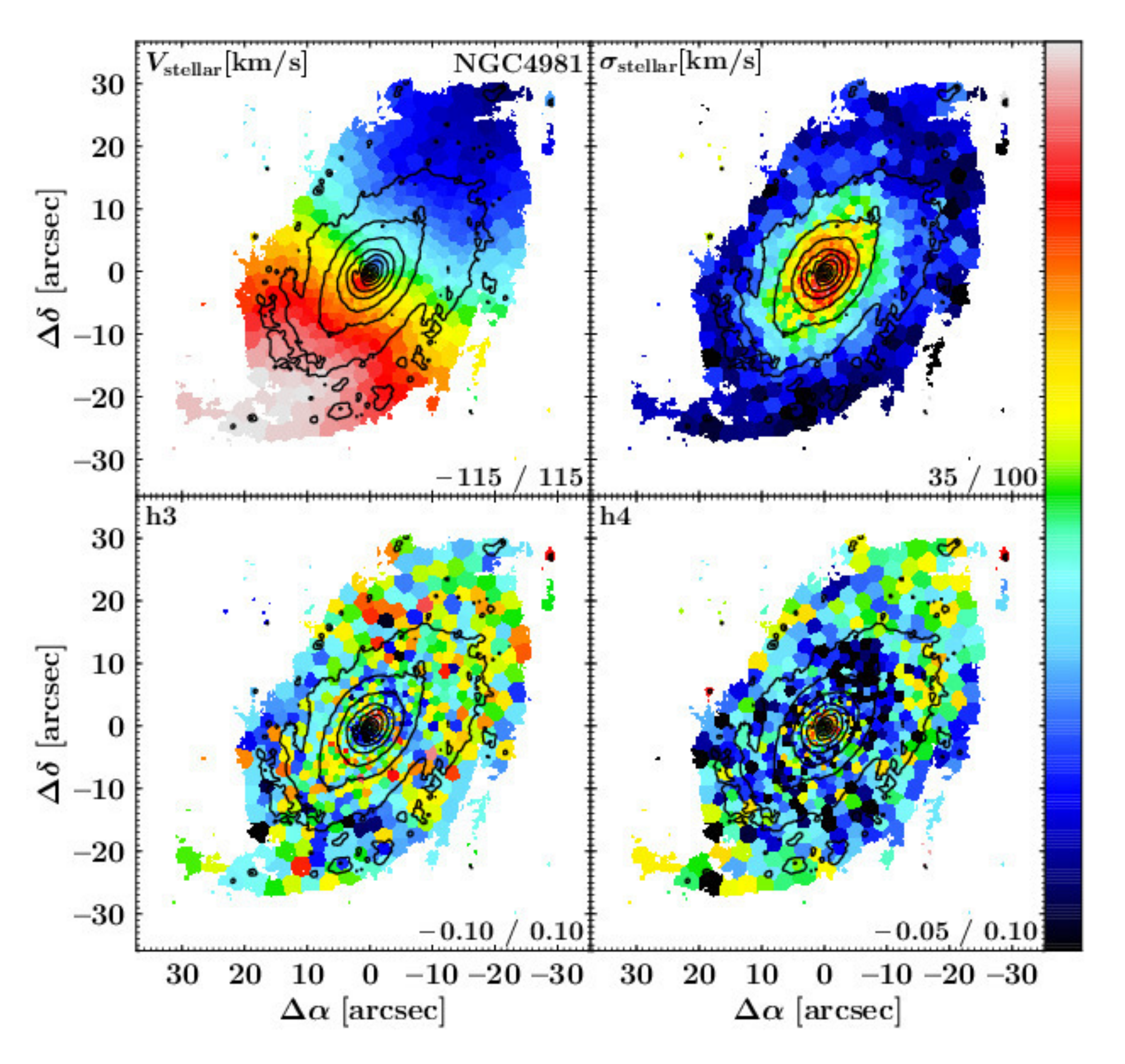}
\end{center}
\caption{Same as Fig.\,\ref{fig:maps} but for the remainder of our sample.}
\label{fig:maps2}
\end{figure*}

\begin{figure*}
\begin{center}
	\includegraphics[clip=true, trim=20 20 20 20, width=0.85\columnwidth]{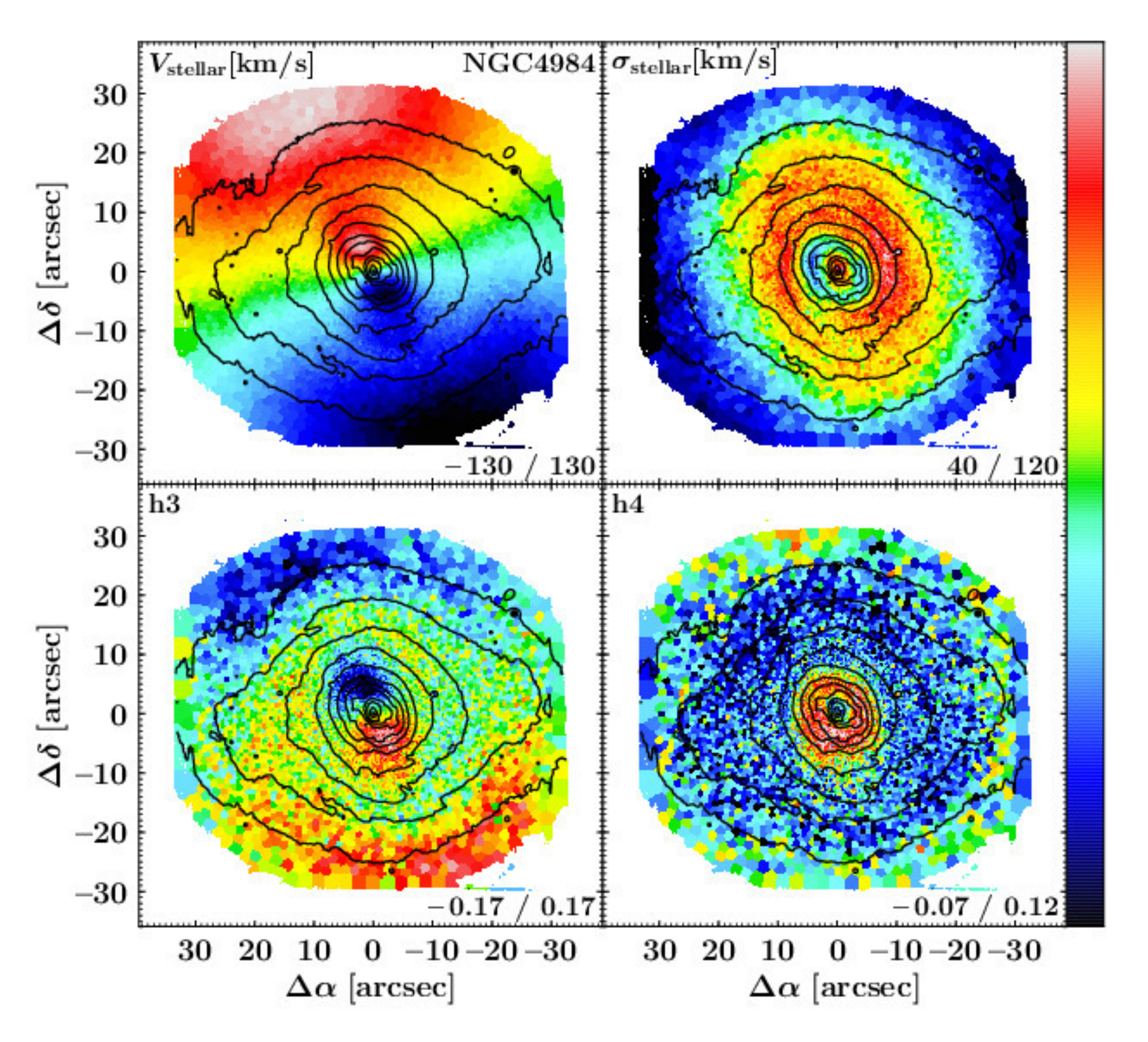}\hskip0.5cm
	\includegraphics[clip=true, trim=20 20 20 20, width=0.8\columnwidth]{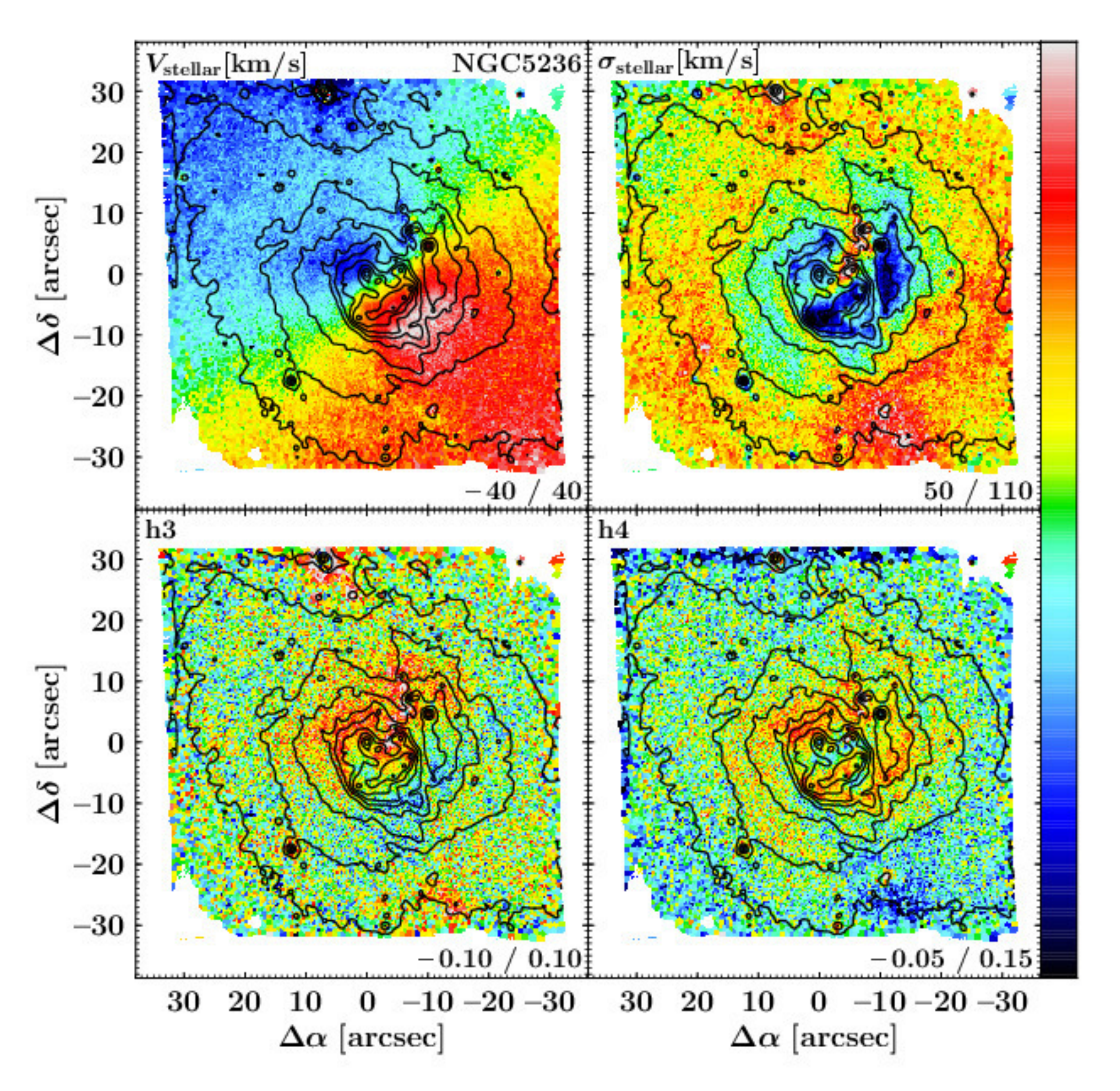}\vskip0.2cm
	\includegraphics[clip=true, trim=20 20 20 20, width=0.85\columnwidth]{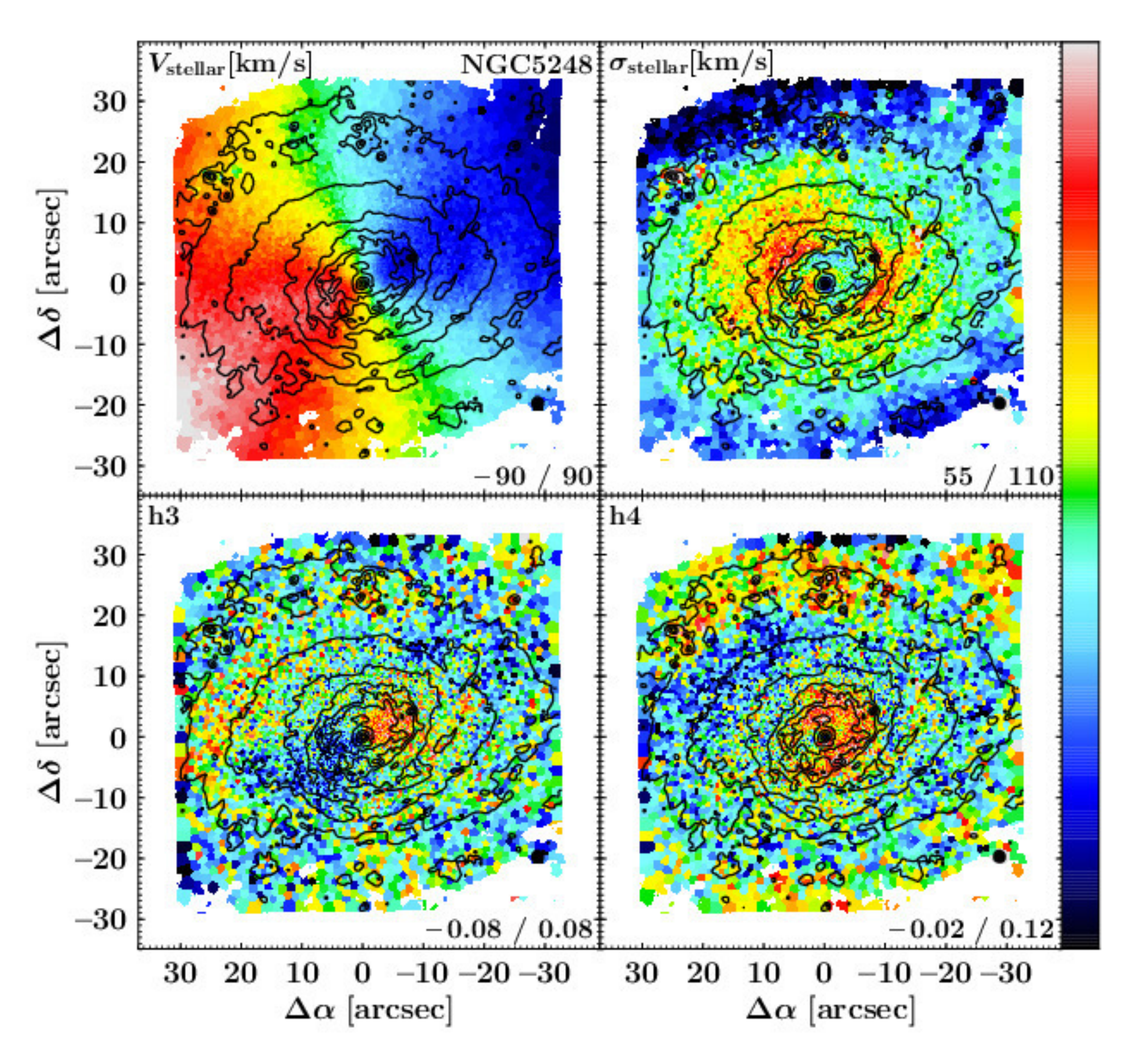}\hskip0.5cm
	\includegraphics[clip=true, trim=20 20 20 20, width=0.8\columnwidth]{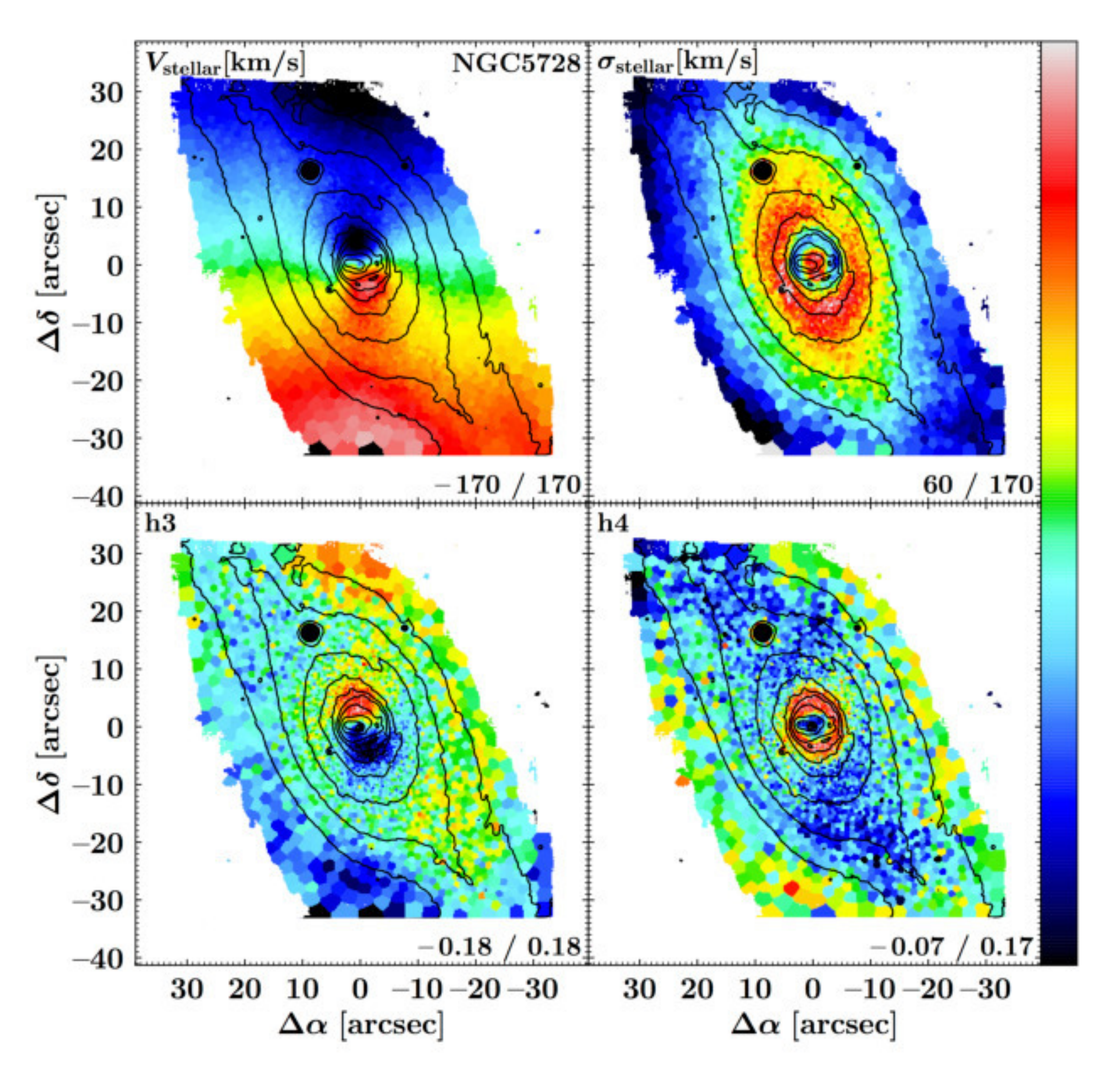}\vskip0.2cm
	\includegraphics[clip=true, trim=20 20 20 20, width=0.85\columnwidth]{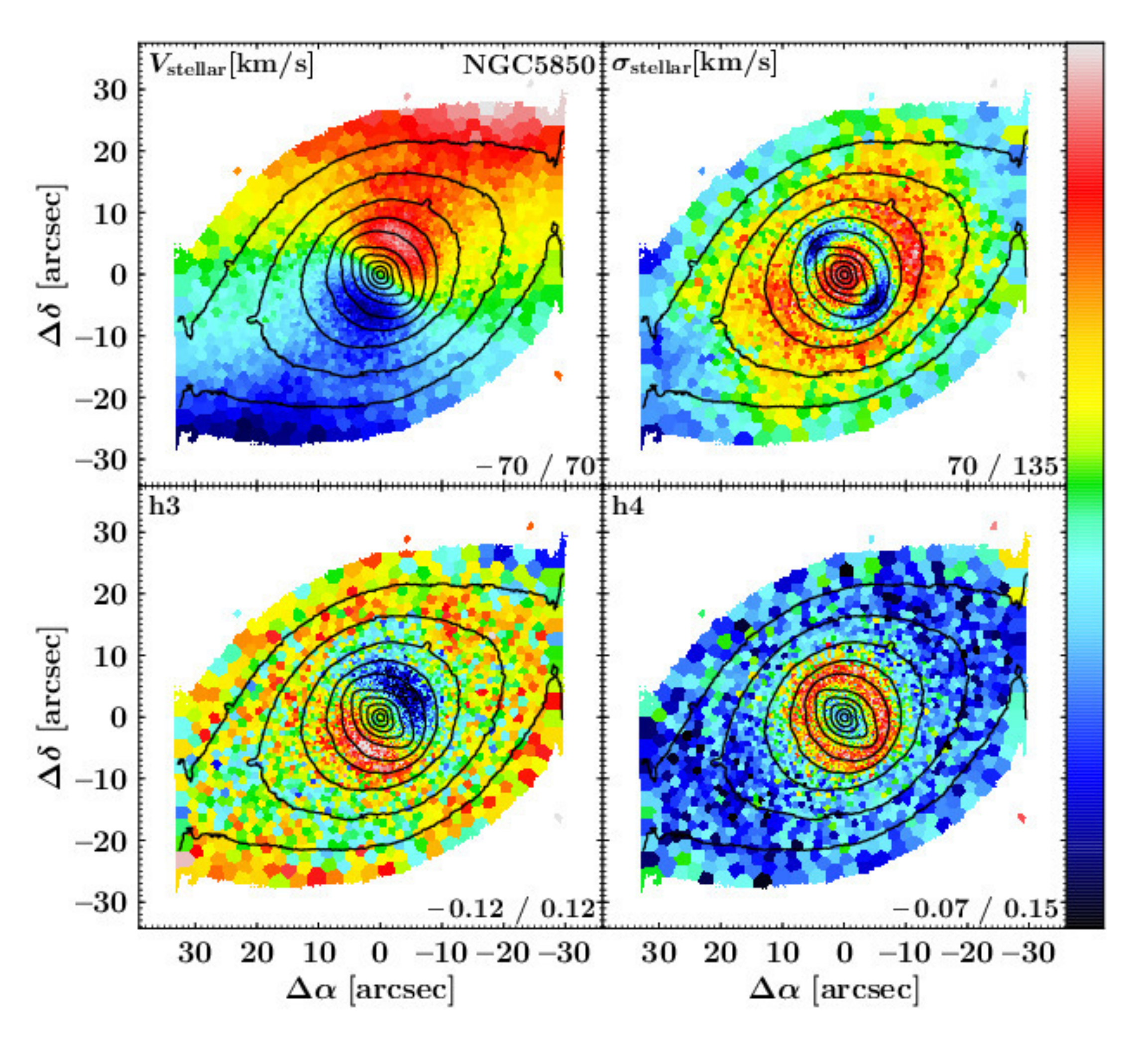}\hskip0.5cm
	\includegraphics[clip=true, trim=20 20 20 20, width=0.85\columnwidth]{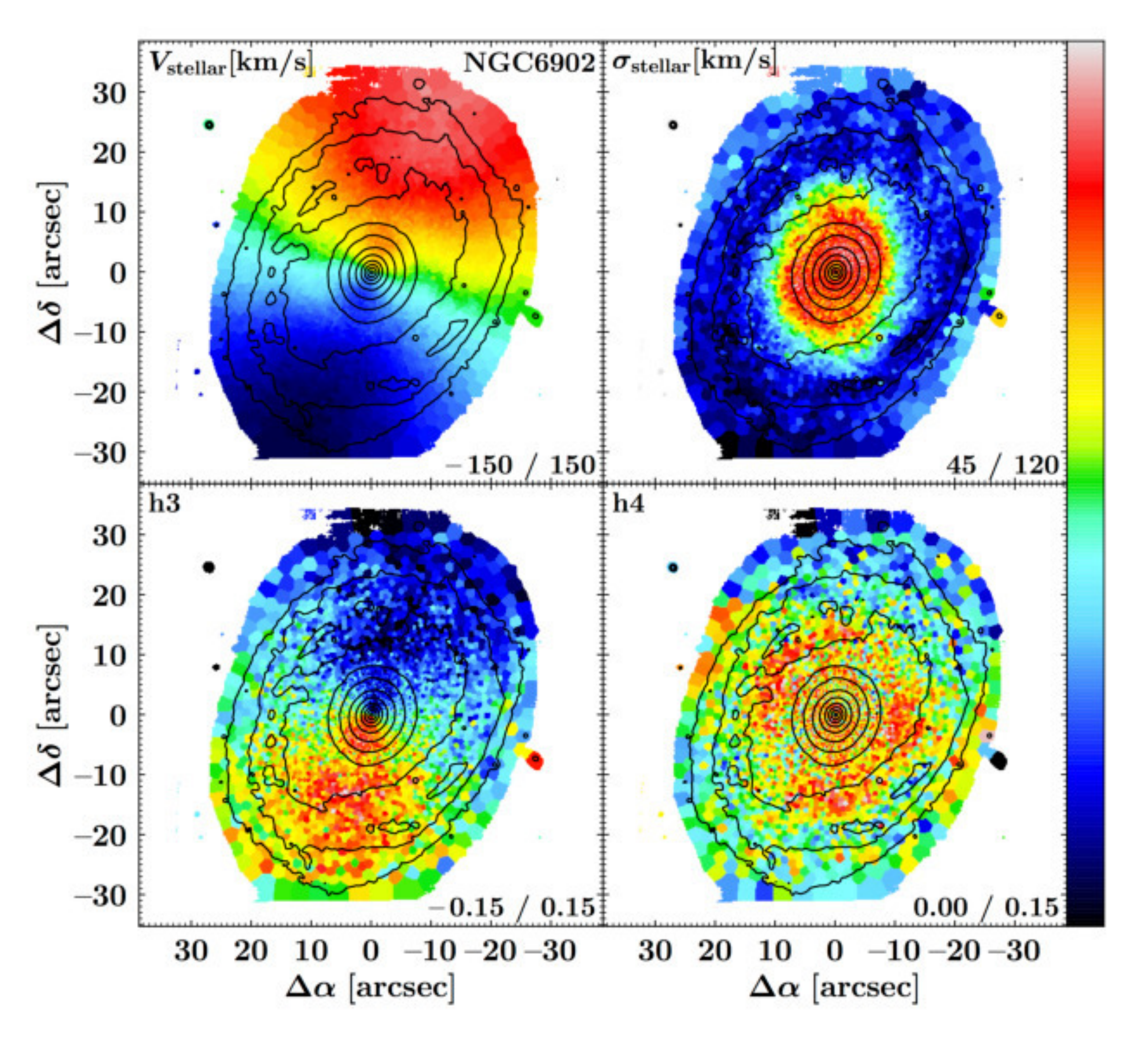}
\end{center}
\addtocounter{figure}{-1}
\caption{Continued.}
\end{figure*}

\begin{figure*}
\begin{center}
	\includegraphics[clip=true, trim=20 20 20 20, width=0.8\columnwidth]{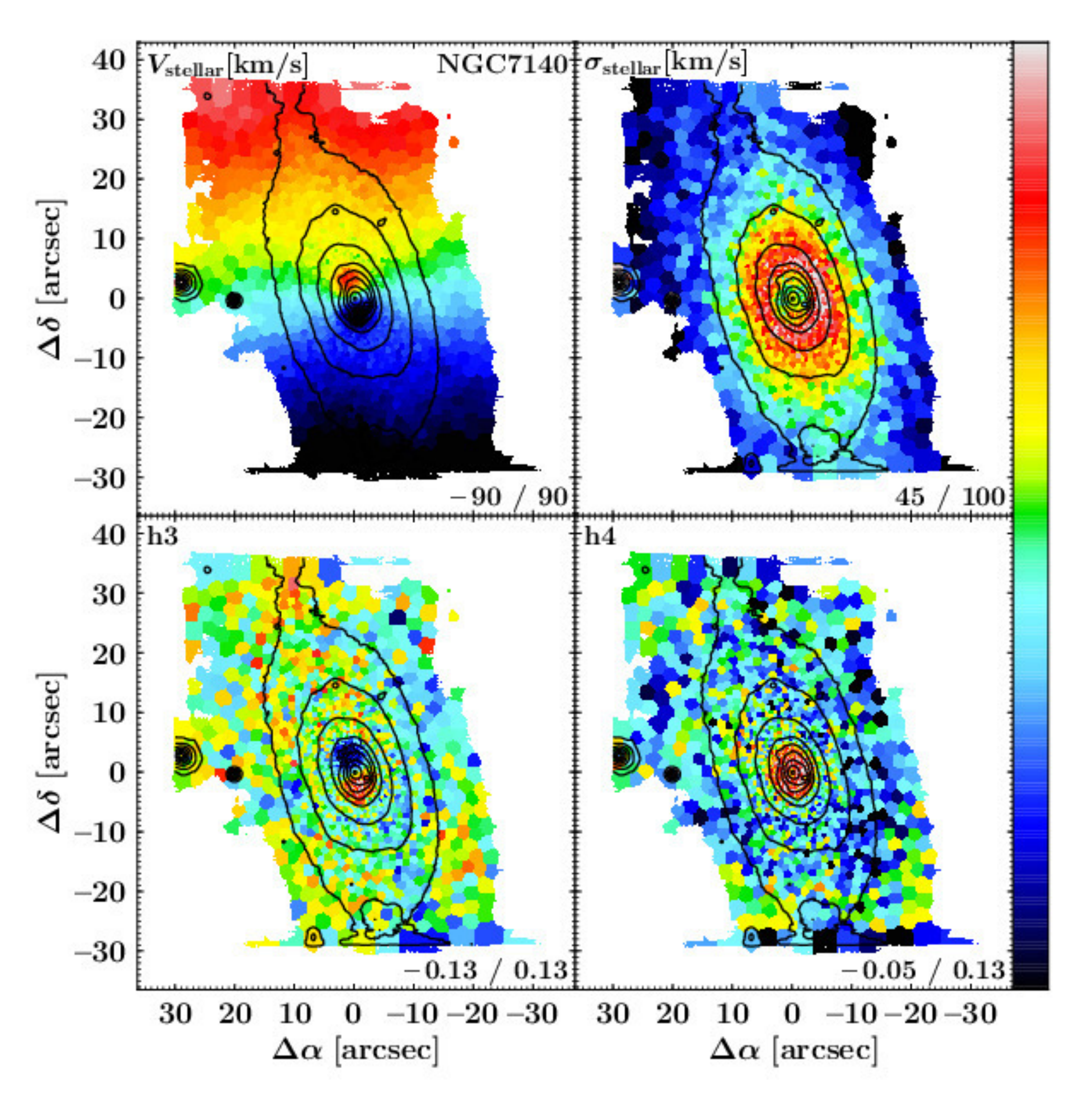}\hskip0.6cm
	\includegraphics[clip=true, trim=20 20 20 20, width=0.87\columnwidth]{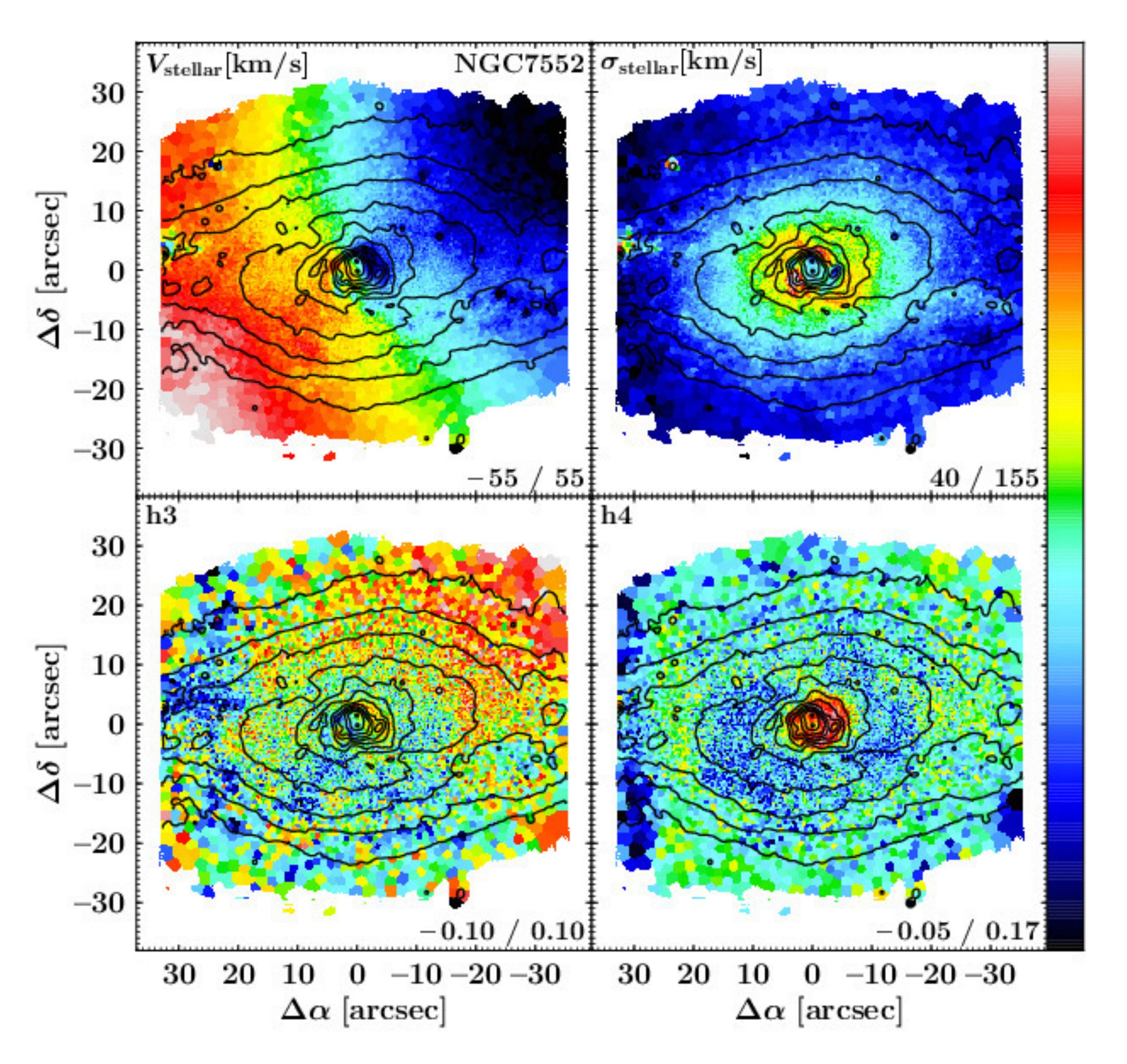}\vskip0.2cm
	\includegraphics[clip=true, trim=20 20 20 20, width=0.85\columnwidth]{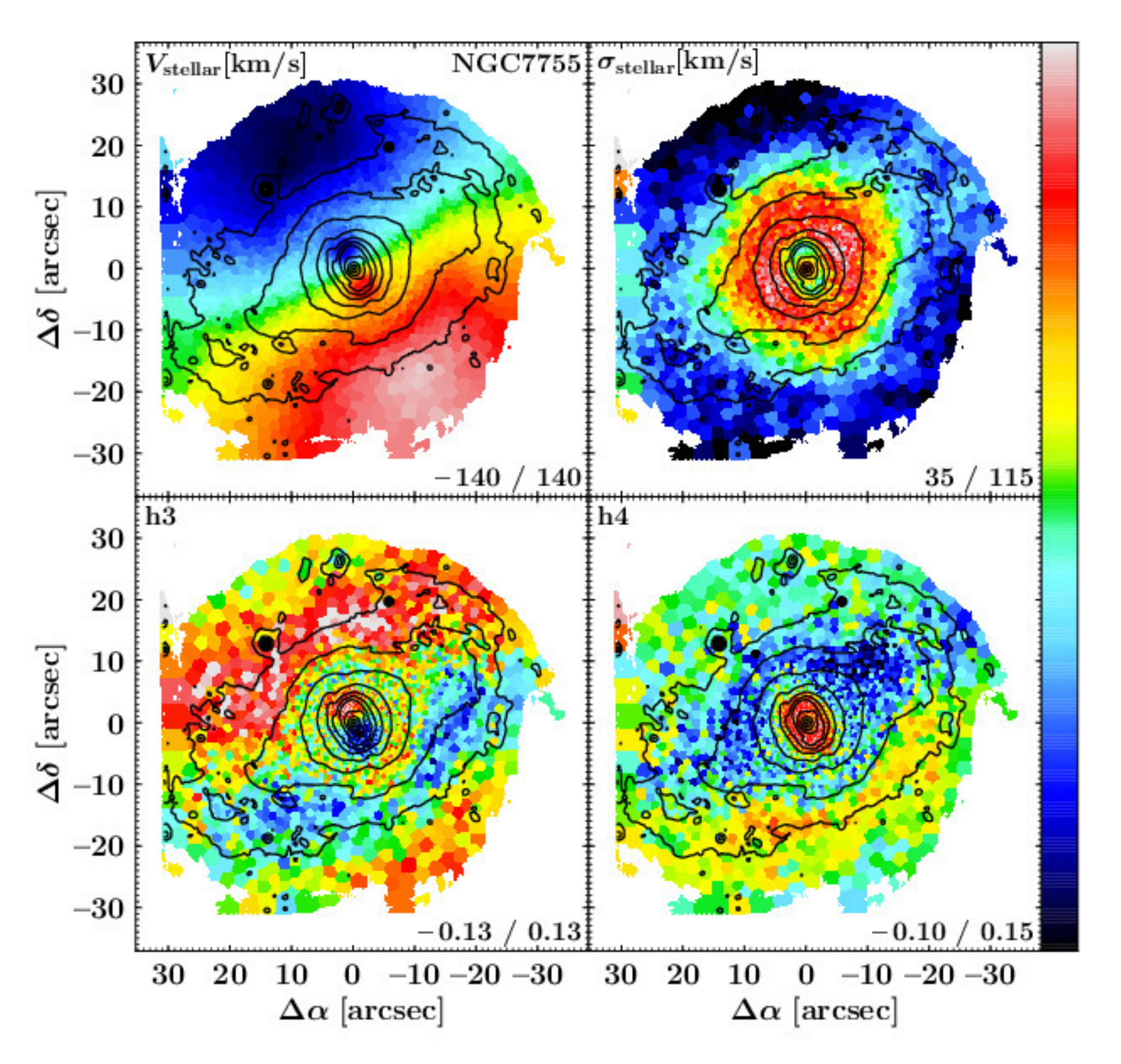}
\end{center}
\addtocounter{figure}{-1}
\caption{Continued.}
\end{figure*}

In this appendix we present in Fig.\,\ref{fig:maps2} the kinematic maps for the remainder of the sample as in Fig.\,\ref{fig:maps}. We also elaborate on exceptions to the general trends discussed in Sects.\,\ref{sec:discs}--\ref{sec:bps}, where we presented kinematic signatures of nuclear discs, bars and box/peanuts, respectively.

\subsection{Nuclear discs}

As discussed in Sect.\,\ref{sec:discs}, we find clear kinematic signatures of the presence of nuclear discs in most galaxies in this study. Apart from NGC\,1291 -- where the fact that the galaxy is very close to face-on prevents us from performing this assessment meaningfully -- there are only two {\em possible} exceptions: NGC\,1365 and NGC\,6902. In NGC\,1365, the maps of $v$ and $v/\sigma$ do not show a discontinuous behaviour and the clear presence of an additional rapidly-rotating nuclear component, even though a nuclear star-forming ring is evident in S$^4$G images, as well as in our MUSE reconstructed images (see Paper I). The galaxy is quite inclined (52$^\circ$) and our MUSE field is heavily affected by dust, which complicates the interpretation of the kinematic maps. Moreover, it hosts an active galactic nucleus with associated large scale outflow and shock-induced emission lines. Nevertheless, it shows elevated values of h$_4$ in the central region, covering about half of the MUSE field\footnote{Note, however, that elevated values of h$_4$ can be produced by radial orbits in non-rotating or slowly-rotating systems \citep[see Fig.\,2 in][]{vanFra93}.}. In addition, it shows a peculiar spiral-shaped region of elevated absolute values of h$_3$ anti-correlated with $v$. Furthermore, an extended region of low $\sigma$ is seen at an angle with the bar, which is consistent with a nuclear, kinematically cold stellar structure seen in projection. It is therefore plausible that the MUSE field is dominated by the nuclear component, which precludes one from seeing the main galaxy disc. The spiral pattern in the h$_3$ map is likely a result from the overly strong dust lanes seen along the leading edges of the bar.

On the other hand, NGC\,6902 is not heavily affected by dust and has a mild inclination angle (37$^\circ$), but it appears to host a rather modest nuclear stellar structure. The $v$ and h$_3$ maps show no clearly distinguished and rapidly-rotating nuclear component, although this is better seen in the $v/\sigma$ map. Nevertheless, the central region of the MUSE field, where the nuclear component resides, is dominated by high values of $\sigma$ rather than low values. In addition, the h$_4$ map shows a ring-shaped region of elevated values {\em around} the central region, not so much in the region dominated by the nuclear component. Therefore, the interpretation of the kinematic maps for NGC\,6902 is not as straightforward as for most of the TIMER sample. We note that NGC\,6902 has the weakest bar of the TIMER sample, in the morphological classification of \citet{ButSheAth15}.

\subsection{Bars}

We do not find the $v-$h$_3$ correlation foreseen in bars in six barred galaxies: NGC\,1097, 1291, 1365, 4371, 5236, and NGC\,6902. Apart from the latter, all these galaxies show prominent bars, but in most of these cases it is likely that the TIMER data do not cover enough of the bar to show this signature. In fact, NGC\,5236, 1291, 1097 and NGC\,1365 are, in this order, the largest projected bars in the TIMER sample, and our MUSE fields cover less than a third of the bar in these cases. The bar of NGC\,4371 is well covered by our MUSE pointing but this is the most inclined galaxy in this study and the bar is close to perpendicular to the line of nodes, and thus projection effects may be preventing us from seeing the $v-$h$_3$ correlation. As mentioned above, NGC\,6902 is the weakest bar in the sample and again appears as an exception.

\subsection{Box/peanuts}

Since our MUSE fields seem to not go much further than any box/peanut vertices, the h$_4$ drops expected for box/peanuts are not clearly present in NGC\,1097, 1365, 1433, and NGC\,3351. We also do not find this box/peanut kinematic signature in NGC\,4371, but in this case this is possibly due to projection effects. The bar is seen inclined around an axis close to its minor axis, and the galaxy inclination angle is relatively large (at 59$^\circ$, this galaxy is the most inclined galaxy in this study).

In NGC\,5236 the signature is not so clear. One sees regions of low h$_4$ on average along the bar major axis but the MUSE field seems to not be large enough to show where h$_4$ would rise again. NGC\,5248 shows h$_4$ minima but it is not clear how they are associated to the rather weak bar, which in addition becomes harder to distinguish due to projection effects.

NGC\,6902 is again an interesting case. As discussed above, it is the weakest bar in TIMER and there are no clear kinematic signatures of the presence of a nuclear disc. Its weak bar also does not show in the kinematic maps. Tentatively, there is weak evidence for a box/peanut, since inside the ring of elevated values of h$_4$ mentioned above one sees dips to values close to zero along the bar major axis.

Finally, NGC\,4643 shows a curious behaviour. Along the bar major axis outwards, h$_4$ drops to a minimum after the central region of elevated values, further out rises again reaching values close to zero, and then drops again to very low values towards the end of the bar. This behaviour would be produced if the galaxy has an inner bar with its own box/peanut (as does NGC\,1291) plus the box/peanut of the primary bar. However, this cannot be the case here as the inner bar should be within the nuclear disc radius, but the first pair of h$_4$ minima is outside that radius. In addition, the two pairs of h$_4$ minima are very well aligned, suggesting that the properties they indicate concern only the (main) bar. To understand the presence of two pairs of h$_4$ minima in NGC\,4643 is beyond the scope of this paper.

\section{Statistical analysis of the trends between r$_{\rm k}$ and selected bar properties (Fig.\,\ref{fig:barsQb})}
\label{app:stats}

\begin{table*}[b]
	\centering
	\caption{Slope (with standard deviation) and correlation coefficient corresponding to the relations shown in Fig.\,\ref{fig:barsQb}, as determined through various statistical methods as indicated (see main text for details).}
	\label{tab:stats}
	\begin{tabular}{lcccccccc}
		\hline
Relation & OLS($Y|X$) & OLS($X|Y$) & OR & OLS bi & RLS & Pearson & Spearman & RLS\\
\omit & slope & slope & slope & slope & slope & CC & CC & CC\\
		\hline
r$_{\rm k}\ \times$ R$_{\rm bar}$       &  0.065$\pm$0.018   &  0.119$\pm$0.020      & 0.065$\pm$0.018     & 0.092$\pm$0.017   & 0.051$\pm$0.009 & 0.74 & 0.77 & 0.91 \\
r$_{\rm k} \times \epsilon_{\rm bar}$  &  -1.170$\pm$0.472  &  -4.027$\pm$1.423     & -3.461$\pm$1.318   & -1.943$\pm$0.495 & -1.170$\pm$0.578 & -0.54 & -0.59 & -0.54 \\
r$_{\rm k}\ \times$ Bar/T                     &  1.339$\pm$0.792   &  5.451$\pm$2.116      & 4.908$\pm$1.776     & 2.292$\pm$0.760  & 1.339$\pm$0.742  & 0.50  & 0.56  & 0.50 \\
r$_{\rm k}\ \times$ Q$_{\rm B}$          &  -0.262$\pm$0.412  &  -11.887$\pm$18.039 & -8.187$\pm$12.157 & -1.189$\pm$0.320 & -0.135$\pm$0.378 & -0.15 & -0.11 & -0.09 \\
r$_{\rm k}\ \times$ A$_2$                   &  0.360$\pm$0.169    &  2.826$\pm$1.764     & 1.023$\pm$0.676     & 1.005$\pm$0.148  & 0.392$\pm$0.149  & 0.36  & 0.44  & 0.57 \\
		\hline
	\end{tabular}
\end{table*}

In order to better understand the statistical significance of the relations presented in Fig.\,\ref{fig:barsQb} and discussed in Sect.\,\ref{sec:origins}, we present in Table\,\ref{tab:stats} the values of the slope and correlation coefficient as determined through different methods for each relation. We employed the SIXLIN IDL implementation of the formulae provided by \citet{IsoFeiAkr90} to calculate the slopes through ordinary least squares regression (OLS) and orthogonal regression (OR). OLS can be calculated with the $Y$ variable (in this case, r$_{\rm k}$) as the dependent variable ($Y|X$) or as the independent variable ($X|Y$). The slope of the bisector of the region between the two OLS lines is also shown.

\citet{IsoFeiAkr90} discussed how, by definition, OLS($Y|X$) and OLS($X|Y$) can often lead to different results, why the former is typically preferred over the latter, and why OR should only be used with scale-free variables. The authors then recommend that the OLS bisector estimate is to be preferred, in particular when the goal is to probe the underlying relation between two variables.

Table\,\ref{tab:stats} also shows the slopes derived using the reweighted least squares regression (RLS) of \citet{Rou84}, which is particularly robust against contamination from outliers. We used the code PROGRESS \citep{RouLer87} to calculate the RLS slopes, as well as three correlation coefficients also shown in Table\,\ref{tab:stats}: the Pearson correlation coefficient, the Spearman rank correlation coefficient and the RLS correlation coefficient.

One sees that the correlation between r$_{\rm k}$ and R$_{\rm bar}$ is strong: the RLS correlation coefficient is particularly strong and the slope derived via the different methods is relatively stable (within the expected differences). The r$_{\rm k}\ \times$ A$_2$ relation also sees a significant increase in the Spearman and RLS correlation coefficients, as compared to the Pearson coefficient. However, the three correlation coefficients are similar for the relations between r$_{\rm k}$ and $\epsilon_{\rm bar}$ and Bar/T. The relatively large variation between the slopes derived with OLS($Y|X$) and the OLS bisector indicates that the correlations between r$_{\rm k}$ and A$_2$, $\epsilon_{\rm bar}$ and Bar/T are only moderately significant. As already mentioned in Sect.\,\ref{sec:origins}, there is no significant correlation between r$_{\rm k}$ and Q$_{\rm B}$, and this is shown again by the results in Table\,\ref{tab:stats}.

We stress again that further work is necessary to fully probe and understand these relations, in particular with larger samples. While the correlation between r$_{\rm k}$ and R$_{\rm bar}$ is strong, and the trends between r$_{\rm k}$ and $\epsilon_{\rm bar}$, Bar/T and A$_2$ do not seem fortuitous, these are not {\em proofs} that nuclear discs are built by bars. However, these relations are consistent with that picture, and there is currently no obvious reason to think that a scenario in which bars are irrelevant to the formation of nuclear discs would produce such relations.

\end{document}